\documentclass[12pt]{article}
\usepackage{amsmath,amscd,amsfonts,eucal,latexsym,amssymb,mathrsfs}
\usepackage{amsxtra, amsthm, calc}
\newlength{\dinwidth}
\newlength{\dinmargin}
 \newtheorem{Definition}{Definition}[section]
 \newtheorem{Theorem}[Definition]{Theorem}
 \newtheorem{Proposition}[Definition]{Proposition}
 \newtheorem{Lemma}[Definition]{Lemma}
 \newtheorem{Corollary}[Definition]{Corollary}

\newsymbol\rest 1316         %% restriction symbol
\def\oere{{o\!\!\!/\hspace*{-0.1pt}}}

\def\cA{{\cal A}}

\def\cD{{\cal D}}

\def\cF{{\cal F}}

\def\cH{{\cal H}}
\def\cI{{\cal I}}

\def\cK{{\cal K}}

\def\cN{{\cal N}}

\def\cP{{\cal P}}

\def\cS{{\cal S}}
\def\cT{{\cal T}}

\def\bC{{\mathbb C}}

\def\bI{{\mathbb I}}

\def\bN{{\mathbb N}}

\def\bR{{\mathbb R}}

\def\bZ{{\mathbb Z}}

\def\a{\alpha}
\def\b{\beta}
\def\g{\gamma}        
\def\d{\delta}        \def\D{\Delta}
\def\eps{\varepsilon}

\def\k{\kappa}
\def\l{\lambda}       \def\L{\Lambda}
\def\m{\mu}
\def\n{\nu}

\def\p{\pi}
\def\r{\rho}

\def\o{\omega}        \def\O{\Omega}

\def\fA{{\mathfrak A}}
\def\fB{{\mathfrak B}}

\def\fF{{\mathfrak F}}
\def\fG{{\mathfrak G}}

\def\fP{{\mathfrak P}}

\def\imply{\Rightarrow}

\def\supp{{\text{supp}}}

\def\Lpo{{\cal L}^{\uparrow}_+} % -- Lorentz group
\def\Lpocov{\tilde{\cal L}^{\uparrow}_+} % -- covering of Lorentz group

\newcommand{\uF}{\underline{F}}
\newcommand{\uA}{\underline{A}}
\newcommand{\uFl}{\underline{F}_{\lambda}}
\newcommand{\uFF}{\underline{\mathfrak F}}
\newcommand{\uAA}{\underline{\mathfrak A}}
\newcommand{\ua}{\underline{\alpha}}
\newcommand{\uaa}{\underline{\alpha}_a}
\newcommand{\ala}{\alpha_{\lambda a}}
\newcommand{\Foi}{\mathcal{F}_{0,\iota}}
\newcommand{\Aoi}{\mathcal{A}_{0,\iota}}
\newcommand{\uo}{\underline{\omega}}
\newcommand{\uol}{\underline{\omega}_{\lambda}}
\newcommand{\ub}{\smash{\underline{\beta}}}
\newcommand{\aoi}{\alpha^{(0, \iota)}}
\newcommand{\boi}{\beta^{(0,\iota)}}
\newcommand{\poi}{\pi_{0,\iota}}
\newcommand{\Hoi}{{\cal H}_{0,\iota}}
\newcommand{\Ooi}{\Omega_{0,\iota}}
\newcommand{\ooi}{\omega_{0,\iota}}
\newcommand{\FFoi}{\mathfrak{F}_{0,\iota}}
\newcommand{\AAoi}{\mathfrak{A}_{0,\iota}}
\newcommand{\Uscoi}{{\mathscr U}_{0,\iota}}
\newcommand{\Uoi}{U_{0,\iota}}
\newcommand{\Noi}{N_{0,\iota}}
\newcommand{\Goib}{G_{0,\iota}^{\bullet}}
\newcommand{\Uoib}{U_{0,\iota}^{\bullet}}
\newcommand{\gb}{g^{\bullet}}
\newcommand{\goib}{g_{0,\iota}^{\bullet}}
\newcommand{\kb}{k^{\bullet}}
\newcommand{\koib}{k_{0,\iota}^{\bullet}}
\newcommand{\fbs}{\boldsymbol{f}}
\newcommand{\Usc}{{\mathscr U}}
\newcommand{\tU}{\tilde{\mathscr U}}
\newsymbol\bt 1202  % \boxtimes
\newcommand{\lcrc}{\mbox{\footnotesize $\,\,\,\circ\,\,\,$}}

\newcommand{\norm}[1]{\left\lVert #1 \right\rVert}
\newcommand{\betr}[1]{\left\lvert #1 \right\rvert}

\newcommand{\hide}[1]{} % make invisible

\newcommand{\gammu}{\gamma^{\,\mu}}

\newcommand{\gamze}{\gamma^{\,0}}
\newcommand{\gamtwo}{\gamma^{\,2}}
\newcommand{\omp}{\o_m(\spp)}
\renewcommand{\slash}[1]{\negthickspace\not\!#1\,}
\newcommand{\ps}{\slash{p}}
\newcommand{\de}{\partial}
\DeclareMathOperator{\Imag}{Im}
\newcommand{\angscalar}[2]{\langle#1,#2\rangle}
\newcommand{\spp}{\boldsymbol{p}}
\newcommand{\abs}{\betr}
\newcommand{\SOo}{\Lpo}
\newcommand{\Pg}{\cP}

\newcommand{\Pport}{\Pg_+^\uparrow}
\newcommand{\rPport}{\tilde{\Pg}_+^\uparrow}

\newcommand{\Vpc}{\overline{V}_+}

\newcommand{\cont}{\subseteq}
\newcommand{\FcO}{\cF(O)}

\newcommand{\unitequiv}{\cong}
\newcommand{\bignorm}[1]{\big\lVert#1\big\rVert}
\newcommand{\bigabs}[1]{\big\lvert#1\big\rvert}
\newcommand{\Bigabs}[1]{\Big\lvert#1\Big\rvert}
\newcommand{\Avac}{\cA_{\text{vac}}}
\newcommand{\AvacO}{\Avac(O)}
\newcommand{\Uvac}{\mathscr{U}_{\text{vac}}}
\newcommand{\fAvac}{\fA_{\text{vac}}}
\newcommand{\Hvac}{\cH_{\text{vac}}}

\newcommand{\fFtoix}{\tilde{\fF}_{0,\iota}^{\times}}
\newcommand{\cFtoix}{\tilde{\cF}_{0,\iota}^{\times}}

\newcommand{\Wop}{W_R}
\newcommand{\LaWo}{\L_{\Wop}}
\newcommand{\Lx}{ (\Lambda , x) }                %(Lambda,x)
\newcommand{\Id}{{\bf 1}}

\newcommand{\poix}{\pi_{0,\iota}^{\times}}
\newcommand{\FtzCC}{\fFtoix(C_1,\dots,C_n)}
\newcommand{\GzCC}{\fG_{0,\iota}^{\times}(C_1,\dots,C_n)}
\newcommand{\Hoix}{\cH_{0,\iota}^{\times}}
\newcommand{\tLWd}{\widetilde{\L_W(\cdot)}}
\newcommand{\tLWt}{\widetilde{\L_W(2\pi t)}}
\newcommand{\nnorm}[1]{\lVert #1 \rVert}
\newcommand{\nabs}[1]{\lvert #1 \rvert}
\newcommand{\sF}{\mathsf{F}}
\newcommand{\sFl}{\sF(\l)}
\newcommand{\uFx}{\underline{\fF}^\times}

\newcommand{\uFCO}{\uFF(C,O)}
\newcommand{\uAl}{\uA_\l}
\newcommand{\Foix}{\cF_{0,\iota}^{\times}}
\newcommand{\Uscoix}{\mathscr{U}_{0,\iota}^{\times}}
\newcommand{\Uoix}{U_{0,\iota}^{\times}}

\newcommand{\aoix}{\a^{(0,\iota)\times}}
\newcommand{\Joix}{J_{0,\iota}^\times}
\newcommand{\Doix}{\D_{0,\iota}^\times}
\newcommand{\Joi}{J_{0\iota}}
\newcommand{\Doi}{\D_{0,\iota}}
\newcommand{\utF}{\smash[b]{\underset{\sim}{\smash{\fF}}}}

\newcommand{\hsF}{\hat{\sF}}
\newcommand{\psC}{\psi^{\hat{C}_1}}
\newcommand{\psbs}{\boldsymbol{\psi}}

\newcommand{\aabs}{\boldsymbol{a}}
\newcommand{\bbs}{\boldsymbol{b}}
\newcommand{\cbs}{\boldsymbol{c}}
\newcounter{propcount}

\newlength{\maxlabelwidth}

\begin{document}
\noindent
\begin{center}
{ \Large \bf Scaling algebras for charged fields and 
short-distance\\[6pt]
 analysis for localizable and topological charges}
\\[30pt]
{\large \sc Claudio D'Antoni$\,{}^*$\footnote{supported by
    MIUR, INDAM-GNAMPA, and the EU}\\[6pt]
   Gerardo Morsella ${}^{**}\,{}^1$\\[11pt]
Rainer Verch${}^{***}$}
\end{center}
${}$\\
                 ${}^*$\,Dipartimento di Matematica,
                 Universit\`a di Roma ``Tor Vergata'',
                 Via della Ricerca Scientifica,
                 I-00133 Roma, Italy ---
                 e-mail: dantoni$@$mat.uniroma2.it
                 \\[10pt]
                 ${}^{**}$\,Dipartimento di Matematica, Universit\`a 
di Roma
                 ``La Sapienza'', P.le Aldo Moro 2, I-00185 Roma, 
Italy --- 
                 e-mail: morsella$@$mat.uniroma1.it
                 \\[10pt]
                 ${}^{***}$\,Max-Planck-Institut for 
                 Mathematics in the Sciences,
                 Inselstr.\ 22,
                 D-04103 Leipzig, Germany ---
                 e-mail: verch$@$mis.mpg.de
${}$\\[26pt]
{\small {\bf Abstract. }     
 The method of scaling algebras, which has been introduced earlier
as a means for analyzing the short-distance behaviour of quantum field
theories in the setting of the model-independent, operator algebraic
approach, is extended to the case of fields carrying superselection
charges.
In doing so, consideration will be given to strictly localizable 
charges
(``DHR-type'' superselection charges) as well as to charges which can 
only
be localized in regions extending to spacelike infinity (``BF-type''
superselection charges).
 A criterion for the preservance of superselection charges in
the short-distance scaling limit is proposed. Consequences of this
preservance of superselection charges are studied. The conjugate
charge of a preserved charge is also preserved, and 
for charges of DHR-type, the preservance of
all charges of a quantum field theory in the scaling limit leads to
equivalence of local and global intertwiners between superselection 
sectors.
 }
${}$\\[10pt]
\section{Introduction}
In an attempt to analyze the short-distance behaviour of quantum field
theories in a completely model-independent manner, and to have a
counterpart of renormalization group analysis at short length scales
in the setting of general quantum field theory, so-called ``scaling
algebras'' have been introduced some time ago \cite{BV1}. The idea of 
this
approach is to associate to a given quantum field theory described in
terms of local observable algebras \cite{HK,Haag} a ``scaling
algebra'' of functions depending on a scaling parameter $\l > 0$ and
taking values in the local observable algebras. These functions are
required to have certain properties regarding their localization and
energy behaviour as $\l$ tends to zero; roughly speaking, the values
of the functions at scale parameter $\l$ should be observables
localized in spacetime regions of extension proportional to $\l$, and
having energy-momentum transfer proportional to $\l^{-1}$.

The collection of all these functions, i.e.\ of all the members of the
scaling algebra, may hence be viewed as ``orbits'' of elements in the
local observable algebras under all possible renormalization group
transformations. By studying the vacuum expectation values of these
functions in the limit $\l \to 0$ (the ``scaling limit''), one can
then analyze the extreme short distance properties of the given
quantum field theory.

This programme, initiated in \cite{BV1}, has been further developed in
\cite{Bu.QGC,BV2,Bu.phsp,Mohrdieck}. It leads to a general
classification of the short distance behaviour of the given theory
which corresponds to the one known in perturbation theory where one
distinguishes theories with stable ultraviolet fixed points under
renormalization group transformations, as opposed to others with
unstable fixed points or no fixed points at all \cite{BV1}.

Moreover, it permits to give a criterion as to when a given quantum
field theory possesses ``confined charges'' which are only visible in
the extreme short distance limit while they are absent at finite
scale, like the colour charge in QCD \cite{Bu.Conf,Bu.QGC}.
According to this criterion  a
 charge is confined if it arises as a
superselection charge in the scaling limit theory of the observables
which is not a scaling limit of the superselection charges of the 
original
theory at scale $\lambda = 1$ (see Sec.\ 5 for discussion).
 The effectiveness of this
criterion has been illustrated in the example of the two-dimensional
Schwinger model \cite{Bu.QGC,BV2}.

However, with the exception of ref.\ \cite{Morsella1}, the
scaling algebra method has up to now only been applied in the setting
of local observable algebras, not in the context of local field
algebras containing charge-carrying local field algebras. In other
words, this method has not yet been applied to studying the 
short-distance behaviour of superselection charges (see
\cite{Rob.lec,Haag} and references cited there) and their
corresponding charge-carrying fields.

In the present work, we generalize the ``scaling algebra'' framework
in the setting of algebraic quantum field theory in the presence of
field operators transforming non-trivially under the action of a
(global) compact gauge group.
We consider separately two cases where the field operators can be
localized (1) in bounded spacetime regions and (2) in regions
extending to spacelike infinity (so-called ``spacelike cones'').
The first case corresponds to superselection charges which can be 
localized
in arbitrary bounded regions of spacetime (``DHR-charges'') while the
second case corresponds to superselection sectors carrying so-called
topological charges (``BF-charges'').\footnote{The acronyms DHR and BF
refer to Doplicher-Haag-Roberts \cite{Doplicher:1969tk,DHR:1971} and
to Buchholz-Fredenhagen \cite{BF}, respectively, who
have introduced and analyzed the corresponding types of superselection
sectors.} In both cases, we will assume that the translations act 
covariantly on the algebras of field operators, and that there is a 
translation-invariant vacuum.
 Our principal interest
lies in the behaviour of the superselection charges in the scaling
limit.

We propose a criterion specifying what it means that a charge 
superselection
sector of the given quantum field theory is ``preserved'' in the
scaling limit.  Then we will show that under
quite general conditions, a superselection charge is preserved in the
scaling limit exactly if this is also the case for the corresponding 
conjugate
charge. As a further application, we extend an earlier result by
Roberts \cite{Rob.dil} (which was obtained for dilation covariant 
quantum
field theories) by showing that in a quantum field theory where all
charges of DHR-type are preserved in the scaling limit, the sets of 
local and
global intertwiners for the superselection charges coincide (see the
first part of Sec.\ 4 for explanation of this terminology). This
amounts to saying that part of the superselection structure is
determined locally if the superselection charges are ultraviolet
stable in the sense of being preserved in the scaling limit. Such a
property is of some relevance in the construction of superselection
theory in a generally covariant setting as recently developed in
\cite{Ver.hab}. 

This article is organized as follows. In Sec.\ 2 we define the quantum
field theories corresponding to case (1), with strictly
localizable field operators, more precisely. We
introduce a class of theories which we call ``quantum field theories
with gauge group action'', abbreviated QFTGA, in the
operator-algebraic setting. This class of theories is slightly more
general than the class of theories obtained via the Doplicher-Roberts
reconstruction from DHR-type superselection charges (which
will be considered in Sec.\ 4). We introduce the scaling algebra for
such QFTGAs, and, in close analogy to \cite{BV1}, we introduce scaling
limit states and scaling limit theories and study their basic
properties.

Then, in Sec.\ 3, we consider QFTGAs with more structure, mainly with
additional Poincar\'e covariance and clustering properties, and study
what additional properties ensue in the scaling limit.

In Sec.\ 4 we introduce ``quantum field systems with gauge symmetry'' 
(QFSGSs)
according to \cite{DRwhy}. These are more special QFTGAs which arise 
by the
Doplicher-Roberts reconstruction theorem from the covariant, strictly
localizable (i.e.\ DHR-type)  superselection sectors with finite 
statistics 
belonging to a quantum field theory of local observables (cf.\ again
\cite{DRwhy}).
Charges of this kind would, e.g., correspond to the flavour charges of
strong interactions.
 The reason why we make a distinction between QFTGA and
QFSGS is that the scaling limit theories of a QFTGA are again of this
type, i.e.\ are QFTGAs. But scaling limit theories of a QFSGS have in
general only the structure of a QFTGA.
 We summarize parts of the
terminology of the theory of superselection sectors and the result on
the existence of a corresponding QFSGS, emphasizing the role played
by the ``field multiplets'' in the local field algebras corresponding
to each superselection charge. 

We will make use of this in Sec.\ 5, where we will state
our criterion of preservance of a charge in the scaling limit in terms
of such field multiplets: Our criterion demands that a charge of 
DHR-type is
preserved in the scaling limit if scaled families of such multiplets
(``scaled multiplets'') have a certain limiting behaviour in the
scaling limit. Then we briefly discuss mechanisms for the 
disappearance of
charges in the scaling limit. Quite generally, a charge may disappear
in the scaling 
limit if it takes typically more energy than proportional to $\l^{-1}$
to create the charge within a spacetime region of extension
proportional to $\l$. 
As will be explained, this may happen if the interaction between
the charges of a quantum field theory is, at extremely short
distances, either strongly binding or strongly repellent.
Moreover, we present some further results on the
structure of superselection charges preserved in the scaling limit,
like the preservance of the conjugate charge.

In Sec.\ 6 we state and prove our result on the equivalence of local
and global intertwiners if all DHR-charges are preserved in the 
scaling limit.

Having up to this point discussed the case (1) of field operators 
localizable
in arbitrary bounded regions, we will turn in Sec.\ 7 to the 
discussion of
QFSGSs where the field operators are only localizable in infinitely 
extended
spacelike cones (corresponding to case (2) alluded to above). Because 
of
the much weaker localization properties of field operators in this 
case, the
scaling algebra method has to be appropriately adapted, and this is 
done by
defining the scaling algebra functions of field operators such that 
the
functions asymptotically (as the scaling parameter $\l$ tends to 0) 
commute
with the scaling algebra functions of the observables (which are
strictly localized) from which the QFSGS is constructed by the 
Doplicher-Roberts
theorem for superselection charges of the BF-type. It will turn out 
that, for
the corresponding definition of the scaling algebra of fields, one 
obtains 
scaling limit QFTGAs where the field operators are strictly 
localizable.

In Sec.\ 8 we will then generalize the criterion for the preservance 
of charges
in the scaling limit to the case of BF-type superselection charges, 
and discuss 
conditions sufficient to conclude that such preserved charges induce 
superselection sectors
of the scaling limit theories of the observables. It will be assumed 
in this discussion that the underlying QFSGS of fields localized in 
spacelike cones is
Poincar\'e covariant. 

There will be two more technical Appendices. In Appendix~\ref{app:example}, we 
present the example
of the Majorana-Dirac field which possesses a $\mathbb{Z}_2$ gauge 
symmetry, and
show that the corresponding charges are preserved in the scaling 
limit according to
our criterion. In Appendix~\ref{app:reehschlieder}, we sketch the proof of a Reeh-Schlieder 
property
for an extended scaling limit field algebra used in the construction 
of the scaling
limit QFTGA in Sec.\ 7.

The article is completed by some concluding remarks.

\section{Quantum field theories with gauge group action and their
  scaling algebras and scaling limits}
\setcounter{equation}{0}
\label{sec:QFTGA}
In the present section we investigate an extension of the
``scaling algebra'' approach of \cite{BV1} to quantum field theories
that include a structure which we will call a {\it normal, covariant
  quantum field theory with gauge group action} (QFTGA) since we will
see that this structure has a counterpart in the scaling limit. In the
next section we add a few more assumptions, such as Lorentz
covariance, geometric modular action, and clustering, but it is not
before Section 4 that we introduce a {\it normal, covariant quantum
  field system with gauge symmetry} according to \cite{DRwhy} which
connects quantum field algebras and superselection sectors, and
explore some properties of the scaling limits for such theories.
\\[10pt]
{\it Notation.} In the following, we consider quantum field theories
on $n$-dimensional Minkowski-spacetime ($n \ge 2$), which will be
identified with $\bR^n$, equipped with the Lorentzian metric $\eta =
(\eta_{\m\n}) = {\rm diag}(1,-1,-1,\ldots,-1)$. We recall that the set
$V_+ := \{(y^0,\ldots,y^{n-1}) \in \bR^n: (y^0)^2 > (y^1)^2 + \ldots +
(y^{n-1})^2,\ y^0 > 0\}$ denotes the open forward lightcone and
$\overline{V_+}$ its closure. A {\it double cone} is any set in
$\bR^n$ of the form $O = x + V_+ \cap y - V_+$
for any pair of $x,y \in \bR^n$ so that $y \in x + V_+$. The set of
all double cones in $\bR^n$ will be denoted generically by $\cK$.

%A {\it spacelike cone} is a set of the form $C = \bigcup_{\l
%      > 0} \l O$ where $O$ is a double one not containing the origin
%    of $\bR^n$.

%
\begin{Definition} {\rm
A quintuple $(\cF,\Usc(\bR^n),U(G),\Omega,k)$ is called a {\em normal,
  covariant quantum field theory with gauge group action} (QFTGA) if
the following properties are fulfilled:
\begin{description}
\item[{\small (QFTGA.1)}] There is a Hilbert-space $\cH$ and a family
  $\{\cF(O)\}_{O \in \cK}$ of von
  Neumann algebras on $\cH$ which is indexed by the members $O$ of the
  set $\cK$ of all double cones in $n$-dimensional Minkowski
  spacetime.
  It will be assumed that isotony holds, i.e.\ 
$$ O_1 \subset O \imply \cF(O_1) \subset \cF(O)\,.$$
Hence, one may form the smallest $C^*$-algebra $\fF :=
\overline{\bigcup_O \cF(O)}^{C^*}$ in $B(\cH)$ containing all local
field algebras $\cF(O)$. (In the above quintuple, $\cF$ is short for
the family $\{\cF(O)\}_{O \in \cK}$.)
\item[{\small (QFTGA.2)}] There is a strongly continuous unitary 
representation
  $\bR^n \owns a \mapsto \Usc(a) \in B(\cH)$ of the group of
  translations $\bR^n$ on $\cH$ whose action on $\{\cF(O)\}_{O \in
    \cK}$ is covariant, i.e.\ 
$$ \Usc(a)\cF(O) \Usc(a)^* = \cF(O + a)\,, \quad a \in \bR^n\,,\ O \in
\cK\,.$$
Moreover, it will be assumed that the relativistic spectrum condition
holds: The joint spectrum of the selfadjoint generators of
$\Usc(\bR^n)$ is cointained in the closed forward lightcone
$\overline{V^+}$.
\item[{\small (QFTGA.3)}] There is a compact group $G$, and a strongly
  continuous,\footnote{whenever this makes sense, i.e.\ when $G$
    possesses continuous parts}
faithful representation $G \owns g \mapsto U(g) \in B(\cH)$ of the
group $G$ on $\cH$. It is assumed that the action of this unitary
representation on $\{\cF(O)\}_{O\in \cK}$ preserves localization,
i.e.\
$$ U(g) \cF(O) U(g)^* = \cF(O)\,, \quad g \in G\,,\ O \in \cK\,,$$
and also that this group representation commutes with the 
translations:
$$ U(g)\Usc(a) = \Usc(a)U(g)\,, \quad g \in G\,,\ a \in \bR^n\,.$$
$G$ will be called the {\em gauge group}.
\item[{\small (QFTGA.4)}] There is a unit vector $\O \in \cH$ which 
is 
  invariant under all $\Usc(a)$, $a \in \bR^n$, and under all $U(g)$, 
$g \in
  G$, and which moreover has the cyclicity property $\overline{\fF \O}
  = \cH$. This vector is called the {\em vacuum vector}.
\item[{\small (QFTGA.5)}] There is an element $k$ contained in the 
centre of
  $G$ and fulfilling $k^2 = 1_G$ (the unit group element) so that,
  upon setting
$$ F_{\pm} := \frac{1}{2}(F \pm U(k)FU(k)^* )\,,$$
the following relations hold whenever $F \in \cF(O_1)$, $F' \in
\cF(O_2)$, and the double cones $O_1$ and $O_2$ are spacelike
sparated:
\begin{equation}
\label{comrel}
F_+F'_+ = F'_+F_+\,, \quad F_+F'_- = F'_- F_+\,, \quad F_- F'_- = -
F'_- F_-\,.
\end{equation}
These properties are referred to as {\it normal commutation 
relations}.
\end{description} }
\end{Definition}
{\it Remark. } It was already mentioned in the introduction that the
definition of a QFTGA is slightly more general than that of a quantum
field system with gauge symmetry (see Sec.\ 4) which is more directly
related to the theory of superselection charges; however, the
differences are minute and mainly of technical nature.
The advantage of working with QFTGAs is that
their structure is stable with respect to passing to scaling limit
theories, as will become clear in the present section. 
\\[10pt]
The next task is to introduce the counterpart of the scaling algebra
for a QFTGA which was defined in \cite{BV1} for quantum field theories
formulated in terms of local observable algebras. To that end, we 
assume that
we are given an arbitrary normal, covariant quantum field theory with
gauge group action $(\cF,\Usc(\bR^n),U(G),\O,k)$ (henceforth called
the ``underlying QFTGA'') and keep it fixed.
It will be convenient to introduce the following notation for the
adjoint actions of translations and gauge group:
$$ \a_a(F) := \Usc(a) F \Usc(a)^*\,, \quad \b_g(F) := U(g)FU(g)^*\,,$$
for all $F \in \fF$, $a \in \bR^n$, $g \in G$. 
\begin{Definition} {\rm
For each $O \in \cK$, we define $\uFF(O)$ as the set of all functions
$\uF: \bR^+ \to \fF$, $\lambda \mapsto \uFl$, having the following
properties:
\begin{itemize}
\item[(a)] $\uFl \in \cF(\l O)$,
\item[(b)] $||\,\uF \,|| := \sup_{\l}\,||\uFl|| < \infty$,
\item[(c)] $||\, \uaa(\uF) - \uF\,|| \to 0$ as $a \to 0$, where
\begin{equation}
\label{lifttrans}
 (\uaa(\uF))_{\l} := \ala(\uFl)\,,
\end{equation}
\item[(d)]
 $||\,\ub_g(\uF) - \uF\,|| \to 0$ as $g \to 1_G$, where
\begin{equation}
\label{liftgauge}
 (\ub_g(\uF))_{\l} := \b_g(\uFl)\,.
\end{equation}
\end{itemize} }
\end{Definition}
In \cite{BV1} the case was considered that $\fF$ is an observable
algebra. In that case, the action of the gauge group $U(G)$ on $\cH$
is trivial, and spacelike commutativity holds for the local algebras
$\cF(O)$, meaning that $\cF(O_1) \subset \cF(O_2)'$ if $O_1$ and $O_2$
are spacelike separated. The motivation for imposing the conditions
(a-d) above is similar as for the scaling algebra in the case that
$\cF$ is an observable algebra discussed in \cite{BV1}. The idea is to
view the $\uFl$ as the image of an element $F \in \fF$ under the
action of any ``renormalization group transformation'' $R_{\l}$ (so
one should think of $\uFl$ as $R_{\l}(F)$). In other words, the
collection of all functions $\l \mapsto \uFl$ with the above stated
properties corresponds to all possible orbits of elements in $\fF$
under all (abstract) renormalization group transformations. The
general properties of renormalization group transformations in the
present, model-independent setting are hence encoded by the conditions
(a-d). We point out that (c) ensures that the energy-momentum
transferred by $\uFl$ scales like const.$\cdot 1/\l$, see \cite{BV1}
for further discussion.

As has been indicated to us by D.\ Buchholz, it should be noted that
there may actually be situations where the lifted action of the gauge
transformations ought to be defined differently than in
\eqref{liftgauge}. This occurs for example if the charges of the
theory have a dimension which isn't independent of length or energy
(in this sense, they are ``dimensionful'' charges), and this can 
happen in
two-dimensional models. For the time being, we neglect this
possibility, but we point out that it deserves attention.

There are some simple consequences of Def.\ 2.2 which we briefly put
on record here, see \cite{BV1} for more details. First, it is easy to
see that each $\uFF(O)$, $O \in \cK$, is a $C^*$-algebra with respect
to the $C^*$-norm introduced in (b) when the algebraic operations are
defined pointwise for each $\l$. Clearly one also has isotony,
$$ O_1 \subset O \imply \uFF(O_1) \subset \uFF(O)\,.$$
One can thus form the $C^*$-algebra $\uFF =
\overline{\bigcup_O\uFF(O)}^{C^*}$. The ``lifted'' actions
$\ua_{\bR^n}$ and $\ub_G$ of translations and gauge group, defined in
\eqref{lifttrans} and \eqref{liftgauge}, respectively, act by
automorphisms on $\uFF$ under preservation of the corresponding
covariance properties, i.e.\
\begin{equation}
\label{liftcov}
\uaa(\uFF(O)) = \uFF(O + a)\,, \quad \ub_g(\uFF(O)) = \uFF(O)\,.
\end{equation}
Moreover, we may define 
$$ \uF_{\pm} = \frac{1}{2}(\uF \pm \ub_k(\uF))$$
and hence obtain relations similar to \eqref{comrel}
for $\uF \in \uFF(O_1)$, $\uF' \in \uFF(O_2)$ and $O_1$ and $O_2$
spacelike separated. Finally we note that one may demonstrate the
existence of a wealth of elements in $\uFF$ as follows. 
Let $\m$ be a left-invariant Borel-measure on $G$ and let $h$ be any
continuous, compactly supported function on $\bR^d \times G$. Pick any
uniformly bounded function $\bR^+ \owns \l \mapsto X_{\l} \in
\fF$ so that $X_{\l} \in \cF(\l O)$ for each $\l$ and some $O \in
\cK$, and define
\begin{equation} \label{smoothing}
 \uFl := \int d^na\,d\m(g)\,h(a,g)\,\ala(\b_g(X_{\l})) 
\end{equation}
where the integral is to be understood in the weak sense. Then it is
easily checked that $\bR^+ \owns \l \mapsto \uFl$ is contained in
$\uFF(O^{\times})$ whenever $O^{\times}$ is any open neighbourhood of
$\overline{O} + \overline{\bigcup_{g \in G}{\rm supp}\,h(\,.\,,g)}$.

Having defined the scaling field algebra $\uFF$ of the underlying
QFTGA, we may associate with any locally normal state $\o'$ on
$\fF$\,\footnote{a state $\o'$ on $\fF$ is called {\em locally normal}
  if $\o' \rest \cF(O)$ is normal for each $O \in \cK$} a parametrized
family $(\uol')_{\l > 0}$ of states on $\uFF$, where
$$ \uol'(\uF) := \o'(\uFl)\,, \quad \uF \in \uFF\,.
   $$
As in \cite{BV1}, we adopt the following definition of scaling limit
states.
\begin{Definition} {\rm
 For each locally normal state $\o'$ on $\fF$, we
  regard the family $(\uol')_{\l > 0}$ as a generalized sequence
  directed towards $\l = 0$. Hence, by the Banach-Alaoglu theorem
  \cite{RS1}, the
  family $(\uol')_{\l > 0}$ on the $C^*$-algebra $\uFF$ possesses
  weak-* limit points. This set of weak-* limit points will be denoted
  by $\{\ooi': \iota \in \bI\}$ where $\bI$ is a suitable index set, 
or
  simply by ${\rm SL}^{\fF}(\o')$. Each $\ooi' \in {\rm 
SL}^{\fF}(\o')$
  is a state on $\uFF$, and is called a {\em scaling limit state of}
  $\o'$. }
\end{Definition}
We note that the definition of weak-* limit points means that there
exists for each label $\iota$ a directed set $K_{\iota}$ together with
a generalized sequence $(\l^{(\iota)}_\k)_{\k \in K_{\iota}}$ of
positive numbers converging to $0$ so that
$$ \ooi'(\uF) = \lim_{\k}\,\uo'_{\l^{(\iota)}_\k}(\uF)\,, \quad \uF 
\in
\uFF\,.$$
Again following \cite{BV1}, we introduce for each scaling limit state
$\ooi' \in {\rm SL}^{\fF}(\o')$ its GNS-representation
$(\poi,\Hoi,\Ooi)$ and define
$$ \Foi(O) := \poi(\uFF(O))''\,, \quad \FFoi := \overline{\bigcup_O
  \Foi(O)}^{C^*}\,.$$
Many of the following results (containing also some new definitions)
concerning the structure of scaling limit states and their associated
GNS-representations in the present setting are generalizations of
similar statements in \cite{BV1}.
\begin{Proposition} \label{SLQFTGA}
\begin{itemize}
\item[1.] For each pair of locally normal states $\o'$ and $\o''$ on
  $\fF$ it holds that 
$$ {\rm SL}^{\fF}(\o') = {\rm SL}^{\fF}(\o'')\,.$$
\item[2.] Let $\o'$ be a locally normal state on $\fF$. Then each 
$\ooi'
  \in {\rm SL}^{\fF}(\o')$ is invariant under the actions of $\uaa$, 
$a
  \in \bR^n$, and $\ub_g$, $g \in G$:
$$ \ooi' \lcrc \uaa = \ooi'\,, \quad \ooi' \lcrc \ub_g = \ooi'\,.$$
Hence, there are unitary group representations
of the translation group and the gauge group on $\Hoi$ which are,
respectively, defined by
$$ \Uscoi(a)\poi(\uF)\Ooi := \poi(\uaa(\uF))\Ooi\,, \quad
\Uoi(g)\poi(\uF)\Ooi := \poi(\ub_g(\uF))\Ooi$$
for all $a\in\bR^n$, $g\in G$, and $\uF \in \uFF$.
\item[3.] The unitary group representations $\Uscoi(a)$, $a \in
  \bR^n$, and $\Uoi(g)$, $g\in G$, are continuous and have the
  properties
$$ \Uscoi(a)\Foi(O)\Uscoi(a)^* = \Foi(O+a)\,, \quad
\Uoi(g)\Foi(O)\Uoi(g)^* = \Foi(O)$$
for all $a\in \bR^n$, $g \in G$ and $O \in \cK$. Moreover, the unitary
translation group $\Uscoi(a)$, $a \in \bR^n$, fulfills the
relativistic spectrum condition.
\item[4.] The set $\Noi$ of all $g \in G$ so that $\Uoi(g)\psi = \psi$
  holds for all $\psi \in \Hoi$ is a closed normal subgroup of
  $G$. Therefore,
\begin{equation}
\label{slgaugegroup}
 \Uoib: \Goib \owns \gb \mapsto \Uoi(g)
\end{equation}
is a continuous faithful representation of the factor group $\Goib =
G/\Noi$. Here, $g \mapsto \gb \equiv \goib$ is the quotient map, and
in \eqref{slgaugegroup}, $g$ is any element in the pre-image of $\gb$
with respect to the quotient map. 
\item[5.] Define for $\fbs \in \FFoi$,
$$ \fbs_{\pm} := \frac{1}{2}(\fbs \pm \Uoib(\kb)\fbs \Uoib(\kb)^*)$$
where $\kb = \koib$. Then the following holds: 
If $O_1$ and $O_2$ are spacelike separated double cones and $\fbs \in
\Foi(O_1)$, $\fbs' \in \Foi(O_2)$, one has the relations
\begin{equation}
\label{slcomrel}
\fbs_+ \fbs'_+ = \fbs'_+ \fbs_+\,,\quad \fbs_+ \fbs'_- = \fbs'_-
\fbs_+\,, \quad \fbs_- \fbs'_- = -\fbs'_- \fbs_-\,.
\end{equation}
\item[6.] The previous statements yield the following corollary: Let
  $\o'$ be a locally normal on $\fF$ (of the underlying QFTGA) and
  $\ooi'\in {\rm SL}^{\fF}(\o')$ an arbitrary scaling limit state, 
then
  the corresponding scaling limit objects

  $(\Foi,\Uscoi(\bR^n),\Uoib(\Goib),\Ooi,\koib)$ form again a normal,
  covariant quantum field theory with gauge group action (which will
  be called a {\em scaling limit QFTGA} of the underlying QFTGA
  corresponding to $\ooi'$).
\end{itemize}
\end{Proposition} 
{\it Proof. } Ad 1.\ The proof is analogous to that in \cite{BV1},
which uses an argument due to Roberts \cite{Rob.dil} showing that
\begin{equation}
\label{distlim}
       ||(\o' -\o'')\rest \cF(\l O) ||  \to 0 \ \ {\rm as} \ \ \l \to 
0
\end{equation}         
holds for any pair of locally normal states $\o'$ and $\o''$ on $\fF$
and $O \in \cK$ as a consequence of
$$ \bigcap_{O \owns 0} \cF(O) = \bC\cdot 1\,.$$
This latter property holds also for the local field algebras owing to
the spectrum condition for the translation group and normal
commutation relations \eqref{comrel}, see \cite{BV1} for details.
\\[6pt]
Ad 2.\ The invariance property is obvious for the case that $\o'$
coincides with the vacuum state $\o(F) = \langle \O,F \O \rangle$
on $\fF$. Then \eqref{distlim} implies the analogous property for any
other locally normal state.
\\[6pt]
Ad 3.\ The continuity follows simply from assumptions (c) and (d) of
Def.\ 2.2. The covariance properties are implied by
\eqref{liftcov}. The spectrum condition for the translations may be
proved as in \cite{BV1}.
\\[6pt]
Ad 4.\ By construction, $\Uoib$ is a faithful unitary representation
of $\Goib$ on $\Hoi$. Continuity follows since the quotient map $g
\mapsto \gb$ is open.
\\[6pt]
Ad 5.\ As indicated above, the relations \eqref{comrel} carry over to
the scaling algebra $\uFF$ by setting $\uF_{\pm} = \frac{1}{2}(\uF \pm
\ub_k(\uF))$. The corresponding relations for the scaling limit
theories follow directly. (It may however happen that $k \in \Noi$; in
this case, the last, ``fermionic'' relation of \eqref{slcomrel} is
absent, and spacelike commutativity holds for the local scaling limit
algebras $\Foi(O)$, $O \in \cK$.)
\\[10pt]
Henceforth, we will (without
restriction of generality in view of 1.\ of Prop.\ \ref{SLQFTGA})
 always consider scaling limit states $\ooi
\in {\rm SL}^{\fF}(\o)$ where $\o(\,.\,) = \langle\O,\,.\,\O\rangle$
denotes the vacuum state.
 
As was done in \cite{BV1}, we will identify scaling limit theories
which are isomorphic in a sense that we will describe next.
\begin{Definition}\label{def:netisomorphism} {\rm
Let 
$$(\Foi,\Uscoi(\bR^n),\Uoib(\Goib),\Ooi,\koib) \quad {\rm and} \quad
(\cF_{0,\g},\Usc_{0,\g}(\bR^n),U^{\bullet}_{0,\g}
(G^{\bullet}_{0,\g}),\O_{0,\g},k^{\bullet}_{0,\g})$$
be two scaling limit theories of an underlying QFTGA. These two
scaling limit theories will be called {\em isomorphic} if there exists
a $C^*$-algebraic isomorphism $\phi: \FFoi \to \fF_{0,\g}$ so that the
following properties hold:
\begin{eqnarray*}
\phi(\Foi(O)) &=& \cF_{0,\g}(O)\,, \quad O \in \cK\,,\\
\phi \lcrc {\rm Ad}\,\Uscoi(a)  &=& {\rm Ad}\,\Usc_{0,\g}(a) \lcrc
\phi\,, \quad a \in \bR^n\,,\\
\phi \lcrc {\rm Ad}\,\Uoi(g) &=& {\rm Ad}\,U_{0,\g}(g)\lcrc
\phi\,,\quad g \in G\,.
\end{eqnarray*} }
\end{Definition}
Note that the last property induces a natural identification between
$\Noi$ and $N_{0,\g}$ and hence a natural identification $\Goib \owns
\goib \mapsto g^{\bullet}_{0,\g} \in G^{\bullet}_{0,\g}$, so that one
obtains, in consequence,
$$ \phi \lcrc {\rm Ad}\,\Uoib(\goib) = {\rm
  Ad}\,U^{\bullet}_{0,\g}(g^{\bullet}_{0,\g}) \lcrc \phi $$
which holds in particular with $\koib$ and $k^{\bullet}_{0,\g}$
inserted for $\goib$ and $g^{\bullet}_{0,\g}$, respectively.

We will moreover say that two isomorphic scaling limit theories have a
{\em unique vacuum structure} if the connecting isomorphism also has 
the
property
$$ \o_{0,\g} \lcrc \phi = \ooi\,.$$
Following once more \cite{BV1}, one may now classify a given
underlying QFTGA according to the following (mutually exclusive)
possibilities:
\begin{itemize}
\item[{\bf (1)}] All scaling limit QFTGAs are isomorphic, and $\FFoi$
  is non-abelian. Then the underlying QFTGA is said to have a {\it
    unique quantum scaling limit}.
\item[{\bf (2)}] All scaling limit QFTGAs are isomorphic, and
  $\FFoi$ is abelian. In this case one says that the underlying QFTGA
  has a {\it classical scaling limit}.
\item[{\bf (3)}] There are scaling limit QFTGAs which are
  non-isomorphic. One then says that the underlying QFTGA has a {\it
    degenerate scaling limit}.
\end{itemize}
The interpretation of these cases is as in the case of observable
algebras \cite{BV1}; see this reference for further discussion.
 The first case would correspond to an underlying
theory which has a single, stable ultraviolet fixed point. The second 
case is
thought to correspond to an underlying theory which has no ultraviolet
fixed point. The third case is in a sense intermediate, the underlying
theory has a very irregular behaviour at small scales and has various,
most likely unstable, ultraviolet fixed points.
\\[10pt]
We next put on record a result from \cite{BV1} connecting the
uniqueness of the scaling limit with the existence of a dilation
symmetry in the scaling limit theories. The proof proceeds exactly as
in the cited reference.
\begin{Proposition}\label{prop:dilation} {\rm \cite{BV1}} Assume that 
all the scaling limit
  QFTGAs 
$$(\Foi,\Uscoi(\bR^n),\Uoib(\Goib),\Ooi,\koib)\,, \quad \iota \in \bI 
\,,$$ 
  of the
  underlying QFTGA are isomorphic, i.e.\ that we are in case {\bf (1)}
  or {\bf (2)} of the just given classification. Then for each $\iota
  \in \bI$ there exists a family $(\d^{(0,\iota)}_{\m})_{\m > 0}$ of
  automorphisms of $\FFoi$ acting as dilations in the corresponding
  scaling limit theory, which means that the following relations hold:
\begin{eqnarray*}
\d^{(0,\iota)}_{\m}(\poi(\uFF(O)) & = & \poi(\uFF(\m O))\,, \quad \m >
0\,,\ O \in \cK\,,\\
 \d^{(0,\iota)}_{\m} \lcrc {\rm Ad}\,\Uscoi(a) &=& {\rm Ad}\,\Uscoi(\m
 a) \lcrc \d^{(0,\iota)}_{\m}\,, \quad a \in \bR^n\,,\ \m > 0\,,\\
\d^{(0,\iota)}_{\m} \lcrc {\rm Ad}\,\Uoi(g) & = & {\rm Ad}\,\Uoi(g)
\lcrc \d^{(0,\iota)}_{\m}\,, \quad g \in G\,,\ \m > 0\,.
\end{eqnarray*}
Furthermore, if the underlying QFTGA also has a unique vacuum
structure in the scaling limit, then it follows that the family of
dilations leaves the scaling limit states invariant: $\ooi
\lcrc \d^{(0,\iota)}_{\m} = \ooi$, $\iota \in \bI$, $\m > 0$.
\end{Proposition}
\section{Scaling limits for QFTGAs with additional properties}
\setcounter{equation}{0}
In the present section we consider an underlying QFTGA with additional
properties, such as Lorentz-covariance, spacelike clustering and
geometric modular action, and we will investigate which further 
properties for
the scaling limit theories ensue. More precisely, let
$(\cF,\Usc(\bR^n),U(G),\O,k)$ be the underlying QFTGA, assumed to
satisfy the conditions (QFTGA.1-5) of Def.\ 2.1. We will consider the
following additional properties:
\begin{description} 
\item[{\small (QFTGA.6)}] (Lorentz covariance) There is a strongly 
continuous 
unitary representation
$\Lpocov \owns L \mapsto \tU(L) \in B(\cH)$ of the covering group of
the proper, orthochronous Lorentz group $\Lpo$ (in $d$ dimensions) on
$\cH$ so that the following relations are fulfilled:
\begin{eqnarray*}
 \tU(L)\Usc(a) &=& \Usc(\L(L)a)\tU(L)\,,\\ \tU(L)U(g) &=&
U(g)\tU(L)\,,\\
 \tU(L)\cF(O)\tU(L)^* &=& \cF(\L(L)O)\,, \quad \quad \tU(L)\O = \O 
\end{eqnarray*}
for all $L \in \Lpocov$, $a\in \bR^n$, $g \in G$ and $O \in \cK$,
where $\Lpocov \owns L \mapsto \L(L) \in \Lpo$ denotes the covering
projection.
\item[{\small (QFTGA.7)}] (Irreducibility) $\fF' = \bC \cdot 1$\,.
\item[{\small (QFTGA.8)}] (Spacelike clustering) We will assume that 
a uniform
  clustering bound holds on the vacuum (for spacetime dimension $d \ge
  3$). To formulate this, we use the following notation. Elements in
  the $x^0 = 0$ hyperplane will be denoted by $\boldsymbol{x} \in
  \bR^{n-1}$ and identified with $(0,\boldsymbol{x})\in \bR^n$. We
  define the derivation
$$ \partial_0(F) := -i\left.\frac{d}{dx^0}\right|_{x^0 =
  0}\alpha_{(x^0,\boldsymbol{0})}(F)$$
on the domain $D(\partial_0)$ of all $F \in \fF$ so that the (weak)
derivative on the right hand side exists as an element in $\fF$. Note
that $D(\partial_0)$ is a weakly dense subset of $\fF$. Then our 
assumption
on the existence of a uniform spacelike clustering bound is: There
exists, for the given underlying QFTGA, a constant $c > 0$ so that for
each double cone $O_r$ having spherical base of radius $r$ in the
$x^0=0$ hyperplane there holds the bound
$$ |\o(F_1\a_{\boldsymbol{x}}(F_2)) -\o(F_1)\o(F_2)|
\le \frac{c
  r^{n-1}}{|\boldsymbol{x}|^{n-2}}(||F_1||\,||\partial_0(F_2)|| +
||\partial_0(F_1)||\,||F_2||)
 $$
for all $F_1,F_2 \in \cF(O_r)\cap D(\partial_0)$ as soon as
$|\boldsymbol{x}| > 3r$.  
\item[{\small (QFTGA.9)}] (Geometric modular action) A {\em wedge 
region} is
  any Poincar\'e-transformed copy of the so-called right wedge $W_R :=
  \{(x^0,\ldots,x^{n-1}): |x^1| < x^0,\ x^0 > 0\}$. For this right
  wedge, we define the wedge-reflection map
 $r_R : \bR^n \to \bR^n$ by
  $$r_R(x^0,x^1,x^2,\ldots,x^{n-1}) := 
(-x^0,-x^1,x^2,\ldots,x^{n-1})\,,$$
  and the Lorentz-boosts
\begin{eqnarray*}
 \lefteqn{\L_R(t)(x^0,x^1,x^2,\ldots,x^{n-1})} \\
 & := & (\cosh(t)x^0 +
 \sinh(t)x^1,\sinh(t)x^0 + \cosh(t)x^1,x^2,\ldots,x^{n-1})\,.
\end{eqnarray*}
For any other wedge-region $W = L W_R$ with a suitable
Poincar\'e-transformation $L$, we define $r_W := L r_R L^{-1}$ and
$\L_W(t) := L \L_R(t) L^{-1}$.
               
For each wedge
  region $W$ in $\bR^n$, the vacuum vector $\O$ of the underlying
  QFTGA is cyclic and separating for the von Neumann algebra $\cF(W) =
  \{\cF(O): \overline{O} \subset W,\ O \in \cK\}''$. Hence, there
  correspond to each wedge region $W$ the Tomita-Takesaki modular
  objects $J_W,\D_W$ associated with $\cF(W),\O$ \cite{TTT}. It will
  then be assumed that, in the presence of (QFTGA.6),
  these modular objects act geometrically in the
  following way:
\begin{alignat}{3}               
 J_W\tU(L)J_W &=\tU(\widetilde{{\rm Ad}r_W}L)\,,&\hspace{0.3cm} J_W 
\Usc(a)J_W &= \Usc(r_W a)\,,&\hspace{0.3cm}&L \in \Lpocov\,,\ a \in
 \bR^n\,,\\
\D_W^{it} &=\tU(\widetilde{\L_W(2\p t)})\,, && &&t \in \bR\,,\\
 J_W \cF(O) J_W &=\cF^t(r_W O)\,, && &&O \in \cK\,. 
\label{eq:geometricmodularJO}
\end{alignat}
In these equations, we have denoted by $\widetilde{{\rm Ad}r_W}$ the
lift of the adjoint action of $r_W$ to $\Lpocov$, and by
$\widetilde{\L_W(t)}$ the lift of $\L_W(t)$ to $\Lpocov$ (both of
which exist, cf.\ \cite{GLchcon}). Moreover, we have introduced the
so-called ``twisted'' local von Neumann algebras
\begin{equation}
\label{twistnet}
 \cF^t(O) := V\cF(O)V^*\,, \quad O \in \cK\,,
\end{equation}
where the twisting operator $V$ is a unitary on $\cH$ defined by
\begin{equation}
\label{twistop}
V := (1 + i)^{-1}(1 + iU(k)) \,.
\end{equation}
Note that the algebras $\cF(O_1)$ and $\cF^t(O_2)$ commute for
spacelike separated $O_1$ and $O_2$ on account of the assumed normal
commutation relations.
\end{description}
We shall continue our investigation of the scaling limit theories of
an underlying QFTGA satisfying some, or all, of the just stated
additional conditions. In order to do that, we have to slightly
re-define the scaling algebras $\uFF(O)$ when the underlying QFTGA
satisfies Lorentz-covariance. For the remaining part of this article 
we
adopt the following
\\[6pt]
{\bf Convention. } Suppose that the underlying QFTGA satisfies also
the condition of Lorentz-covariance (QFTGA.6). In this case, the local
scaling algebras $\uFF(O)$, $O \in \cK$, are defined as in Def.\ 2.2
but demanding in addition that the elements $\uF \in \uFF(O)$ fulfill 
the
also the condition
\begin{itemize}
\item[(e)] ${}$ \quad $||\,\tilde{\ua}_L(\uF) - \uF\,|| \to 0$ as $L
  \to 1_{\Lpocov}$
\end{itemize}
where
$$ (\tilde{\ua}_L(\uF))_{\l} := \tU(L)\uFl\tU(L)^*\,.$$
${}$\\
Again, it is not difficult to demonstrate that, with that convention,
the $\uFF(O)$ are $C^*$-algebras containing plenty of elements, and
$\ua_{\bR^n}$, $\ub_G$ and $\tilde{\ua}_{\Lpocov}$ act as strongly
continuous groups of automorphisms on $\uFF$ with the covariance
properties \eqref{liftcov} and, in addition,
$$ \tilde{\ua}_L(\uFF(O)) = \uFF(\L(L)O)\,, \quad L \in \Lpocov\,,\ 
O\in
\cK\,.$$
The following statement is again essentially a transcription of
analogous results established for observable algebras in \cite{BV1}. 
\begin{Proposition}\label{prop:QFTGAadd}
Suppose that the underlying QFTGA fulfills the conditions of Def.\
2.2.
\begin{itemize}
\item[1.] If the underlying QFTGA fulfills also Lorentz-covariance
  (QFTGA.6), then this property holds also for all scaling limit
  QFTGAs.
\item[2.] If the underlying QFTGA fulfills (QFTGA.$6\, \&\, 7$) and $n
  \ge 3$, then all
  scaling limit QFTGAs fulfill (QFTGA.$6\, \&\, 7$).
\item[3.] If the underlying QFTGA fulfills (QFTGA.8) and $n \ge 3$,
  then all scaling
  limit QFTGAs fulfill (QFTGA.$7$).
\item[4.] If the underlying QFTGA fulfills (QFTGA.$6\,\&\,9$), then
  all scaling limit QFTGAs fulfill (QFTGA.$6\,\&\,9$), too.
\end{itemize} 
\end{Proposition}
{\it Proof. } Ad 1.\ This statement is proved in complete analogy to
the corresponding statement in \cite{BV1}; we note that for any
scaling limit state $\ooi \in {\rm SL}^{\fF}(\o)$ (where $\o$ is any
locally normal state on $\fF$) there holds $\ooi \lcrc \tilde{\ua}_L =
\ooi$ and hence one obtains a unitary representation of $\Lpocov$ on
$\Hoi$ via setting
$$ \tU_{0,\iota}(L)\poi(\uF)\Ooi := \poi(\tilde{\ua}_L(\uF))\Ooi\,,
\quad L \in \Lpocov\,, \ \uF \in \uFF\,.$$
It is also easily checked that this unitary representation has all the
properties analogous to those listed in (QFTGA.6) with respect to the
scaling limit theory.  
\\[6pt]
Ad 2.\ If the underlying theory has the additional properties
(QFTGA.$6\,\&\,7$), 
then this entails that the underlying theory also has the property
(QFTGA.8) according to a result by Araki, Hepp and Ruelle \cite{AHR};
cf.\ also the proof of Lemma 4.3 in \cite{BV1}. The statement then
follows from 1. and 3.
\\[6pt]
Ad 3.\ Let $\uF'{}^{(1)},\uF'{}^{(2)}\in\uFF(O_{r'})$ and define, for
some $h \in C^{\infty}_0(\bR^n)$, 
$$ \uF^{(j)} := \int d^na\,h(a)\uaa(\uF'{}^{(j)})\,, \quad j =1,2\,.$$
Then there is some $r > r'$ so that $\uF^{(j)}\in \uFF(O_r)$, and
clearly $\uFl^{(j)} \in \cF(\l O) \cap D(\partial_0)$. We apply the
uniform clustering bound to obtain, for each $\l > 0$ and
$|\boldsymbol{x}| > 3r$,
\begin{eqnarray*}
\lefteqn{|\uol(\uF^{(1)}\ua_{\boldsymbol{x}}(\uF^{(2)})) -
  \uol(\uF^{(1)})\uol(\uF^{(2)})|} \\
&= & |\o(\uFl^{(1)}\a_{\l \boldsymbol{x}}(\uFl^{(2)}))
-\o^{\O}(\uFl^{(1)})\o(\uFl^{(2)})|\\
&\le & \frac{c (\l r)^{n -1}}{|\l \boldsymbol{x}|^{n-2}}
 (||\uFl^{(1)}||\,||\partial_0(\uFl^{(2)})|| +
 ||\partial_0(\uFl^{(1)})||\,||\uFl^{(2)}||) \\
&\le&  \frac{c r^{n -1}}{|\boldsymbol{x}|^{n-2}}
 (||\,\uF^{(1)}\,||\,||\,\underline{\partial}_0(\uF^{(2)})\,|| +
 ||\,\underline{\partial}_0(\uF^{(1)})\,||\,||\,\uF^{(2)}\,||)\,,
\end{eqnarray*}
where we have defined $\underline{\partial}_0(\uF^{(j)}) :=-i
\left.\frac{d}{dx^0}\right|_{x^0=0}\ua_{(x^0,\boldsymbol{0})}(\uF^{(j)})$

and used the fact that 
\\
$||\partial_0(\uFl^{(j)})|| \le
\l^{-1}||\,\underline{\partial}_0(\uF^{(j)})\,||$. Now 
$||\,\underline{\partial}_0(\uF^{(j)})\,|| < \infty$ by the definition
of the $\uF^{(j)}$, and taking the $\limsup_{\l}$ on the left-hand
side of the last inequality, one concludes that asymptotic spacelike
clustering holds on the vacuum of each scaling limit theory since
$\uF^{(j)}$ approaches $\uF'{}^{(j)}$ in the scaling algebra norm for
$h \to \delta$. Because
of normal commutation relations in each scaling limit QFTGA, this
entails that $\FFoi' = \bC\cdot 1$ holds in all scaling limit 
theories.
\\[6pt]
Ad 4.\ The proof proceeds analogously to the proof of Lemma 4.3 in
\cite{BV1}.
\hfill $\Box$
\\[10pt]
There is another result worth mentioning here which also generalizes a
corresponding result established for observable algebras in \cite{BV1}
and connects a duality condition in scaling limit theories with the
type of the local von Neumann algebras of the underlying QFTGA. 
\begin{Theorem} \label{typeoflocalg}
Suppose that the underlying QFTGA fulfills the assumptions of Def.\
2.1. Moreover, suppose that there exists a scaling limit QFTGA
$$(\Foi,\Uscoi(\bR^n),\Uoib(\Goib),\Ooi,\koib)$$ 
having the property of
``twisted wedge duality'',
$$   \Foi(W)' = \Foi^t(r_W(W)) $$
for some wedge region $W$ in $\bR^d$  (with the definition of the
twisted local von Newmann algebras analogous to \eqref{twistnet} and
\eqref{twistop} with respect to the corresponding objects in the
scaling limit QFTGA); moreover, suppose that $\FFoi 
\ne \bC \cdot 1$. In this case it holds that the local von Neumann
algebras $\cF(O)$ are of type ${\rm III}_1$ for each double cone $O
\subset W$ 
whose boundary intersects $\overline{W} \cap \overline{r_W(W)}$, and
for all translates of such double cones $O$.
If twisted wedge duality holds for all wedge regions in some scaling
limit QFTGA, then one concludes that $\cF(O)$ is of type ${\rm III}_1$
for all double cones.
\end{Theorem}
We refer to Prop.\ 6.4 in \cite{BV1} for a proof of this statement. We
note also that according to the previous Proposition, the validity of
conditions (QFTGA.$6\,\&\,7\,\&\,9$) in the underlying theory implies
that the assumptions of Thm.\ \ref{typeoflocalg} are fulfilled.  
\section{Quantum Field Systems with Gauge Symmetry}
\setcounter{equation}{0}
\label{sec:QFSGS}
We now wish to investigate the scaling limits of QFTGAs that really
correspond to superselection charges of a system of observables. Such
QFTGAs are, more specifically, quantum field systems with gauge
symmetry in the terminology of Doplicher and Roberts \cite{DRwhy}.
In order to summarize their definition here, and also for later
reference, we first recapitulate some concepts of the
Doplicher-Haag-Roberts approach to superselection theory, mainly from
the sources \cite{Haag,Rob.lec,DRwhy}.

This approach starts from the assumption that one is given an
observable quantum system in a vacuum representation together with a
further, distinguished set of representations modelling localized
charges. The structure of an observable quantum system in a vacuum
representation is described in terms of a collection of objects
$(\cA_{\rm vac},\Usc_{\rm vac}(\bR^n),\O_{\rm vac})$ whose properties 
are
assumed to be as follows.
\begin{itemize}
\item[(a)] $\cA_{\rm vac}$ symbolizes a family $\{\cA_{\rm 
vac}(O)\}_{O\in\cK}$ of von
  Neumann algebras in a separable 
  Hilbert space $\cH_{\rm vac}$, subject to
  conditions of isotony (see above) and duality,
$$ \cA_{\rm vac}(O)' = \cA_{\rm vac}(O') :=
 \{\cA_{\rm vac}(O_1): \overline{O_1} \subset O'\,,\ O_1
\in \cK\}''\,,$$
where $O'$ denotes the open causal complement of $O$. Setting moreover
$\fA_{\rm vac} := \overline{\bigcup_O \cA_{\rm vac}(O)}^{C^*}$,
 it is assumed that $\fA_{\rm vac}' =
\bC \cdot 1$.
\item[(b)] $\Usc_{\rm vac}(a)$, $a \in \bR^n$, is a strongly
  continuous unitary
  representation of the translation group on $\cH_{\rm vac}$, acting
  covariantly on the family $\{\cA_{\rm vac}(O)\}_{O\in\cK}$, and 
fulfilling
  the spectrum condition (see above). Furthermore, $\O_{\rm vac} \in
  \cH_{\rm vac}$ is a unit vector which is let invariant by the action
  of $\Usc_{\rm vac}(a)$, $a \in \bR^n$.
\end{itemize}
{\it Remark.} Usually, also the assumption is made that the family
$\{\cA_{\rm vac}(O)\}_{O \in\cK}$ has the Borchers property
(``Property B''). This property says that given $O,O_1 \in \cK$ with
$\overline{O} \subset O_1$ and a non-zero projection $E \in \cA(O)$,
then there is $V \in \cA(O_1)$ with $VV^* = E$ and $V^*V =
1$. However, Roberts has shown \cite{Rob82} that this property can
already be deduced from the other assumptions (essential being
separability of $\cH_{\rm vac}$ and the spectrum condition).
\\[10pt] 
Given an observable 
quantum system $(\cA_{\rm vac},\Usc_{\rm vac}(\bR^n),\O_{\rm
  vac})$, one may look for representations of $\fA_{\rm vac}$ 
describing the
presence of charges. Following Doplicher, Haag and Roberts, one may
consider the set $\fP^{\rm DHR}$ of representations $\pi$ of 
$\fA_{\rm vac}$ which are
unitarily equivalent to the vacuum representation in restriction to the causal
complement of any double cone. That means, if $\fA_{\rm vac}(O')$
is defined as the $C^*$-algebra generated by all
$\cA_{\rm vac}(O_1)$ where $\overline{O_1} \subset O'$,
then $\pi$ is in $\fP^{\rm DHR}$ if $\pi \rest \fA_{\rm vac}(O')$ is 
unitarily equivalent to the identical representation of $\fA_{\rm vac}(O')$ on $B(\cH_{\rm
  vac})$ for 
each $O \in \cK$. Such representations describe superselection charges
which are strictly localizable, see \cite{Haag,Rob.lec} for further
discussion. We shall be interested only in the subset $\fP_{\rm 
cov}^{\rm DHR}$
of those $\pi$ in $\fP^{\rm DHR}$ which are translation-covariant, 
meaning that
there is a strongly continuous representation $\Usc_{\pi}(a)$, $a \in
\bR^n$, of the translation group on the representation-Hilbertspace of
$\pi$ fulfilling the spectrum condition and the intertwining
property
\begin{equation}\label{repcovar} 
{\rm Ad}\,\Usc_{\pi}(a)(\pi(A)) = {\rm 
Ad}\,\Usc_{\rm vac}(a)(A)\,,\quad a \in \bR^n,\ A \in \fA_{\rm vac}\,.
\end{equation}
By identifying the representation-Hilbertspace $\cH_{\pi}$ with
$\cH_{\rm vac}$, the set $\fP_{\rm cov}^{DHR}$ may alternatively (and
equivalently) be described in terms of the set $\D_t^{\rm cov}$ of
covariant, localized and transportable endomorphisms of $\fA_{\rm 
vac}$. Here,
an endomorphism $\rho: \fA_{\rm vac} \to \fA_{\rm vac}$ is
 called localized in $O \in \cK$
if $\rho(A) = A$ holds for all $A \in \fA_{\rm vac}(O')$. It is called
transportable if, given an arbitrary region $O_1 \in \cK$, there
exists a unitary $V$ so that $V\rho(\,.\,)V^*$ is an endomorphism of
$\fA$ localized in $O_1$; one can show that $V$ may be chosen as an
element of $\fA$.

An element $\rho \in \D^{\rm cov}_t$ is called {\it irreducible} if
$\rho(\fA_{\rm vac})' = \bC \cdot 1$, and the set ${\rm Sect}^{\rm 
cov}$ of
all equivalence classes
$$ [\rho] := \{V\rho(\,.\,)V^* : V^* = V^{-1} \in \fA_{\rm vac}\}$$
for irreducible $\rho \in \D^{\rm cov}_t$ is called the set of {\em
  translation-covariant superselection sectors} of the given
observable quantum system $(\cA_{\rm vac},\Usc_{\rm
  vac}(\bR^n),\O_{\rm vac})$.

If $\rho,\rho' \in \D^{\rm cov}_t$, one defines by $\cI(\rho,\rho')$
the set of ${\it intertwiners}$ between $\rho$ and $\rho'$ as the set
of all $T \in \fA_{\rm vac}$ which satisfy
$$ T\rho(A) = \rho'(A)T\,, \quad A \in \fA_{\rm vac}\,.$$
Strictly speaking, one should refer to $\cI(\rho,\rho')$ as the set of
{\bf global intertwiners} between $\rho$ and $\rho'$. Given $O_1 \in
\cK$ and $\rho,\rho' \in \D_t^{\rm cov}$ localized in $O_1$, one can
introduce $\cI(\rho,\rho')_{O}$, the set of {\bf local intertwiners}
with respect to the localization region $O \supset O_1$, as consisting
of all $T \in \fA_{\rm vac}$ fulfilling
$$ T \rho(A) = \rho'(A)T\,, \quad A \in \cA_{\rm vac}(O)\,.$$
Hence it is obvious that $\cI(\rho,\rho')_O \supset \cI(\rho,\rho')$
for all $O \in \cK$, and in Sec.\ 6 we will link the question if local
and global intertwiners are equivalent, i.e.\ if $\cI(\rho,\rho')_O =
\cI(\rho,\rho')$ holds for all $O \in \cK$, to the preservance of
charges in the scaling limit.

Presently, we need to very briefly summarize some further concepts of
charge superselection theory (see, e.g.\ \cite{Rob.lec} for a more
detailed account). First, one can introduce for $T_1 \in
\cI(\rho_1,\rho_1')$ and $T_2 \in \cI(\rho_2,\rho_2')$ a product
operation $T_1 \times T_2$ yielding an element in
$\cI(\rho_1\rho_2,\rho_1'\rho_2')$. There is then a distinguished
family of intertwiners $\epsilon(\rho_1,\rho_2) \in
\cI(\rho_1\rho_2,\rho_2\rho_1)$, for irreducible $\rho_1,\rho_2\in
\D^{\rm cov}_t$, characterized by the property that it describes the
exchange in the intertwiner product according to
$$ (T_2 \times T_1) \epsilon(\rho_1,\rho_2) =
\epsilon(\rho_1',\rho_2') (T_1 \times T_2)\,, \quad T_j \in
\cI(\rho_j,\rho_j')\,,$$
together with the properties $\epsilon(\rho_1,\rho_2) =
1_{\rho_1\rho_2}$ if the localization regions of $\rho_1$ and $\rho_2$
are spacelike separated, and
$\epsilon(\rho_2,\rho_1)\epsilon(\rho_1,\rho_2) =
1_{\rho_1\rho_2}$. Moreover, one can show that each irreducible $\rho
\in \D^{\rm cov}_t$ posseses a left inverse $\varphi_{\rho}$, i.e.\ a
positive linear map on $\fA_{\rm vac}$ which preserves the unit and
fulfills $\varphi_{\rho}(A\rho(B)) = \varphi_{\rho}(A)B$. Then there
is for $\rho$ a number $\l_{\rho}$ so that
$$ \varphi_{\rho}(\epsilon(\rho,\rho)) = \l_{\rho}1.$$
The number $\l_{\rho}$ depends only on the equivalence class $[\rho]$
of $\rho$ and is called the {\it statistics parameter} of the
corresponding superselection sector. If $\l_{\rho} \ne 0$, then the
superselection sector is said to have {\it finite statistics}. We
define by ${\rm Sect}^{\rm cov}_{\rm fin}$ the set of all
translation-covariant superselection sectors of the underlying
observable quantum system which have finite statistics, and by
$\D^{\rm cov}_{\rm fin}$ the set of all endomorphisms $\rho$ with
$[\rho] \in {\rm Sect}^{\rm cov}_{\rm fin}$.

Finally, we need to recollect the notion of a conjugate charge. One
can show (cf.\ e.g.\ \cite{Rob.lec})
 that for each $\rho \in \D^{\rm cov}_{\rm fin}$ localized in
$O \in \cK$ there is some $\overline{\rho} \in \D^{\rm cov}_{\rm
  fin}$, also localized in $O$, together with isometries $R$ and
$\overline{R}$ in $\cA(O)$ which intertwine the endomorphisms
$\overline{\rho}\rho$ and $\rho\overline{\rho}$, respectively, with
the identical endomorphism of $\fA_{\rm vac}$, that is,
$$ \overline{\rho}(\rho(A))R = RA \quad {\rm and} \quad
\rho(\overline{\rho}(A))\overline{R} = \overline{R}A\,, \quad A \in
\fA_{\rm vac}\,.$$
In this case, one calls $[\overline{\rho}]$ the conjugate
superselection sector of $[\rho]$ or, synonymously, the conjugate
charge of $[\rho]$.
\\[10pt]
Doplicher and Roberts \cite{DRwhy} have shown that one can construct 
from 
$\D^{\rm cov}_{\rm fin}$ and the interwiners a system of local field 
algebras,
 acted upon by a faithful unitary representation of a compact group
--- called the gauge group --- such that the local algebras of the 
initially
given observable quantum system are embedded in the local field 
algebras as 
exactly containing the invariant elements under the gauge group 
action. In more
precise terms, they have shown that one can associate with
$(\cA_{\rm vac},\Usc_{\rm vac},\O_{\rm vac})$ a {\it quantum field 
system with
gauge symmetry} (QFSGS), defined as follows:
\begin{Definition} {\rm 
$(\cF,\Usc(\bR^n),U(G),\O,k)$ is a QFSGS for
 $(\cA_{\rm vac},\Usc_{\rm vac},\O_{\rm vac})$ and $\D^{\rm cov}_{\rm 
fin}$
if the following conditions hold:
\begin{description}
\item[{\small (QFSGS.1)}] $(\cF,\Usc(\bR^n),U(G),\O,k)$ is a QFTGA; 
the Hilbert space
on which the von Neumann algebras $\cF(O)$ 
of $\cF = \{\cF(O)\}_{O\in\cK}$ act will be denoted by
$\cH$. Moreover, $\fF' = \mathbb{C}{\bf 1}$.
\item[{\small (QFSGS.2)}] There is a $C^*$-algebraic monomorphism
$$ \pi: \fA_{\rm vac} \to \fF$$
containing the vacuum representation (i.e., the identical 
representation
of $\fA_{\rm vac}$ on $\cH_{\rm vac}$) as a sub-representation, and
such that $\pi(\cA_{\rm vac}(O))$ consists exactly of all $A \in 
\cF(O)$ having the
property $U(g)AU(g)^* = A$  for all $g \in G$. 
We will use the shorter notation
$$ \cA(O) :=  \pi(\cA_{\rm vac}(O))\,.$$
Moreover, the sub-Hilbert-space $\cH_0$ of $\cH$ which is generated 
by all vectors $\cA(O)\O$, as
$O$ ranges over the double cones, is cyclic for the algebras 
$\cF(O)$. 
\item[{\small (QFSGS.3)}] Let $[\rho] \in {\rm Sect}^{\rm cov}_{\rm 
fin}$ be a
superselection sector. Then there exists a finite dimensional,
irreducible, unitary 
representation
$$ v_{[\rho]} = (v_{[\rho]ji})_{i,j=1}^d$$
of $G$ (acting as a matrix representation for some suitable $d =
d_{[\rho]}$) so that, 
for each $O \in \cK$, there is a multiplet
$\psi_1,\ldots,\psi_d$ of elements in $\cF(O)$ having the
following properties:
\begin{eqnarray}
\label{1st}
 & & U(g) \psi_i U(g)^* = \sum_{j=1}^d \psi_j v_{[\rho]ji}(g)\,, \\
\label{2nd}
 & & \psi^*_i \psi_j = \d_{ij}{\bf 1}\,, \quad \ \ \ \sum_{j=1}^d
 \psi_j\psi_j^* = 
  {\bf 1}\,, \\
\label{3rd}
 & & \pi \lcrc \rho_O(A) = \sum_{j=1}^d\psi_j \pi(A)\psi_j^*\,, \quad 
  A \in \fA_{\rm vac}\,,\\
 & & \mbox{for\ some\ representer} \ \, \rho_O\ \,\mbox{of}\ \, 
[\rho]\ \,
 \mbox{localized\ in} \ \, O\,. \nonumber
\end{eqnarray}
These properties fix $v_{[\rho]}$ to within unitary equivalence.
\item[{\small (QFSGS.4)}] $\cF(O)$ is generated by $\cA(O)$ and all
  multiplets $\psi_j$, $j=1,\ldots,d_{[\rho]}$, with the properties
\eqref{1st},\eqref{2nd}, \eqref{3rd}, as $[\rho]$ ranges over all
superselection 
sectors in ${\rm Sect}^{\rm cov}_{\rm fin}$. For each
finite-dimensional, irreducible, 
unitary representation $v$ of $G$ there is some superselection sector
$[\rho] \in  
{\rm Sect}^{\rm cov}_{\rm fin}$ so that $v = v_{[\rho]}$ where
$v_{[\rho]}$ has the 
properties of (QFSGS.3).
\end{description} 
     }
\end{Definition} 
The conditions for a QFSGS associated with
 $(\cA_{\rm vac},\Usc_{\rm vac}(\bR^n),\O_{\rm vac})$
and $\D^{\rm cov}_{\rm fin}$ are given here in a form slightly
different from the 
statement in \cite{DRwhy}; however, the present formulation is
convenient for our 
purposes. 

It is plain that a QFSGS is a QFTGA fulfilling additional properties.
Condition (QFSGS.4) states, in particular, that ${\rm Sect}^{\rm
  cov}_{\rm fin}$ can be  
identified with the dual, $\widehat{G}$, of the gauge group
$G$. The connection 
between field algebra and superselection sectors is essentially
expressed through 
the multiplet operators $\psi_1,\ldots,\psi_d$ with the properties
listed in (QFSGS.3). 
In fact, the occurrence of such ``charge multiplets'' associated with
the superselection sector $[\rho]$ is equivalent to the presence of
the corresponding charge in the QFSGS $(\cF,\Usc(\bR^n),U(G),\O,k)$.
This will, basically, be our starting point for formulating criteria
that express  
``preservation of a charge'' in the scaling limit.

\section{Preservance of Charges in the Scaling 
Limit}\label{sec:DHRpreservance}
\setcounter{equation}{0}
Let us now discuss the problem of characterizing ``preservation of 
charges in the scaling
limit'' in greater detail. To this end, let 
$(\cF,\Usc(\bR^n),U(G),\O,k)$ be a QFSGS
associated with  $(\cA_{\rm vac},\Usc_{\rm vac}(\bR^n),\O_{\rm vac})$
and $\D^{\rm cov}_{\rm fin}$. Since $(\cF,\Usc(\bR^n),U(G),\O,k)$ is 
a QFTGA, we can
form the corresponding scaling algebra $\uFF$ as in Sec.\ 2. We may 
then define
$$ \underline{\fA}(O) =\{\underline{A} \in \uFF(O): 
\underline{A}{}_{\l} \in 
\cA(\l O)\}\,,$$
and it is not difficult to see that $\underline{\fA}(O)$ consists 
exactly of the
$\underline{A} \in \uFF(O)$ so that
$$ \underline{\beta}{}_g(\underline{A}) = \underline{A} $$
for all $g\in G$.

Now let $\ooi \in {\rm SL}^{\fF}(\o)$ be a scaling limit state on 
$\uFF$, and
denote by 
$$ (\FFoi,\Uscoi(\bR^n),\Uoib(\Goib),\Ooi,\koib) $$
the corresponding scaling limit QFTGA.
Let us also denote by 
$$ \cA_{0,\iota}(O) = \poi(\underline{\fA}(O)) \,, \quad O \in \cK\,,
  $$
the von Neumann algebra formed by the scaling limits of the
observables of the underlying QFSGS, and define by 
$$ \Foi(O)^{\Goib} = \
 \{\fbs \in \Foi(O): \Uoib(\gb)\fbs = \fbs \Uoib(\gb)\ \, \forall\ 
\gb \in \Goib\} $$
the fixed point algebra of the gauge group action in the scaling
limit. With this notation, and recalling that $\Hoi =
\overline{\FFoi\Ooi}$, we find:
\begin{Lemma}
\label{AandF}
 (i) $\cA_{0,\iota}(O) = \Foi(O)^{\Goib}$, $O \in \cK$.\\[4pt]
(ii) Suppose that $\Ooi$ is the unique (up to a phase) unit vector in
$\Hoi$ which is invariant under $\Uscoi(\bR^n)$ (equivalently, $\FFoi'
= \bC \cdot 1$). If $\fA_{0,\iota} =
\overline{\bigcup_O \cA_{0,\iota}(O)}^{C^*}$ is abelian, then $\FFoi =
\mathbb{C}\cdot 1$ and hence, $\Hoi = \mathbb{C}\Ooi$.
\end{Lemma}
{\it Proof. } (i) Clearly, one has $\cA_{0,\iota}(O) \subset
\Foi(O)^{\Goib}$. To show that the reverse inclusion holds, let $\fbs
\in \Foi(O)^{\Goib}$. Denote by $m_{0,\iota}(\boldsymbol{h}) = \int_G
d\mu(g)\,U_{0,\iota}(g) \boldsymbol{h} U_{0,\iota}(g)^*$,
$\boldsymbol{h} \in \FFoi$, the mean over the action of $G$ on
$\FFoi$. We have $m_{0,\iota}(\fbs) = \fbs$. Let
$\underline{F}^{(n)}$, $n \in \bN$, be a sequence of elements in
$\underline{\fF}(O)$ so that
$w$-$\lim_{n\to\infty}\poi(\underline{F}^{(n)}) = \fbs$. Such a
sequence exists because, by a Reeh-Schlieder argument, $\Ooi$ is
separating for $\Foi(O)$. Using this separating property of $\Ooi$
once more, also $m_{0,\iota}(\poi(\underline{F}^{(n)}))$ approximates
$\fbs$ weakly. On the other hand,
$$ m_{0,\iota}(\poi(\underline{F}^{(n)})) = \int_G
d\mu(g)\,\poi(\underline{\beta}_g(\underline{F}^{(n)})) = \poi(\int_G
d\mu(g)\,\underline{\beta}_g(\underline{F}^{(n)})) \,,$$
where we made use of the continuity of $\underline{\beta}_G$ in norm 
on
the scaling algebra to interchange representation and integration. 
Since 
$\int_G d\mu(g)\,\underline{\beta}_g(\underline{F}^{(n)})$ is
contained in $\underline{\fA}(O)$, we see that $\fbs$ is weakly
approximated by elements in $\cA_{0,\iota}(O)$ and hence is itself
contained in $\cA_{0,\iota}(O)$.
\\[6pt]
(ii) Under the given hypotheses, a result by Buchholz (Lemma 3.1 in
\cite{Bu.phsp}) shows that $\fA_{0,\iota} = \bC \cdot 1$. Hence, the
strongly continuous group $\beta_g^{(0,\iota)} = {\rm
  Ad}\,U_{0,\iota}(g)$, $g \in G$, of automorphisms on $\FFoi$ acts
ergodically, meaning that $\beta_g^{(0,\iota)}(\fbs) = \fbs$ for all
$g \in G$ implies $\fbs \in \bC \cdot 1$. Using Thm.\ 4.1 in
\cite{HLS}, it follows that the unique ergodic state for
$\beta_G^{(0,\iota)}$ on $\FFoi$ is a trace. The scaling limit vacuum
$\langle \Ooi,\,.\,\Ooi\rangle$ is a pure
$\beta_G^{(0,\iota)}$-invariant state on $\FFoi$ and hence is a
trace. (Purity of this state holds since the space of
translation-invariant vectors in $\Hoi$ is one-dimensional.) 
This implies 
$$ \langle \Ooi, \fbs^*\Uscoi(x)\fbs \Ooi \rangle = \langle \Ooi, \fbs
\Uscoi(-x) \fbs^*\Ooi\rangle $$
for each $\fbs \in \Foi(O)$, $O \in \cK$, and all $x \in
\bR^n$. Arguing with spectrum condition and clustering (as a
consequence of the assumption that every translation-invariant vector
in $\Hoi$ is a multiple of $\Ooi$) in the same manner as in the proof
of Lemma 3.1 in \cite{Bu.phsp}, one concludes that $\fbs \in \bC \cdot
1$. Hence $\FFoi = \bC \cdot 1$.
 \hfill $\Box$ 
\\[10pt]
The Lemma shows that all charges of the underlying QFSGS disappear in
a scaling limit theory once
the scaling limit theory is known to be classical for the observables,
provided the underlying theory satisfies very general conditions such
as clustering (QFTGA.8) or (for $n \ge 3$) Lorentz-covariance 
(QFTGA.6).

At this point, we should emphasize the distinction between charges in 
the scaling limit
QFTGA which are ``scaling limits of charges of the underlying 
QFSGS'', and ``charges arising
as superselection sectors of the scaling limit theory'', as was first 
discussed by
D.\ Buchholz \cite{Bu.Conf}. Charges of the first mentioned type 
correspond to the
situation that $\Goib$ is non-trivial and hence $\Uoib(\Goib)$ acts 
non-trivially
(and faithfully) on $\FFoi$. In this case, the action of 
$\Uoib(\Goib)$ on the elements
of $\FFoi$ may be seen as a short-distance remnant of the action of 
$U(G)$ on $\fF$
so that, correspondingly, the members of the spectrum
$\widehat{G}{}^{\bullet}_{0,\iota}$ of 
$\Goib$ may be viewed as representing short-distance remnants of the
charges in $\widehat{G}$ 
of the underlying QFSGS. It is important to note that, to some 
extent, these charges
of the scaling limit theory have been present in the underlying 
QFSGS. We will discuss this
case in more detail below.

The second type of charges in the scaling limit arises in a different 
way. One may consider
the scaling limit theory (induced by $\ooi \in {\rm 
SL}^{\fF}(\omega)$)
$$ (\cA_{0,\iota},\Uscoi(\bR^n),\Ooi)$$
which is gained from the observables of the underlying QFSGS as a new 
observable quantum
system in its own right
(provided it fulfills the assumptions of irreducibility).
 Then one can assign a set of superselection sectors
${\rm Sect}^{\rm cov}_{\rm fin} = {\rm Sect}^{\rm cov}_{\rm
  fin}(\cA_{0,\iota})$ to this observable quantum system, and by the 
Doplicher-Roberts reconstruction theorem\footnote{Provided that $ 
(\cA_{0,\iota},\Uscoi(\bR^n),\Ooi)$
fulfills all conditions for an observable quantum system in vacuum 
representation. See our
discussion after Prop.\ 5.6, and Prop.\ 5.7.}, we can now  associate  
to these data a QFSGS, which
we may denote by
$$
(\cF^{(0,\iota)},\Usc^{(0,\iota)}(\bR^n),U^{(0,\iota)}(G^{(0,\iota)}),\O^{(0,\iota)}, 
 k^{(0,\iota)})\,.$$
Thus, this QFSGS contains the superselection charges which arise in 
the scaling limit
theory of the {\it observables} of the underlying QFSGS.
In general, it may occur that $\FFoi$ is properly contained 
in $\fF^{(0,\iota)}$ and that 
$G_{0,\iota}^{\bullet}$ is a factor group of $G^{(0,\iota)}$ by some
non-trivial normal subgroup, so that the QFTGA associated with 
$\FFoi$ may 
be viewed as a proper subtheory (in the sense of \cite{DRwhy}) of the
QFSGS associated with $\fF^{(0,\iota)}$.
Buchholz \cite{Bu.Conf} 
proposed to consider such a case as a criterion for confinement,
 since it 
models  the situation where charges appear as superselection sectors 
of the (observables')
scaling limit
theory which do not arise as scaling limits of charges that occur as 
superselection
sectors in the underlying QFSGS.
We refer to \cite{Bu.Conf,Bu.QGC} for further discussion, and we note 
that examples
for superselection charges of this second type have been constructed 
for the Schwinger model
in two spacetime dimensions \cite{Bu.QGC,BV2}.

In the present work, we shall restrict attention solely to charges in
the scaling limit 
QFTGAs of an underlying QFSGS of the first mentioned type, i.e.\ 
which arise
as ``scaling limits'' of charges present in the underlying
QFSGS. Having clarified this 
basic point, we must find criteria which express that a charge of the
underlying QFSGS has a non-trivial scaling limit. There are some
prefatory observations which may be helpful as a guideline. We have
already seen that the gauge group $G^{\bullet}_{0,\iota} =
G_{0,\iota}/N_{0,\iota}$ of a scaling limit QFTGA is a factor group of
$G_{0,\iota}$ which is itself a copy of $G$, the gauge group of the
underlying QFSGS. It may in general happen that the normal subgroup
$N_{0,\iota}$ is non-trivial, and hence that $G_{0,\iota}^{\bullet}$
is ``smaller'' than $G$. In this situation, certainly not all the
charges of the underlying QFSGS will have counterparts in the scaling
limit QFTGA. Thus, we will in general be confronted with a situation
which is in a sense complementary to that of $\FFoi \subset
\fF^{(0,\iota)}$ mentioned just before and where, morally, the scaling
limit QFTGA associated with $\FFoi$ corresponds to a subtheory of the
underlying QFSGS, at least as far as the charge structure is
concerned.\footnote{The dynamics of the theories corresponding to
  $\FFoi$ and $\fF$ are expected to be different and so the former
  can't be a subtheory of the latter in the full sense of the 
definition.}
 However, since there is no inclusion of
$\FFoi$ into $\fF$, we need to establish a correspondence between
elements in $\FFoi$ and in $\fF$ which allows to decide if charges
present in the underlying QFSGS are also present in the scaling limit.

As we have mentioned above, the presence of a superselection charge 
in the underlying
QFSGS manifests itself through the presence of charge multiplets
$\psi_1,\ldots,\psi_d \in \fF$ which transform under a finite 
dimensional, irreducible, 
unitary representation $v_{[\rho]}$ as described in (QFSGS.3). 
This will be the starting point for our criterion of charge 
preservance in the
scaling limit. To fix ideas,
 let $(\cF,\Usc(\bR^n),U(G),\O,k)$ denote the underlying QFSGS, and 
let
$[\rho] \in {\rm Sect}^{\rm cov}_{\rm fin}$
 be one of its superselection  sectors, and pick some
arbitrary $O\in \cK$. Then there is
a finite-dimensional, irreducible, unitary representation $v_{[\rho]}$
of $G$ and, for each $\lambda > 0$, a multiplet of elements
$\psi_1(\l),\ldots,\psi_d(\l)$ in $\cF(\l O)$ having the properties of
(QFSGS.3) with respect to the localization region $\l O$. We will
refer to any such multiplet family 
$\{\psi_1(\l),\ldots,\psi_d(\l)\}_{\l > 0}$
as a {\bf scaled multiplet} for $[\rho]$.
 The principal idea is now
to view the functions $\l \mapsto \psi_j(\l)$ as ``would-be'' 
elements of
$\uFF(O)$ and to follow their fate as $\l$ approches $0$.
However, these functions won't satisfy  the
``phase-space constraint'' condition (c) of Def.\ 2.2 which is
essential in order to interpret them as orbits of field algebra
elements under (abstract) renormalization group transformations.
Hence, if $\ooi$ is a scaling limit state, in general one can't form
$\poi(\psi_j(\,.\,))$ since $\psi_j(\,.\,)$ won't belong to the 
scaling algebra 
$\uFF$. But one can still check if, in the scaling limit, scaled
multiplets become close to elements of $\poi(\uFF)$ so that they can 
effectively
be regarded as representing elements in the scaling limit von Neumann
algebras $\Foi(O) = \poi(\uFF(O))''$. We will introduce some new
terminology which gives this idea a more precise shape.
\begin{Definition} \label{asympcont} {\rm 
Let $\ooi \in {\rm SL}^{\fF}(\o)$ be a scaling limit state of the 
underlying
QFSGS.\\
 Then we say that a family ${\sf F} = \{{\sf F}(\l)\}_{\l > 0}$ 
fulfilling (i) ${\sf F}(\l)
\in \fF(\l O_1)$ for some $O_1 \in \cK$, (ii) $\sup_{\l > 0}\,||{\sf 
F}(\l)|| <
\infty$,
  and (iii) $\sup_{\l}\,||{\beta}_g({\sf F}(\l))-{\sf F}(\l)|| \to 0$ 
for $g \to 1_G$, is 
 {\em asymptotically
  contained} in $\Foi(O)$ if the following holds:
\\[4pt] For each given $\epsilon > 0$ there are elements
$\uF$ and $\uF'$ in
$\uFF(O)$ such that
\begin{equation} \label{ststascon}
 \limsup_{\k}\,\left(||({\sf F}(\l_{\k}) - \uF_{\l_\k})\O|| +
   ||({\sf F}(\l_{\k}) -\uF'_{\l_{\k}})^*\O|| \right) <
 \epsilon\,,
\end{equation}
where the net $\{\l_{\k}\}_{\k \in K}$ of positive numbers
converges to $0$, with $\ooi = \lim_{\k}\,\underline{\o}_{\l_{\k}}$ on
$\uFF$. }
\end{Definition}
We will next collect some immediate consequences of this definition; 
this requires yet some
further notation. Given a finite-dimensional Lie group $X$ endowed 
with a Borel measure
$\mu$ which is invariant under group transformations (for our 
purposes, $X = \bR^n$ or
$X = \widetilde{\cP}^{\uparrow}_+$), we call a sequence of functions 
$\{h_{\nu}\}_{\nu \in \bN}$
of class $L^1(X,\mu) \cap C_0^\infty(X)$ a {\it $\delta$-sequence} if 
${\rm supp}\,h_{\nu + 1} \subset {\rm supp}\,h_{\nu}$, 
$\bigcap_{\n} {\rm supp}\,h_{\n} = 1_X$,
$\sup_{\nu}||h_{\n}||_{L^1} < \infty$, 
and if $\int_X h_{\nu} \chi\,d\mu$ converges to $\chi(1_X)$ as $\nu 
\to \infty$ 
for all continuous functions $\chi$ on $X$. Here, $1_X$ is the group 
unit element; note that $1_{\bR^n} = 0$. 
\begin{Lemma} \label{alliseq}
Let $\ooi$ be a scaling limit state of the underlying QFSGS, and
suppose that ${\sf F} = \{{\sf F}(\l)\}_{\l > 0}$ is a family of
elements in $\fF$ with the properties as in the previous
definition. Then the following 
statements are equivalent:
\begin{itemize}
\item[$(a)$] ${\sf F}$ is  asymptotically
  contained in $\Foi(O)$ for all $O \supset \overline{O_1}$,
\item[$(b)$] In the scaling limit, ${\sf F}$ is approached in
  the $*$-strong topology by elements in $\poi(\uFF(O))$ in the
  following sense: Whenever $O \supset \overline{O_1}$, $\epsilon > 0$
  and finitely many 
  $\uF^{(1)},\ldots,\uF^{(N)} \in \uFF$ are given, then there is an
  $\uF \in \uFF(O)$ fulfilling $||\uF|| \le \sup_{\l}\,||{\sf 
F}(\l)||$ and
$$ \limsup_{\k}\,\left(||({\sf F}(\l_{\k}) - 
\uF_{\l_{\k}})\uF^{(j)}_{\l_{\k}}\O|| +
   ||({\sf F}(\l_{\k}) -\uF_{\l_{\k}})^*\uF^{(j)}_{\l_\k}\O|| \right) 
<
 \epsilon\,, \quad j = 1,\dots,N\,,$$
where $\{\l_{\k}\}_{\k \in K}$ is as in the previous
definition,
\item[$(c)$] Given any $\delta$-sequence $\{h_{\nu}\}$ on $\bR^n$, 
there holds
\begin{equation} \label{asapprox}
  \lim_{(\k,\n)}\,\left(||\,(\, (\underline{\alpha}_{h_{\n}}{\sf 
F})(\l_{\k})
    - {\sf F}(\l_{\k})\,)\O\,|| + ||\,(\, 
(\underline{\alpha}_{h_{\n}}{\sf F})(\l_{\k})
    - {\sf F}(\l_{\k})\,)^*\O\,|| \right) = 0 \,,
\end{equation}
where the limit is taken with respect to the partial ordering
on $K \times \bN$ given by $(\k,\n) > (\k',\n')$ $ :\Leftrightarrow$
$\k > \k'$ and $\n > \n'$, and
$$ (\underline{\alpha}_h{\sf F})(\l) = \int d^nx\, h(x) \alpha_{\l
  x}({\sf F}(\l))\,, \quad \l > 0\,, \ h \in L^1(\mathbb{R}^n)\,.$$
(The latter integral is to be interpreted in the weak topology on
$\fF$; $\{\l_{\k}\}_{\k \in K}$ is as before).
\end{itemize}
\end{Lemma}
{\it Proof. } $(a) \Rightarrow (c)$.
Let $\epsilon > 0$ be arbitrary. Then we must show that there exist
$\k_{\epsilon} \in K$ and $\n_{\epsilon} \in \mathbb{N}$ so that
\begin{equation} \label{atoc}
 ||\,(\,(\underline{\alpha}_{h_{\n}}{\sf F})(\l_{\k}) - {\sf 
F}(\l_{\k})\,)\O\,||
  + ||\,(\,(\underline{\alpha}_{h_{\n}}{\sf F})(\l_{\k}) - {\sf 
F}(\l_{\k})\,)^*\O\,|| <
 \epsilon 
\end{equation}
holds for all $(\k,\n) > (\k_{\epsilon},\n_{\epsilon})$.
Writing $(\underline{\alpha}_h\uF)_{\l} = \int d^nx\,h(x)\alpha_{\l
  x}(\uF_{\l})$, we consider the estimate
\begin{eqnarray} \label{theestimate}
||(\,(\underline{\alpha}_{h_{\nu}}{\sf F})(\l_{\k}) -
  {\sf F}(\l_{\k})\,)^{\sharp}\O|| 
& \le & ||(\,(\underline{\alpha}_{h_{\nu}}{\sf F})(\l_{\k})-
(\underline{\alpha}_{h_{\nu}}\uF)_{\l_{\k}}\, )^{\sharp}\O|| 
 + ||(\,(\underline{\alpha}_{h_{\nu}}\uF)_{\l_{\k}} -
\uF_{\l_{\k}}\,)^{\sharp} \O||\nonumber \\
& & + \ ||(\uF_{\l_{\k}} - {\sf F}(\l_{\k}))^{\sharp}\O||\,, 
\end{eqnarray}
where $(\dots)^{\sharp}$ stands for either $(\dots)$ or $(\ldots)^*$. 
Now we use the fact that, owing to the definition of asymptotic 
containment, one may choose $\uF,\uF'$ in such a way that there is
some $\k_{\epsilon}$ with $(\sup_{\nu}||h_{\nu}||_{L^1}||(f(\l_{\k}) 
-\uF_{\l_{\k}}^{\sharp})^{\sharp}\O|| 
< \epsilon/6$ for all $\k > \k_{\epsilon}$, where $\uF^{\sharp} = 
\uF$ or
$\uF^{\sharp} = \uF'$ according if $(\ldots)^{\sharp} = (\ldots)$ or
$(\ldots)^{\sharp} = (\ldots)^*$.
Denoting by $\hat{h}$ the Fourier transform of $h$ and by $P =
(P_{\nu})_{\nu = 0}^{n-1}$ the selfadjoint generators of the unitary
translation group of the underlying QFSGS, the first term on the right
hand side of \eqref{theestimate} is seen to equal
$$ ||\hat{h}_{\nu}(P)({\sf F}(\l_{\k}) - 
\uF_{\l_{\k}}^{\sharp})^{\sharp}\O|| \le
\sup_{\n}||h_{\n}||_{L^1}||({\sf F}(\l_{\k}) - 
\uF_{\l_{\k}}^{\sharp})^{\sharp}\O|| < \epsilon/6 $$
for all $\k > \k_{\epsilon}$. 
The second term on the right hand side
of \eqref{theestimate} can be estimated by
$$\sup_{\n}||h_{\n}||_{L^1} \sup_{x \in {\rm 
supp}\,h_{\n}}\,||\underline{\alpha}_{
  x}(\uF^{\sharp}) - \uF^{\sharp} || $$
and using the continuity of $\uF^{\sharp}$ with respect to
$\underline{\a}_x$, this quantity may be made smaller than 
$\epsilon/6$
for all $\n$ smaller than some suitable $\n_{\epsilon}$. Summing up,
we obtain \eqref{atoc} for all $(\k,\n) > 
(\k_\epsilon,\n_{\epsilon})$.
\\[2pt]
${}$ \quad $(c) \Rightarrow (b)$. 
It holds that $\l \mapsto \underline{\Phi}_{\l} =
(\underline{\alpha}_{h_{\nu}}{\sf F})(\l)$ is contained in
$\uFF(O_{\times})$ where $O_{\times}$ is any double cone containing
$O_1 + {\rm supp}\,h_{\nu}$. 
A standard Reeh-Schlieder argument shows that, if
$W$ is any wedge region in the causal complement of $O_{\times}$, 
then $\Foi(W)\Ooi$ is dense in $\Hoi$ hence $\Foi^{t}(W)\Ooi$ is 
dense in $\Hoi$.
As a
consequence, there is for given $\uF^{(j)} \in \uFF$ and given $\eta >
0$ some $B^{(j)} \in \uFF(W)$ so that, if $V_{0,\iota}$ is the natural twist on $\Foi$,
$$ ||(\poi(\uF^{(j)}) - V_{0,\iota}\poi(\underline{B}^{(j)}))\Ooi|| = \lim_\k||(\uF^{(j)}_{\l_\k} - V\underline{B}^{(j)}_{\l_\k})\O||< \eta.\,$$
Thus, making first $\eta$ and then $\n^{-1}$ small enough, one can 
arrange
that
\begin{eqnarray*}
\lefteqn{ \limsup_{\k}\,\left( ||(\underline{\Phi}_{\l_{\k}} -
    {\sf F}(\l_{\k}))\uF^{(j)}_{\l_{\k}}\O|| 
+||(\underline{\Phi}_{\l_{\k}} -
    {\sf F}(\l_{\k}))^*\uF^{(j)}_{\l_{\k}}\O|| \right)  }
\\
& \le & \limsup_{\k}\,\left( ||(\underline{\Phi}_{\l_{\k}} -
    {\sf F}(\l_{\k}))V\underline{B}^{(j)}_{\l_{\k}}\O|| 
+||(\underline{\Phi}_{\l_{\k}} -
    {\sf F}(\l_{\k}))^*V\underline{B}^{(j)}_{\l_{\k}}\O||\right) + 
4\eta \sup_{\k}\,||{\sf F}(\l_{\k})||
\\
& = & \limsup_{\k}\,\left( 
||\underline{B}^{(j)}_{\l_{\k}}V^*(\underline{\Phi}_{\l_{\k}} -
    {\sf F}(\l_{\k}))\O|| 
+||\underline{B}^{(j)}_{\l_{\k}}V^*(\underline{\Phi}_{\l_{\k}} -
    {\sf F}(\l_{\k}))^*\O|| \right)+ 4\eta \sup_{\k}\,||{\sf 
F}(\l_{\k})||
\\
& \le & \limsup_{\k}\,||\underline{B}^{(j)}||\left( 
||(\underline{\Phi}_{\l_{\k}} -
    {\sf F}(\l_{\k}))\O|| +||(\underline{\Phi}_{\l_{\k}} -
    {\sf F}(\l_{\k}))^*\O||\right)+ 4\eta \sup_{\k}\,||{\sf 
F}(\l_{\k})||
\end{eqnarray*}
can be made smaller than any given $\epsilon > 0$; then, for
a sufficiently large $\nu$, $\underline{\Phi}$ can be taken as the
$\uF$ required in $(b)$. Note that in
passing from the second line to the third we have used that
$V\underline{B}^{(j)}_{\l}V^*$ commutes with $(\underline{\Phi}_{\l} -
    {\sf F}(\l))$ and its adjoint, 
     because of the localization
    properties of the operators involved.

The implication $(b) \Rightarrow (a)$ is obvious. \hfill $\Box$
\\[10pt]
{\it Remark.} 
In view of statement (b) of the previous Lemma, one might refer to our
notion of asymptotic containment more precisely as $*$-strong
asymptotic containment.
It should then be obvious how to introduce, e.g., the notion of
strong or weak asymptotic containment in $\Foi(O)$ for families
${\sf F} = \{{\sf F}(\l)\}_{\l> 0}$ fulfilling the properties as in 
\ref{asympcont}. 
One could also drop condition (iii) on ${\sf F}$ in the
definition of asymptotic containment, then having to define in Lemma
\ref{alliseq}
  $\underline{\alpha}_h{\sf F}$ differently, cf.\ \eqref{smoothing}.
\\[10pt]
After these preparations, we can now present our criterion for
preservance of charges in the scaling limit.
\begin{Definition}  
\label{def:DHRpreservance} 
{\rm Let $\ooi \in {\rm SL}^{\fF}(\o)$ be a scaling limit state of 
the underlying
QFSGS, and let $[\rho] \in {\rm Sect}^{\rm cov}_{\rm fin}$ be a
superselection sector. Then we say that the charge $[\rho]$ is {\em
  preserved in the scaling limit QFTGA of $\ooi$} if, for each $O_1
\in \cK$, there is some scaled multiplet
$\{\psi_1(\l),\ldots,\psi_d(\l)\}_{\l > 0}$ for $[\rho]$ with
$\psi_j(\l) \in \cF(\l O_1)$ such that all families
$\{\psi_j(\l)\}_{\l > 0}$, $j = 1,\ldots,d$, are asymptotically 
contained
in $\Foi(O)$ if $O \supset \overline{O_1}$. }
\end{Definition}
Let us briefly convince ourselves that each family $\{\psi_j(\l)\}_{\l
  > 0}$ of a scaled multiplet satisfies the assumptions (i)--(iii) of
Def.\ \ref{asympcont}. Clearly, only condition (iii) need be checked,
and one has
$$ \sup_{\l}\,|| \beta_g(\psi_j(\l)) - \psi_j(\l) || =
\sup_{\l}\,||\sum_{i = 1}^d \psi_i(\l)(v_{[\rho]ij}(g) -
\delta_{ij})||
\le d \max_{i,j}\,|v_{[\rho]ij}(g) -\delta_{ij}|
 $$
where the last term tends to $0$ if $g \to 1_G$ if $G$ is a continuous
group.
\\[6pt]
We remark that, in view of part (c) of Lemma \ref{alliseq}, a similar
criterion has been used recently by Morsella \cite{Morsella1}. Part
(c) of Lemma \ref{alliseq} also provides some insight into the basic 
mechanism
which might cause charges to disappear in the scaling limit.
To elaborate on that, we consider a scaled multiplet
$\{\psi_1(\l),\ldots,\psi_d(\l)\}_{\l > 0}$ for the charge
$[\rho]$. Moreover, for $h \in L^1(\bR^n)$ with compact support and $h
\ge 0$, $\int d^nx\,h(x) = 1$, we define
$$ \underline{\Phi}^{(h,j)}_{\l} = 
(\underline{\alpha}_h\psi_{j})(\l)\,, \l >
0\,.$$
Now by Lemma \ref{alliseq} it follows that for the charge
$[\rho]$ to be preserved in the scaling limit QFTGA of $\ooi$, one
must be able to choose a scaled multiplet and $h$ in such a way that
$||\poi(\underline{\Phi}^{(h,j)})\Ooi||$ comes arbitrarily close to 
1. It
could however happen that for all scaled multiplets and any choice of
$h$  one ends up with 
$$ ||\poi(\underline{\Phi}^{(h,j)}{}^*)\Ooi|| = 0\,,$$
which also implies $\poi(\underline{\Phi}^{(h,j)}) = 0$ since
$\Ooi$ is separating for the local field algebras of the scaling 
limit QFTGA.
We can interpret this as follows.
 The convolution of the scaled charge multiplets
 $\psi_j(\l)$ with respect to the scaled action of the translations,
 which produces elements $\underline{\Phi}^{(h,j)}$ in
 $\underline{\fF}$,  results in an energy damping of the charged
 states that are 
 obtained by applying the $\underline{\Phi}^{(h,j)*}_{\l}$ to the
 vacuum vector $\O$. This energy damping scales inversely, that is,
 proportional to $\l^{-1}$, to the localization scale of the
 $\underline{\Phi}^{(h,j)}_{\l}$. Depending on the dynamics of the
 underlying QFSGS, it may happen that the amount of energy-momentum
 required to create the charged vectors $\psi_j(\l)^*\O$
 from the vacuum in a small
 region of scale $\l$ is typically larger than $\sim \l^{-1}$, e.g.\
 of the type $\sim \l^{-q}$ with some $q > 1$. In this case, the
 energy damping leads to a ``blotting out'' of the charged 
contributions of
 $\underline{\Phi}^{(h,j)}_{\l}\O$, resulting in the vanishing of
the norm of these vectors as $\l$ approaches $0$. 

In other words, our preservance criterion amounts essentially to requiring that the energy-momentum needed to localize the considered charges is only restricted by Heisenberg's principle, and, also in view of the specific phase space properties of renormalization group orbits  encoded in the scaling algebra construction, it is evident that a condition of this kind is needed in order to single out ``elementary, pointlike'' charges, which survive the scaling limit (cf.\ also mechanism (B) below).

Let us sketch two --- quite distinct --- physical mechanisms that may 
account for the
disappearance of charges in the scaling limit.

\begin{itemize}
\item[(A)] There is a strongly attractive force between the charges
at short distances. This may have the effect
that certain ``compounds'' of charges are dynamically more
favourable than single charges. That is to say, it may
cost far less energy to create a compound of several charges at small
scales than the single charges contained in the compound. In this
case, the compound charges could survive the scaling limit (i.e.\ be
preserved), while certain single charges disappear since their
creation costs too much energy at small scales. The compound charges
preserved in the scaling limit could then well be invariant under some
normal subgroup of the gauge group of the underlying quantum field
theory. In a sense, this mechanism is complementary to that of
confinement at finite distances of charges which would be viewed as
``free'' charges in the short-distance scaling limit (asymptotic
freedom) as in QCD. There, one expects that the colour charges
correspond to charges which are present as superselection charges of a
scaling limit quantum field theory (corresponding to field multiplets
in $\cF^{(0,\iota)}$, not in $\Foi$), while in the underlying quantum
field theory, at finite scale, only colour-neutral compounds of the
colour-charges appear. 
\item[(B)] The charges are strongly repellent at short 
distances.\footnote{This
mechanism has been pointed out to us by D.\ Buchholz} In this 
situation, in order to
localize two or more charges in a small spacetime region of scale 
$\l$, one
requires more energy than of the order of $\l^{-1}$. This 
will typically
lead to disappearance of certain charges in the scaling limit in the 
following
way. If $\{\psi_{1}(\l),\ldots,\psi_d(\l)\}_{\l > 0}$ is a scaled 
multiplet for
some charge $[\rho]$, then there acts the tensor product 
representation
$\overline{v}_{[\rho]} \otimes \overline{v}_{[\rho]}$ of the gauge group on the space 
of vectors
spanned by $\psi_j^*(\l)\psi_k^*(\l)\O$, for each $\l > 0$. This 
tensor product
representation can be decomposed into a sum of irreducibles, so that 
$$ \psi_j^*(\l)\psi_k^*(\l)\O = \sum_{\ell,m}a_{\ell,m}(\l) 
{\psi}_{\ell,m}^*(\l)\O$$
with suitable coefficients $a_{\ell,m}(\l)$, where the 
$\{{\psi}_{1,m}(\l),\ldots,{\psi}_{d_m,m}(\l)\}_{\l>0}$ are scaled 
multiplets
corresponding to the charges $[\rho_m]$ labelled by $m$ which appear 
in the decomposition
of $\overline{v}_{[\rho]} \otimes \overline{v}_{[\rho]}$. Assuming now that the 
interaction between charges
of type $[\rho]$ is strongly repellent at short distances this will, 
in keeping with our
discussion on the energy damping caused by applying 
$\underline{\a}_h$, typically result in 
$$ \poi(\underline{\a}_h(\psi_j^*\psi_k^*))\Ooi = 0$$
for some indices $j,k$, and therefore in
$$ \poi(\underline{\a}_h({\psi}_{\ell,m}^*))\Ooi = 0 $$
for some indices $\ell,m$.
\end{itemize} 
Concerning the question whether our criterion for preservance of
charges is fulfilled in certain quantum field models, we note that
the charges of the Majorana-Dirac field satisfy indeed this criterion 
in
all scaling limit states (see Appendix~\ref{app:example}). We also remark that all charges in a dilation covariant theory complying with the hypotheses of~\cite{Rob.dil} are
preserved in all scaling limit QFTGAs: if for a multiplet $\psi_j\in \cF(O)$, $j=1,\dots,d$, associated to a given sector $[\rho] \in {\rm Sect}^{\rm cov}_{\rm fin}$, we define $\psi_j(\l)= D(\l)\psi_j D(\l)^{-1}$, $j=1,\dots,d$, $D(\l)$ being the unitary
implementation of dilations, we get a scaled
multiplet $\{\psi_1(\l),\ldots,\psi_d(\l)\}_{\l > 0}$ for $[\rho]$ such that $\psi_j(\cdot) \in \uFF(O)$, as can be easily derived from the commutation relations between
dilations and translations, and then in particular $\psi_j(\cdot)$ is asymptotically contained in every $\Foi(O)$. (This proves incidentally that a dilation covariant theory admits no classical
scaling limit as soon as it has multiplets with dimension $d> 1$).  

Our criterion of charge preservance not only bars the situation
of charge disappearance,
but it even implies that the limits of 
$\poi(\underline{\Phi}^{(h,j)})$,
$j =1,\ldots,d$, as $h$ tends to the $\delta$-measure, yield 
charge multiplets corresponding to the
charge $[\rho]$ with respect to their transformation behaviour under 
the
scaling limit gauge group. This is the content of the following
statement.
\begin{Proposition}\label{prop:limitmultiplet}
Suppose that the charge $[\rho]$ is preserved in the scaling limit
QFTGA of $\ooi$. Let $\{\psi_1(\l),\ldots,\psi_d(\l)\}_{\l > 0}$ be a
scaled multiplet for $[\rho]$ which is asymptotically contained in
$\Foi(O)$, and let $\underline{\Phi}^{(h,j)}$ be defined as before
with respect to the $\{\psi_j(\l)\}_{\l > 0}$. 

Then for any $\d$-sequence $\{h_{\n}\}$ on $\bR^n$, the limit
operators 
\begin{equation} \label{SLmulti}
 \boldsymbol{\psi}_j = s\mbox{-}\lim_{\nu \to \infty}\,
\poi(\underline{\Phi}^{(h_{\mu},j)}) \quad {\it and}\quad
 \boldsymbol{\psi}_j^* = s\mbox{-}\lim_{\nu \to \infty}\,
\poi(\underline{\Phi}^{(h_{\mu},j)})^* 
\end{equation}
exist, are independent of the chosen $\d$-sequence and are contained 
in $\Foi(\hat{O})$
whenever $\hat{O} \supset \overline{O}$. Furthermore,
$\boldsymbol{\psi}_1,\ldots, \boldsymbol{\psi}_d$ forms a multiplet
transforming under the adjoint action of $\Uoib(\Goib)$ according to
the irreducible, unitary representation $v_{[\rho]}$. More precisely,
denoting by $G \owns g \mapsto g^{\bullet} \in \Goib$ the quotient
map, there is a finite-dimensional, irreducible, unitary
representation $v_{[\rho]}^{\bullet}$ of $\Goib$ so that
$v_{[\rho]}^{\bullet}(g^{\bullet}) = v_{[\rho]}(g)$ for all $g\in G$
and 
$$ \Uoib(g^{\bullet})\boldsymbol{\psi}_j \Uoib(g^{\bullet})^* =
\sum_{i=1}^d \boldsymbol{\psi}_i
v_{[\rho]ij}^{\bullet}(g^{\bullet})\,, \quad g^{\bullet} \in
\Goib\,.$$
\end{Proposition}
{\it Proof. } First we need to establish existence of the limit.
 Let $\{h_{\n}\}$ and $\{\tilde{h}_{\tilde{\n}}\}$ be
$\d$-sequences on $\bR^n$. Choose any $\epsilon > 0$. Then one can 
find $\nu_0
> 0$ so that
\begin{eqnarray*} 
\lefteqn{||(\,\poi(\underline{\Phi}^{(h_{\nu},j)})-
  \poi(\underline{\Phi}^{(\tilde{h}_{\tilde{\nu}},j)}\,)\Ooi|| =
 \lim_{\k}\,||(\underline{\Phi}^{(h_{\nu},j)}_{\l_{\k}}
 -\underline{\Phi}^{(\tilde{h}_{\tilde{\nu}},j)}_{\l_{\k}})\O   ||   
  } \\
& \le & \limsup_{\k}\, \left(
 ||(\underline{\Phi}^{(h_{\nu},j)}_{\l_{\k}} - \psi_j(\l_{\k}))\O|| +
 ||(\underline{\Phi}^{(\tilde{h}_{\tilde{\nu}},j)}_{\l_{\k}} -
 \psi_j(\l_{\k}))\O|| 
\right) < \epsilon
\end{eqnarray*}
if $\nu,\tilde{\nu} > \nu_0$. This shows that
$\poi(\underline{\Phi}^{(h_{\nu},j)})\Ooi$ is a Cauchy sequence in
$\nu \to \infty$ and hence has a limit in $\Hoi$; it shows also that 
the
limit is independent of the chosen $\d$-sequence. Since $\Ooi$ is 
separating for the local
scaling limit field algebras and $||\underline{\Phi}^{(h_{\nu},j)}||$
is bounded uniformly in $\nu$, one can thus conclude that 
$\poi(\underline{\Phi}^{(h_{\nu},j)})$ converges strongly to some
$\boldsymbol{\psi}_j$ which is contained in $\Foi(\hat{O})$ if
$\hat{O} \supset \overline{O}$. Similarly one argues that
$\poi(\underline{\Phi}^{(h_{\nu},j)})^*$ converges strongly to
$\boldsymbol{\psi}_j^*$. 

Next we demonstrate $\boldsymbol{\psi}_j^*\boldsymbol{\psi}_k =
\delta_{jk}1$. To this end, we observe that for any $\uF \in \uFF$
there holds the following chain of equations,
\begin{eqnarray*}
\lefteqn{\langle
  \poi(\uF)\Ooi,(\boldsymbol{\psi}^*_j\boldsymbol{\psi}_k -
  \delta_{jk}1)\Ooi\rangle} \\
&=& \lim_{\nu \to \infty}\,\lim_{\k}\,\boldsymbol{[}\,
 \langle
 \underline{\Phi}^{(h_{\nu},j)}_{\l_{\k}}\uF_{\l_{\k}}\O,
\underline{\Phi}^{(h_{\nu},k)}_{\l_{\k}}\O\rangle - \delta_{jk}\langle
\uF_{\l_{\k}}\O,\O\rangle  
 \, \boldsymbol{]} \\
&=& \lim_{\nu\to \infty}\,\lim_{\k}\, \boldsymbol{[}\,\langle
\psi_j(\l_{\k})\uF_{\l_{\k}}\O,\psi_k(\l_{\k})\O\rangle -
\delta_{jk}\langle \uF_{\l_{\k}}\O,\O\rangle \\
& & \quad \quad \ \ \ \ \ \  + \
\langle(\underline{\Phi}^{(h_{\nu},j)}_{\l_{\k}} -
\psi_j(\l_{\k}))\uF_{\l_{\k}} \O,\psi_k(\l_{\k})\O\rangle \\
& & \quad \quad \ \ \ \ \ \  + \ \langle
\underline{\Phi}^{(h_{\nu},j)}_{\l_{\k}}\O,
 (\underline{\Phi}^{(h_{\nu},k)}_{\l_{\k}} -\psi_k(\l_{\k}))\O\rangle
 \, \boldsymbol{]}\,.
\end{eqnarray*}
The expression on the third to last line is equal to $0$ since
$\psi_j(\l)^*\psi_k(\l) = \delta_{jk}1$ by assumption, and the limits
of the
expressions on the last two lines vanish by the argument having led to
the conclusion $(c) \Rightarrow (b)$ in the proof of Lemma
\ref{alliseq}. This proves $\boldsymbol{\psi}^*_j\boldsymbol{\psi}_k =
\delta_{jk}1$ by the separating property of $\Ooi$ for the local field
algebras in the scaling limit.

The proof of $\sum_{j = 1}^d \boldsymbol{\psi}_j\boldsymbol{\psi}^*_j
= 1$ is completely analogous. 

For the last part of the statement, we observe that 
$$ \Uoi(g)\boldsymbol{\psi}_j \Uoi(g)^* = \sum_{k=1}^d 
 \boldsymbol{\psi}_{k}v_{[\rho]kj}(g)\,,
\quad g \in G\,,$$
is simply a consequence of
$$ \underline{\b}_g(\underline{\Phi}^{(h,j)}) =
\sum_{k=1}^d\underline{\Phi}^{(h,k)}v_{[\rho]kj}(g)\,, \quad g \in
G\,;$$
this, in turn, can be seen from $\underline{\Phi}^{(h,j)} =
\underline{\alpha}_h\psi_j$ and the commutativity of
$\underline{\beta}_g$ and $\underline{\alpha}_x$.

On the other hand, from the definition of $\Noi$ one obtains
$$ \boldsymbol{\psi}_j\Ooi = \Uoi(n)\boldsymbol{\psi}_j\Ooi =
\sum_{k=1}^d\boldsymbol{\psi}_kv_{[\rho]kj}(n)\Ooi $$
for all $n \in \Noi$, and multiplying by $\boldsymbol{\psi}^*_i$ from
the left yields $\delta_{ij}\Ooi = v_{[\rho]ij}(n)\Ooi$ for all $n \in
\Noi$. This shows $v_{[\rho]}(n) = 1$ (the unit matrix) for all
$n\in\Noi$ and hence there is an irreducible, unitary representation
$v^{\bullet}_{[\rho]}$ of $\Goib$ so that
$v_{[\rho]}^{\bullet}(g^{\bullet}) = v_{[\rho]}(g)$ for all $g \in G$,
proving the last part of the statement.
 ${}$ \hfill $\Box$
\\[10pt]
 There is an obvious connection between the scaling limits of scaled
multiplets for a charge $[\rho]$ and the
scaling limits of endomorphisms induced by
scaled multiplets in case that $[\rho]$ is preserved in a scaling
limit state. While fairly immediate, we put the corresponding result
on record here.
\begin{Proposition}\label{prop:SLendo}
Let $\ooi \in {\rm SL}^{\fF}(\o)$ and let $[\rho] \in {\rm Sect}^{\rm
  cov}_{\rm fin}$ be a charge of the underlying QFSGS which is
preserved in the scaling limit QFTGA of $\ooi$. Moreover, let
$\{\psi_1(\l),\ldots,\psi_d(\l)\}_{\l > 0}$ be a scaled multiplet for
$[\rho]$ asymptotically contained in $\Foi(O)$ and let, with respect
to this scaled multiplet,
$\boldsymbol{\psi}_1,\ldots,\boldsymbol{\psi}_d$ be defined as in
\eqref{SLmulti}.

Then for each $\uA \in \underline{\fA}$ the family
$\{\rho(\uA)(\l)\}_{\l > 0}$ defined by
$$ \rho(\uA)(\l) = \sum_{j=1}^d \psi_j(\l)\uA_{\l}\psi_j(\l)^* $$
is asymptotically contained in $\AAoi$. Furthermore, for each 
$\d$-sequence
$\{h_{\n}\}$ on $\bR^n$ there holds
\begin{equation} 
s\mbox{-}\lim_{\nu \to 
\infty}\,\poi(\underline{\alpha}_{h_\nu}\rho(\uA)) = 
 \sum_{j=1}^d \boldsymbol{\psi}_j\poi(\uA)\boldsymbol{\psi}_j^*\,,
 \quad \underline{A} \in \uAA\,;
\end{equation}
and $\boldsymbol{\rho}$ defined by
\begin{equation} \label{SLendo}
 \boldsymbol{\rho}(\boldsymbol{a}) = \sum_{j=1}^d
\boldsymbol{\psi}_j
 \boldsymbol{a} \boldsymbol{\psi}_j^*\,, \quad \boldsymbol{a} \in
 \AAoi\,,
\end{equation}
is a localized, transportable, irreducible endomorphism of $\AAoi$
which is moreover covariant and has finite 
statistics.
\end{Proposition}
{\it Proof. } The asymptotic containment in $\AAoi$ of
$\{\rho(\uA)(\l)\}_{\l > 0}$ is simply a consequence of the asymptotic
containment of each $\{\psi_j(\l)\}_{\l > 0}$ in $\Foi(O)$ and the
fact that $\rho(\uA)(\l) \in \cA(\l(O_1 \cap O_2))$ for $\uA \in
\underline{\fA}(O_2)$, with the conventional assumption that
$\psi_j(\l) \in \cF(\l O_1)$. Owing to \eqref{SLendo} and the
properties of a multiplet, $\boldsymbol{\rho}$ is clearly a localized,
irreducible endomorphism of $\AAoi$. The transportability may be seen 
as
follows. According to the definition of preserved charge, there is for
any double cone $O_{\times}$ different from $O$ a scaled multiplet for
$[\rho]$, $\{\tilde{\psi}_1(\l),\ldots,\tilde{\psi}_d(\l)\}_{\l > 0}$,
which is asymptotically contained in $\Foi(O_{\times})$. In the same
way as the $\{\psi_{j}(\l)\}_{\l>0}$ lead to multiplet operators
$\boldsymbol{\psi}_d$ in
$\Foi(\hat{O})$ for all $\hat{O} \supset \overline{O}$, the
$\{\tilde{\psi}_j(\l)\}_{\l > 0}$ lead to multiplet operators
$\tilde{\boldsymbol{\psi}}_j$ contained in $\Foi(\hat{O}_{\times})$
for all $\hat{O}_{\times} \supset \overline{O_\times}$. For the
corresponding endomorphism $\tilde{\boldsymbol{\rho}}$ it then holds
that $\boldsymbol{T} \tilde{\boldsymbol{\rho}}(\,.\,) =
\boldsymbol{\rho}(\,.\,) \boldsymbol{T}$ with the unitary intertwiner
$\boldsymbol{T} = \sum_{j=1}^d
\boldsymbol{\psi}_j\tilde{\boldsymbol{\psi}}{}^*_j$. Now it is easy to
see that the family $\{T(\l)\}_{\l> 0}$ defined by $T(\l) =
\sum_{j=1}^d \psi_j(\l)\tilde{\psi}_j^*$ 
is asymptotically contained in $\AAoi(O_*)$ for some double cone 
$O_*$,
and by an argument by now familiar, $\boldsymbol{T} =
s\mbox{-}\lim_{\nu \to \infty} \poi(\underline{\alpha}_{h_{\nu}}T)$ 
showing
that $\boldsymbol{T}$ is
contained in $\Aoi(\hat{O}_*)$ for $\hat{O}_* \supset \overline{O_*}$.
Covariance follows from a general argument: Given a multiplet
$\boldsymbol{\psi}_1,\ldots,\boldsymbol{\psi}_d$, it holds that
$\boldsymbol{\rho}(U\boldsymbol{a}U^*) = W\boldsymbol{\rho}(a)W^*$ for
each unitary $U$ with $W =
\sum_{j=1}^d\boldsymbol{\psi}_jU\boldsymbol{\psi}_j^*$ which is itself
unitary. Moreover, if a continuous unitary group $a \mapsto U(a)$, $a \in \bR^n$,
fulfills the spectrum condition, then $a \mapsto W(a) =\sum_{j=1}^d\boldsymbol{\psi}_jU(a)\boldsymbol{\psi}_j^*$ 
is clearly also a continuous unitary group fulfilling the spectrum condition.
That $\boldsymbol{\rho}$ has finite statistics follows from the
finiteness of the dimension $d$ of the multiplet.
${}$ \hfill $\Box$
\\[10pt]
Thus we have now  seen that a preserved charge gives rise to 
localized, transportable
endomorphisms of the observable quantum system in the scaling limit. 
However,
it is not clear if the scaling limit theories 
$(\Aoi,\Uscoi(\bR^n),\Ooi)$ 
gained from the observables of the underlying QFSGS satisfy all the 
technical
properties assumed to hold for an observable quantum system in vacuum 
representation.
Namely, the following two conditions can not be asserted for 
$(\Aoi,\Uscoi(\bR^n),\Ooi)$
from the general assumptions made so far: (1) separability of the 
Hilbert-space
$\Hoi^{(0)} = \overline{\fA_{0,\iota}\Ooi}$ and (2) Haag-duality.
Without these two conditions, one isn't really in the situation where 
standard superselection
theory applies, and so it is not really clear if one can construct a 
QFSGS for
$(\Aoi,\Uscoi(\bR^n),\Ooi)$ that could be compared to the scaling 
limit QFTGA of the
underlying theory.

Separability
of the vacuum Hilbert-space of an observable quantum system is 
expected to hold
quite generally for physically realistic theories, and moreover can 
be concluded
for scaling limit theories from a decent behaviour of the 
energy-level density
of the states of the underlying quantum field theory at short scales 
\cite{Bu.phsp}. 
On the other hand, the difficulty with the possible failure of 
Haag-duality can
be overcome by passing to the ``dual net'' to $\Aoi$. This is the 
family $\Aoi^d =\{\Aoi^d(O)\}_{O\in\cK}$
of von Neumann algebras defined by
$$ \Aoi^d(O) := \Aoi(O')' $$
where $O'$ denotes the causal complement of $O$ and $\Aoi(O')$ is the 
von Neumann algebra
generated by the $\Aoi(O_1)$ with $O_1 \subset O'$. Note that with 
this definition, since
the family $\Aoi$ satisfies the condition of locality, it holds that 
$\Aoi(O) \subset \Aoi^d(O)$.
Quite obviously, $\Uscoi(\bR^n)$ extends to a covariant action of the 
translations
on $\Aoi^d$ fulfilling the spectrum condition.
Now it is known \cite{Rigotti:1978} that, if the family $\Aoi$ fulfills the 
condition of geometric
modular action analogous to condition (QFTGA.9), then the condition 
of Haag duality is fulfilled
for the dual net $\Aoi^d$. In turn, if the observable system 
$(\cA_{\rm vac},\mathscr{U}_{\rm vac}(\bR^n),\O_{\rm vac})$
of the underlying QFSGS is Lorentz covariant (QFTGA.6) and fulfills 
geometric modular action (QFTGA.9),
then --- observing the convention stated in Sec.\ 3 --- also each 
scaling limit theory
$(\Aoi,\Uscoi(\bR^n),\Ooi)$ fulfills QFTGA.9 (cf.\ Prop.\ 3.1,4). 
Moreover, it was shown in
Sec.\ 3.4 of \cite{Rob.lec} that any localized, transportable 
endomorphism of $\Aoi$ extends
to a localized, transportable endomorphism of $\Aoi^d$ provided the 
latter fulfills the
condition of Haag-duality. Therefore we have deduced the following 
result:
\begin{Proposition} Let the dimension of spacetime $n$ be 3 or higher.
Assume that the observable quantum system of the underlying QFSGS 
fulfills Lorentz covariance
(QFTGA.6) and geometric modular action (QFTGA.9) and, moreover, that 
the scaling limit Hilbert-space
$\Hoi^{(0)} = \overline{\fA_{0,\iota}\Ooi}$ is separable. Then 
$(\Aoi^d,\Uscoi(\bR^n),\Ooi)$ is
an observable quantum system in vacuum representation fulfilling the 
conditions {\rm (a)} and
{\rm (b)} at the beginning of Sec.\ 4. If a superselection charge 
$[\rho]$ 
of the underlying QFSGS is preserved in this scaling limit, then the 
corresponding 
$\boldsymbol{\rho}$ defined in \eqref{SLendo} extends to a localized, 
transportable endomorphism
of $\fA_{0,\iota}^d$ which is covariant and has finite statistics.
\end{Proposition}
The last result presented in this section concerns the preservance of
the conjugate charge of a preserved charge. To this end, let us assume
for the remainder of this section that the underlying QFSGS fulfills
also the condition of geometric modular action as formulated in
(QFTGA.9) in Sec.\ 3. (We note that this can be deduced already if a
similar form of geometric modular action is initially only assumed to
hold for the underlying observable quantum system provided it fulfills
some mild additional conditions. We refer to
\cite{GLchcon,GLspst,Kuckert} for discussion of this issue.) In this
case, let $\psi_1,\ldots,\psi_d$ be a multiplet for the charge $[\rho]
\in {\rm Sect}^{\rm cov}_{\rm fin}$, with all $\psi_j$ contained in
$\cF(O)$ for some $O \in \cK$, and assume that $W$ is a wedge region
containing $O$. Let $J_W$ denote the Tomita-Takesaki modular
conjugation associated with $\cF(W)$ and the vacuum vector $\O$. Then
one can take an arbitrary multiplet $\psi'_1,\ldots,\psi'_d$ for
$[\rho]$ with all $\psi_j' \in \cF(r_WO)$, and define a new multiplet
of operators $\overline{\psi}_j \in \cF(O)$, $j = 1,\ldots,d$, by
$$  \overline{\psi}_j = J_WV \psi'_j V^*J_W $$
where $V$ is the ``twist'' operator defined in \eqref{twistop}. It is
easy to check that the $\overline{\psi}_j$ indeed form a multiplet,
i.e.\ $\sum_{j=1}^d \overline{\psi}_j\overline{\psi}{}_j^* = 1$ and
$\overline{\psi}{}_j^*\overline{\psi}_k = \delta_{jk}1$; however, 
since
$J_W$ is antilinear, this multiplet transforms under the gauge group
action according to the conjugate representation
$\overline{v}_{[\rho]}$ of $v_{[\rho]}$,
$$ U(g)\overline{\psi}_jU(g)^* = \sum_{i=1}^d \overline{\psi}_i
\overline{v}_{[\rho]ij}(g)\,, \quad g \in G\,,$$
if $\psi_1,\ldots,\psi_d$ transforms under the gauge group according
to $v_{[\rho]}$. This indicates that
$\overline{\psi}_1,\ldots,\overline{\psi}_d$ is a multiplet of the
conjugate sector $[\overline{\rho}]$ of $[\rho]$. Indeed, writing
$$ \rho(A) = \sum_{j=1}^d \psi_j A \psi_j^*\,, \quad
\overline{\rho}(A) = \sum_{j=1}^d \overline{\psi}_j A
\overline{\psi}{}_j^*\,, \quad R = \frac{1}{\sqrt{d}}\sum_{j=1}^d
\overline{\psi}_j\psi_j\,, \quad \overline{R} = \frac{1}{\sqrt{d}}
\sum_{j=1}^d\psi_j\overline{\psi}_j \,,$$
one can easily check that $R$ and $\overline{R}$ are isometries in
$\cA(O)$ and moreover, there holds
$$ \overline{\rho}(\rho(A))R = RA  \quad {\rm and} \quad
\rho(\overline{\rho}(A))\overline{R} = \overline{R}A\,, \quad A \in 
\fA\,.$$
(This can actually also be deduced from a rather more general argument
of \cite{GLchcon}.)

Equipped with these observations, we can now state the result.
\begin{Theorem} \label{PCT}
Suppose that the underlying QFSGS fulfills the conditions
of Poincar\'e covariance (QFTGA.6) and of geometric
modular action (QFTGA.9), and let $\ooi \in {\rm SL}^{\fF}(\o)$ be one
of its scaling limit states. Then a charge $[\rho] \in {\rm Sect}^{\rm
  cov}_{\rm fin}$ is preserved in the scaling limit state $\ooi$
 if and only if also the conjugate
charge $[\overline{\rho}]$ is preserved.
\end{Theorem}
{\it Proof. }
Assume that $[\rho]$ is preserved in the scaling limit state $\ooi$
and let $O \in \cK$. Then for $O_1 \in \cK$ with $\overline{O_1}
\subset O$ there is a scaled multiplet
$\{\psi_1'(\l),\ldots,\psi'_d(\l)\}_{\l > 0}$ for $[\rho]$, contained
in $\cF(\l r_WO_1)$ and asymptotically contained in $\Foi(r_W O_1)$. Let
$\overline{\psi}_j$ be defined by
$$ \overline{\psi}_j(\l) = J_WV \psi'_j(\l) V^* J_W $$
where $V$ is the ``twist'' operator (cf.\ eq.\ \eqref{twistop}) and
$J_W$ is the modular conjugation associated with $\cF(W)$ and the
vacuum vector $\O$. Then
$\{\overline{\psi}_1(\l),\ldots,\overline{\psi}_d(\l)\}_{\l > 0}$ is a
scaled multiplet for the conjugate charge $[\overline{\rho}]$ and each
$\overline{\psi}_j(\l)$ is contained in $\cF(\l O_1)$. Moreover, for
compactly supported $h \in L^1(\bR^n)$ it holds that 
$$ (\underline{\alpha}_h\overline{\psi}_j)(\l) = J_W V
(\underline{\alpha}_{h \circ r_W} \psi'_j)(\l) V^*J_W $$
and this shows that the $\{\overline{\psi}_j(\l)\}_{\l > 0}$ are
asymptotically contained in $\Foi(O)$. ${}$ \hfill $\Box$
\\[10pt]
{\it Remarks.} (i) Note that under the conditions of Thm.\ \ref{PCT}
one also obtains asymptotic scaling limit versions of the isometries
which intertwine $\boldsymbol{\rho}$ and
$\boldsymbol{\overline{\rho}}$. More precisely, suppose that a charge
$[\rho]$ is preserved in the scaling limit state $\ooi$, and let
$\boldsymbol{\psi}_1,\ldots,\boldsymbol{\psi}_d$ be a corresponding
multiplet contained in $\Foi(O)$ induced by a scaled multiplet
$\{\psi_1(\l),\ldots,\psi_d(\l)\}_{\l> 0}$. As the previous Theorem
shows, there is then a conjugate multiplet
$\boldsymbol{\overline{\psi}}_1,\ldots,\boldsymbol{\overline{\psi}}_d$ 
in $\Foi(O)$
induced by a scaled multiplet
$\{\overline{\psi}_1(\l),\ldots,\overline{\psi}_d(\l)\}_{\l > 0}$, 
and it
is straightforward to show that $\boldsymbol{R} =  d^{-1/2} \sum_{j=1}^d
\boldsymbol{\overline{\psi}}_j\boldsymbol{\psi}_j$ and
$\boldsymbol{\overline{R}} = d^{-1/2}\sum_{j=1}^d
\boldsymbol{\psi}_j\boldsymbol{\overline{\psi}}_j$ are given by
$$ \boldsymbol{R} = s\mbox{-}\lim_{\nu\to
  \infty}\,\frac{1}{\sqrt{d}}\sum_{j=1}^d\poi(\underline{\alpha}_{h_\nu}\overline{\psi}_j\psi_j)
 \quad {\rm and} \quad \boldsymbol{\overline{R}} = 
s\mbox{-}\lim_{\nu\to 
  \infty}\,\frac{1}{\sqrt{d}}\sum_{j=1}^d\poi(\underline{\alpha}_{h_\mu}\psi_j\overline{\psi}_j)$$
where $\{h_{\n}\}$ is any $\d$-sequence on $\bR^n$. Using this, one 
deduces
$$
\boldsymbol{\overline{\rho}}(\boldsymbol{\rho}(\boldsymbol{a}))\boldsymbol{R}
= \boldsymbol{Ra} \quad {\rm and} \quad
\boldsymbol{\rho}(\boldsymbol{\overline{\rho}}(\boldsymbol{a}))
\boldsymbol{\overline{R}}  
= \boldsymbol{\overline{R}a}\,, \quad  \boldsymbol{a} \in
\fA_{0,\iota}\,,$$
where  $\boldsymbol{\rho}$ and $\boldsymbol{\overline{\rho}}$ relate
to the $\boldsymbol{\psi}_j$ and $\boldsymbol{\overline{\psi}}_j$,
respectively, as in \eqref{SLendo}.  
\\[6pt]
(ii) In addition, one obtains also the following: Let $J_{W\,0,\iota}$ and 
$V_{0,\iota}$ denote the analogous objects to $J_W$ and $V$
 in the scaling limit theory
of $\ooi$, then a conjugate charge multiplet
$\boldsymbol{\overline{\psi}}_1,\ldots,\boldsymbol{\overline{\psi}}_d$ 
to
$\boldsymbol{\psi}_1,\ldots,\boldsymbol{\psi}_d$ is obtained by
$$ \boldsymbol{\overline{\psi}}_j = J_{W\,0,\iota}V_{0,\iota}
\boldsymbol{\psi}_j'
  V_{0,\iota}^*J_{W\,0,\iota} $$
whenever $\boldsymbol{\psi}'_1,\ldots,\boldsymbol{\psi}_d'$ is a 
multiplet
equivalent to $\boldsymbol{\psi}_1,\ldots,\boldsymbol{\psi}_d$ 
localized
in $r_WO$. 

\section{On Equivalence of Local and Global Intertwiners}
\setcounter{equation}{0}
In the present section we will address the question of equivalence of
local and global intertwiners of superselection sectors. We shall
extend an argument of Roberts \cite{Rob.dil} who considered the
setting of dilation covariant quantum field theories, showing that the
preservance of all charges in some scaling limit theories\footnote{As already remarked in sec.~\ref{sec:DHRpreservance}, all charges are preserved in all scaling limits of dilation covariant theories.} is, together
with the assumption that the local field algebras $\cF(O)$ are
factors, sufficient for the equivalence of local and global
intertwiners.
Our main technical result is stated in the following Lemma.
\begin{Lemma} \label{slc}
 Let $[\rho] \in {\rm Sect}^{\rm cov}_{\rm fin}$ be a
  superselection sector of the underlying QFSGS, let $O\in\cK$, and
  suppose that there are (i) a scaling limit state $\ooi \in {\rm
    SL}^{\cF}(\o)$, (ii) a scaled multiplet
  $\{\psi_1(\l),\ldots,\psi_d(\l)\}_{\l > 0}$ for $[\rho]$ with
  $\psi_j(\l) \in \cF(\l O)$, (iii) some compactly supported,
  non-negative $h \in L^1(\bR^n)$, such that
  $$ ||\poi(\underline{\Phi}^{(h,j)})\Ooi || > 0\,, \quad
  j=1,\ldots,d\,,$$
where $\underline{\Phi}^{(h,j)}_{\l} =
(\underline{\alpha}_h\psi_j)(\l)$.
Then for all unitaries $U \in \cA(O)' \cap \cF(O)$ and all multiplets
$\tilde{\psi}_1,\ldots,\tilde{\psi}_d \in \cF(O)$ for $[\rho]$ ($O \in
\cK$) there holds
\begin{equation}\label{ciao}
\begin{alignedat}{2}
 \o(\tilde{\psi}_j^*U^*\tilde{\psi}_kU) &= \delta_{jk}\, 
 &\quad &\text{if } \b_k(\tilde{\psi}_i)=\tilde{\psi}_i 
\\
\o(\tilde{\psi}_j^* \b_k(U^*)\tilde{\psi}_kU) &= \delta_{jk}\,
&&\text{if }\b_k(\tilde{\psi}_i)= -\tilde{\psi}_i 
\end{alignedat}
\end{equation}

%$$ \o(\tilde{\psi}_j^*U^*\tilde{\psi}_kU) = \delta_{jk}\,.$$
\end{Lemma}
{\it Proof. }
We will treat explicitly the  ``even'' case in (\ref{ciao}), the 
``odd'' case being completely analogous.
First we note that $||\poi(\underline{\Phi}^{(h,j)})\Ooi
|| > 0$ for any of the $j=1,\ldots,d$ implies that the
$\poi(\underline{\Phi}^{(h,j)})\Ooi$, $j = 1,\ldots,d$, are linearly
independent. To see this, note that the contrary assumption of linear
dependence implies that there is an invertible $d \times d$ matrix
$(u_{j\ell})$ so that $\sum_{j}\poi(\underline{\Phi}^{(h,j)})\Ooi
u_{j\ell} = 0$ for some $\ell$. But this implies
$$ 0 = \Uoi(g)\sum_j\poi(\underline{\Phi}^{(h,j)})\Ooi u_{j\ell} =
      \sum_{j,k}\poi(\underline{\Phi}^{(h,k)})\Ooi
      v_{[\rho]kj}(g)u_{j\ell}$$
for all $g \in G$ and hence, since $v_{[\rho]}$ is irreducible,
$\poi(\underline{\Phi}^{(h,j)})\Ooi = 0$ for all $j$. 

We further observe that it constitutes no restriction of generality
to prove the statement of the
theorem only for $O \in \cK$ which contain the origin $0 \in \bR^n$
 in their spacelike
boundary (i.e.\ the origin is contained both in the boundary of $O$
and in the boundary of its spacelike complement)
 since the underlying QFSGS is translation covariant. Thus we continue
 to prove the statement for an arbitrary $O$ of this type.   

We begin by noting that
from our observation above, the
$\poi(\underline{\Phi}^{(h,j)})\Ooi$, $j = 1,\ldots,d$, span a
$d$-dimensional subspace of $\Hoi$. Now let $W \supset O$ be a wedge 
region
containing the origin in its spacelike boundary.
Then let $W'$ be the wedge which is the causal complement of $W$, and
let $W'_h$ be a copy of $W'$ shifted by some suitable spacelike vector
into the interior of $W'$ such that $W_h'$ lies in the causal
complement of $O + 
{\rm supp}\,h$. By a standard Reeh-Schlieder argument $\Foi(W'_h)\Ooi$
is dense in $\Hoi$   and hence, choosing some $\epsilon > 0$
arbitrarily, there will be some double cone $\hat{O} \subset W'_h$ and
$\uF^{(1)},\ldots,\uF^{(d)} \in \uFF(\hat{O})$ such that
$$
| \langle
  \poi(\uF^{(j)})^*\Ooi,\poi(\underline{\Phi}^{(h,k)})\Ooi\rangle
  -\delta_{jk}|
 = |\ooi(\uF^{(j)}\underline{\Phi}^{(h,k)}) - \delta_{jk}| <
\epsilon\,.
$$
Now let $(\l_{\k})_{k \in K}$ be a subnet of the positive reals,
converging to $0$, with $\ooi = \lim_{\k} \underline{\o}_{\l_{\k}}$ on
$\uFF$. Since $\uF^{(j)}_{\l_{\k}}\underline{\Phi}_{\l_{\k}}^{(h,k)}$
converges weakly to a multiple of $1$ owing to $\bigcap_{O \owns
  0}\cF(O) = \mathbb{C}1$ (see \cite{Rob.dil}), we obtain 
$$ \ooi(\uF^{(j)}\underline{\Phi}_{\l_{\k}}^{(h,k)}) = \lim_{\k}\,
   \o(\uF^{(j)}_{\l_{\k}}\underline{\Phi}_{\l_{\k}}^{(h,k)}) =
   \lim_{\k}\,\o'(\uF^{(j)}_{\l_{\k}}\underline{\Phi}_{\l_{\k}}^{(h,k)})
  $$
for each locally normal state $\o'$ on $\fF$, and this implies
$$
|\lim_{\k}\,\o'(\uF^{(j)}_{\l_{\k}}\underline{\Phi}_{\l_{\k}}^{(h,k)})
- \delta_{jk}| < \epsilon $$
whenever $\o'$ is locally normal.
On the other hand, since $||(\alpha_{\l x}(U) - U)\O|| \to 0$ as $\l
\to 0$ uniformly for $x$ ranging over compact sets, it follows that
\begin{eqnarray*}
 \lefteqn{\o(U^*\uF^{(j)}_{\l_{\k}}\underline{\Phi}_{\l_{\k}}^{(h,k)}U)}\\
&=& \int d^nx \,h(x)
\o(\alpha_{\l_{\k}x}(U^*)\uF^{(j)}_{\l_{\k}}
\alpha_{\l_{\k}x}(\psi_k(\l_{\k}))U) + o(\l_{\k}) \\
&=& \int d^nx \,h(x)
\o(U^*\alpha_{-\l_{\k}x}(\uF^{(j)})_{\l_{\k}}
\psi_k(\l_{\k})\alpha_{-\l_{\k} x}(U)) + o(\l_{\k})\,,
\end{eqnarray*}
where $o(\l)$ tends to $0$ for $\l \to 0$, and we have used invariance
of the vacuum state $\o$ under the action of the translations 
$\alpha_x$. 
We have also inserted the definition of the
$\underline{\Phi}^{(h,k)}$, so that the scaled multiplets $\psi_k(\l)$
appear here.

Next we write $\psi_j(\lambda =1) = \psi_j$, and we notice that
$\psi_j(\l) = T_{\l}\psi_j$ where $T_{\l} = \sum_{j=1}^d
\psi_j(\l)\psi_j^*$ is contained in $\cA(O)$, and thus commutes with
$U \in \cA(O)'\cap \cF(O)$. We note also that for every $B \in \fF$
we have, 
%denoting by $V$ the ``twist'' operator of \eqref{twistop},
%\begin{eqnarray*}
$$\o(U^*\alpha_{-\l_{\k} x}(\uF^{(j)}_{\l_{\k}})B) =
 % \o(VU^*V^*V\alpha_{-\l_{\k} x}(\uF^{(j)}_{\l_{\k}})V^*VBV^*)\\& = &
 % \o(V\alpha_{-\l_{\k} x}(\uF^{(j)}_{\l_{\k}})V^*(VU^*V^*)VBV^*) = 
\o(\alpha_{-\l_{\k} x}(\uF^{(j)}_{\l_{\k}})U^*B)$$
%\end{eqnarray*}
for $\l_{\k} \le 1$ and $x \in {\rm supp}\,h$ since then 
$\alpha_{-\l_{\k} x}(\uF^{(j)}_{\l_{\k}}) \in \cF(W')$ and $U^* \in
\cA(O)'\cap \cF(O) \subset \cF(W)$. Hence we get for $\l_{\k} \le 1$,
\begin{eqnarray*}
\lefteqn{\int d^nx \,h(x)
\o(U^*\alpha_{-\l_{\k}x}(\uF^{(j)})_{\l_{\k}}
 \psi_k(\l_{\k})\alpha_{-\l_{\k} x}(U)) }\\
& = & \int d^nx\, h(x)\o(\alpha_{-\l_{\k}
  x}(\uF_{\l_{\k}}^{(j)})T_{\l_{\k}}
(\sum_{i=1}^d\psi_i\psi_i^*)U^*\psi_k\alpha_{-\l_{\k}x}(U)) \\
& = & 
  \sum_{i=1}^d\int
  d^nx\,h(x)\o(\alpha_{-\l_{\k}x}(\uF^{(j)}_{\l_{\k}})
T_{\l_{\k}}\psi_i\alpha_{-\l_{\k}x}(\psi^*_iU^*\psi_kU) ) + p(\l_{\k})
\\
& = &   
\sum_{i=1}^d\o(\uF_{\l_{\k}}^{(j)}
\underline{\Phi}_{\l_{\k}}^{(h,i)}\psi^*_iU^*\psi_kU)  + p(\l_{\k})
\end{eqnarray*}
with some function $p(\l)$ tending to $0$ as $\l \to 0$, where we used
that
$$ \lim_{\l \to 0}\,|| (\psi^*_iU^*\psi_k\alpha_{-\l x}(U) - 
\alpha_{-\l
  x}(\psi^*_iU^*\psi_kU))\O ||  = 0 $$
uniformly for $x$ ranging over compact sets. Also we used the
translational invariance of $\o$ again. Summing up these findings we
have for $\l_{\k} \le 1$,
$$ \o(U^*\uF_{\l_{\k}}^{(j)}\underline{\Phi}^{(h,k)}_{\l_{\k}}U)  
   =\sum_{i=1}^d\o(\uF_{\l_{\k}}^{(j)}
\underline{\Phi}_{\l_{\k}}^{(h,i)}\psi^*_iU^*\psi_kU) + o(\l_{\k}) + 
p(\l_{\k})
\,.$$
Making now use of the fact that for all normal states $\o'$ it holds
that
$$ \lim_{\k}\,|\o'(\uF_{\l_{\k}}^{(j)}
\underline{\Phi}_{\l_{\k}}^{(h,i)}) - \delta_{ji}| < \epsilon\,,$$
the previous equation yields, upon taking the limit over $\k$,
$$ |\o(\psi_j^*U^*\psi_kU) - \delta_{jk}| < (d+1)\epsilon\,.$$
Here $\epsilon > 0$ was arbitrary, and hence we conclude that
$$   \o(\psi_j^*U^*\psi_kU) = \delta_{jk} $$
holds for all unitary $U \in \cA(O)' \cap \cF(O)$ and the special
multiplet $\psi_j = \psi_j(\lambda =1)$. However, given any
other multiplet $\tilde{\psi}_j$ in $\cF(O)$ for the charge $[\rho]$,
there is the unitary $T = \sum_{j=1}^d \tilde{\psi}_j\psi_j^*$ in
$\cA(O)$ so that $\tilde{\psi}_j = T\psi_j$, and thus we obtain, for
each unitary $U \in \cA(O)' \cap \cF(O)$,
$$ \o(\tilde{\psi}_j^*U^*\tilde{\psi}_kU) = 
\o(\psi_j^*T^*U^*T\psi_kU) =
\omega(\psi_j^*U^*\psi_kU) = \delta_{jk}$$
since $U$ and $T$ commute. 
${}$ \hfill $\Box$ 
\\[10pt]
Now we make use of the following result  which has been proved in
\cite{Rob.dil} (using also \cite{DR.CMP28}):  If, for some $O \in 
\cK$,
there holds (\ref{ciao}) for all charge multiplets $\tilde{\psi}_j$ (of all superselection
sectors) contained in $\cF(O)$ and for all unitaries $U$ contained in
$\cA(O)' \cap \cF(O)$, then
$$ \cA(O)' \cap \cF(O) = \cF(O)' \cap \cF(O)\,.$$ 
If moreover the local field algebras of the underlying QFSGS are 
factors, i.e.\
if
\begin{equation} \label{factor}
\cF(O) \cap \cF(O)' = \bC {\rm 1}\,, \quad O \in \cK\,,
\end{equation}
then  equivalence of local and
global intertwiners ensues: Given $[\rho]$ and $[\rho']$ in ${\rm
  Sect}^{\rm cov}_{\rm fin}$ it holds that
\begin{equation} \label{loctoglob}
 \cI(\rho,\rho')_O = \cI(\rho,\rho') \quad \mbox{for\ $\rho,\rho'$\
   localized\ in}\ O\,.
\end{equation}
(Cf.\ Sec.\ 4 for the definition of $\cI(\rho,\rho')_O$ and 
$\cI(\rho,\rho')$.)
\begin{Corollary}
Suppose that all local field algebras of the underlying QFSGS are
factors, i.e.\ that \eqref{factor} holds for all $O \in \cK$. 
Moreover,
suppose that for each charge $[\rho] \in {\rm Sect}^{\rm cov}_{\rm
  fin}$ there is some scaling limit state $\ooi \in {\rm 
SL}^{\fF}(\o)$
(which may depend on $[\rho]$) such that $[\rho]$ is preserved in that
scaling limit state. Then in the underlying QFSGS there holds
the equivalence of local and global intertwiners \eqref{loctoglob}.
\end{Corollary}
The factorial property of the local field algebras has been checked in
free field models. Assuming that this is a general feature of quantum
field theories, the assertion of the Corollary shows that part of the
charge superselection structure is determined entirely locally if all
charges are preserved in suitable scaling limit states; in other
words, if the charges are, in this (somewhat generalized) sense,
ultraviolet stable. For further discussion as to how much else of the
superselection structure may be determined locally, we refer to
\cite{Rob.dil}.
%%%%%%%%%%%%%%%%%%%%%%%%%%%%%%%%%
%% start of new material: Sect. 7 %%
%%%%%%%%%%%%%%%%%%%%%%%%%%%%%%%%%
%%
%%
\section{Scaling Algebras for Quantum Field Systems Localized in
Spacelike Cones}\label{sec:scalingcone}
\setcounter{equation}{0}
Up to this point, we have considered quantum fields localizable in 
arbitrary
bounded open regions, corresponding by the Doplicher-Roberts 
reconstruction theorem
to superselection charges of the DHR-type. There are more general 
types superselection
charges whose localization properties with respect to the vacuum 
representation of
the observables are weaker. Before we enter into discussion of this 
fact, let us first introduce the 
relevant terminology.

As before, we identify $n$-dimensional Minkowski spacetime with 
$\mathbb{R}^n$.
Following \cite{DRwhy}, the timelike hyperbolic submanifold 
$\mathscr{D} =\{s \in \mathbb{R}^n: \eta_{\m\n}s^{\m}s^{\n} = -1\}$
will be taken to represent all points at spacelike infinity of 
$n$-dimensional Minkowski spacetime
since each $s\in\mathscr{D}$ represents a spacelike direction of unit 
Minkowskian length. Let 
a pair of points $s_+,s_-$ in $\mathscr{D}$ be given, where $s_+ \in  
(s_- + V_+)$, then we call
$D(s_+,s_-) = (s_+ + V_-) \cap (s_- + V_+) \cap \mathscr{D}$ a {\bf 
double cone at spacelike infinity}
with future direction $s_+$ and past direction $s_-$. Then a {\bf 
spacelike cone} is a set of the
form  
$$  C = a + \{\l D(s_+,s_-): \l > 0\}\,$$
where $a$ is any element of $\mathbb{R}^n$ and $D(s_+,s_-)$ is any 
double cone at spacelike
infinity. A spacelike cone is thus a conic set extending to spacelike 
infinity having its apex
at $a$; it can be viewed as the set of points lying in a certain 
opening angle around a spacelike
direction. With this definition, each spacelike cone is causally 
complete, i.e.\ taking its
double causal complement reproduces each spacelike cone. For further 
properties of spacelike
cones, we refer the reader to the discussion in the appendix of 
\cite{DRwhy}. We will denote
by $\mathcal{S}$ the set of all spacelike cones.

A profound analysis by Buchholz and Fredenhagen \cite{BF}
has shown that in a theory with no
massless excitations a general superselection charge is generically
localized in spacelike cones. Namely, they have proven that for any massive
single particle representation $\pi$ of $\cA_{\text{vac}}$ (i.e. $\pi$ is a translation covariant representation having no translation invariant vector and the single particle states are separated from the continuum by a gap in the spectrum of the corresponding translations representation $\Usc_\pi$), there exists an irreducible vacuum representation $\tilde{\pi}_{\text{vac}}$ (i.e. a translation covariant representation with a translation invariant vector) such that,
 for any spacelike cone $C$,
$\pi\rest\fA_{\text{vac}}(C')$ is unitarily equivalent to $\tilde{\pi}_{\text{vac}}\rest\fA_{\text{vac}}(C')$, $\fA_{\text{vac}}(C')$ being the C$^*$-algebra generated by all $\cA_{\text{vac}}(O)$ with $\overline{O} \subset C'$.

Then, even in absence of massless particles, as in massive non-abelian gauge
theories, DHR localization could be too strong a requirement. That such sectors really should arise in this kind of theories is suggested by the fact that spacelike cones can be thought as idealizations of flux
tubes joining pairs of infinitely separated opposite gauge charges.

While in the case of strictly localizable field operators it was 
meaningful to define
the scaling algebra of a QFTGA and to study the corresponding scaling 
limit theories
even for the case that the QFTGA has not all the features of a QFSGS, 
the situation
is somewhat different for the case of field operators which are only 
localizable in
spacelike cones. To illustrate this, consider the case that one is 
given a collection
of von Neumann algebras $\{\cF(C)\}_{C\in\mathcal{S}}$ indexed by 
double cone (to be
viewed as localization regions). Then one may fix, say, some 
spacelike cone $C$ having
its apex at the origin, and consider uniformly norm-bounded functions 
$\uF:\bR^+ \to \cF(C)$.
In order to take up the ideas that led to the definition of the
 scaling algebra for strictly localizable fields, however, one would 
now have to further restrict
these functions in a manner expressing that $\uF_{\l}$ becomes 
localized near the origin
as $\l$ tends to $0$. But since $\l C = C$ for all $\l > 0$, it is 
obvious that imposing
$\uF_{\l} \in \cF(\l C)$ leads to no restriction in localization at 
all, and hence the
localization constraint
has to be implemented by making use of additional structure. And this 
can be achieved if it is
assumed that the collection of von Neumann algebras  
$\{\cF(C)\}_{C\in\mathcal{S}}$ belongs to
a QFSGS corresponding to BF-type superselection charges, where one 
can exploit the strict
localization properties of the quantum system of observables.

Therefore, let us now sketch, following \cite{DRwhy},
 the Doplicher-Roberts reconstruction theorem for the case of
charges of BF-type, which is very much in parallel to the discussion 
of the DHR case in
Sec.\ 4. The starting point is again an observable quantum system 
$(\cA_{\rm vac},\mathscr{U}_{\rm vac}(\bR^n),
\O_{\rm vac})$ fulfilling the properties listed at the beginning of 
Sec.\ 4. Then one may, as in
\cite{BF}, consider the set $\fP^{\rm BF}$ of representations which, upon restriction to the spacelike 
complements of 
arbitrary spacelike cones, 
are unitarily equivalent to some fixed vacuum representation, which may be assumed to be the identical one. That means, in view of the above discussion, $\pi$ is in $\fP^{\rm BF}$ iff
$\pi \rest \fA_{\rm vac}(C')$ is unitarily equivalent to the 
identical representation of
$\fA_{\rm vac}(C')$ on $B(\cH_{\rm vac})$. Then we restrict to the 
subset 
$\fP^{\rm BF}_{\rm cov}$ of such representations
which are translation-covariant, where the definition of covariance 
is exactly as in Sec.\ 4
(cf.\ eqn.\  \eqref{repcovar}). It can again be shown that the set 
$\fP^{\rm BF}_{\rm cov}$
is in one-to-one correspondence with the set of $\D_t^{\rm BF,cov}$ 
of covariant, localized, transportable
morphisms of $\fA_{\rm vac}$, now taking values in $B(\cH_{\text{vac}})$, which are defined similarly as in 
Sec.\ 4 with the difference
that these morphisms are no longer localized in double cones but 
in spacelike cones. The concepts
of translation-covariant superselection sectors and of (global) 
intertwiners carry over literally form
the situation of Sec.\ 4. This is likewise the case for the 
intertwiner product, the notion of statistics
and of conjugate charge, keeping in mind that all morphisms are 
now localized in spacelike cones
and that all properties referring to localization and spacelike 
separation must take this into
account. This understood, one is led to defining the set of 
transportable, irreducible, covariant morphisms $\D_{\rm 
fin}^{\rm BF,cov}$ of BF-type that have
finite statistics, and their corresponding unitary equivalence 
classes collected
 in ${\rm Sect}_{\rm fin}^{\rm BF,cov}$
representing the superselection charges of BF-type of the theory. 
\begin{Definition}\label{def:coneQFSGS}{\rm
One says that a collection of objects 
$(\cF,\mathscr{U}(\bR^n),U(G),\O,k)$ is a QFSGS associated
with $(\cA_{\rm vac},\mathscr{U}_{\rm vac}(\bR^n),\O_{\rm vac})$ and 
$\D_{\rm fin}^{\rm BF,cov}$
if the following holds:
\begin{itemize}
\item[\mbox{\small $(*)$}] $\cF$ denotes a family of von Neumann 
algebras (called field algebras) 
$\{\cF(C)\}_{C\in\cS}$ on a  separable Hilbert-space indexed by 
spacelike
cones. The family satisfies the analogues of conditions (QFTGA.1-5) 
when changing double cone localization
regions of the field algebras to spacelike cones, and observing the 
following alterations: Since the
set $\cS$ isn't directed with respect to inclusion, there is no 
counterpart to the quasilocal algebra $\fF$,
and the assumption of cyclicity has to be altered to demanding that 
the space generated by $\cF(C)\O$, where
$C$ ranges over all of $\cS$, is dense in $\cH$.
\item[\mbox{\small $(**)$}] The analogue of (QFSGS.2) holds when 
replacing double cone localization regions by spacelike cones,
more precisely, there is a $C^*$-algebraic monomorphism $\pi:\fA_{\rm 
vac} \to B(\cH)$ so that
$$\cA(O) := \pi(\cA_{\rm vac}(O)) \subset \cF(C)$$
holds for all double cones $O$ and all spacelike cones $C \supset O$. 
Moreover, $\pi$ contains the
vacuum representation of $\fA_{\rm vac}$ on $\cH_{\rm vac}$ as a 
sub-representation, and 
$$ \cA(C):= \bigvee_{O \subset C} \cA(O)
 = \cF(C)^G\,.$$ 
Furthermore, if one denotes, for $D(s_+,s_-)$ any double cone at 
infinity, the $C^*$-algebra
generated by all $\cF(a + \bR^+D(s_+,s_-))$, $a \in \bR^n$, by 
$\cF[D(s_+,s_-)]$, then there holds
$$ \pi(\fA_{\rm vac})' \cap \cF[D(s_+,s_-)] = \bC {\bf 1}\,.$$
\item[$(^{{}\,*}_{**})$] \label{coneQFSGS3}The analogue of (QFSGS.3) 
holds, with the difference that the multiplets
$\psi_1,\ldots,\psi_d$ are now elements of the $\cF(C)$, and 
correspondingly one has to change
$\rho_O$ to $\rho_C$, a representer of $[\rho]$ localized in $C$, in 
(4.3).
\item[$(_{**}^{**})$] The analogue of (QFSGS.4) holds upon replacing 
double cones $O$ by spacelike
cones $C$.
\end{itemize}}
\end{Definition}
As Doplicher and Roberts \cite{DRwhy} have shown, there is for each 
observable quantum system
$(\cA_{\rm vac},\mathscr{U}_{\rm vac}(\bR^n),\O_{\rm vac})$ together 
with $\D_{\rm fin}^{\rm BF,cov}$
an associated QFSGS. We will now assume that we are given such a 
QFSGS corresponding to BF-type
superselection charges, and construct a scaling algebra for it. 

In order to do so, we have to introduce some notation. First of all, 
we recall that $\a_a$ stands for
the adjoint action of $\mathscr{U}(a)$, and $\b_g$ stands for the 
adjoint action of $U(g)$. Then, let us denote by 
$B_{\b}(\bR^+,B(\cH))$ the $C^*$-algebra of all
bounded functions ${\sf F}: \bR^+ \to B(\cH)$ with $C^*$-norm given by
$$ ||\,{\sf F}\,|| := \sup_{\l > 0}\,||\,{\sf F}(\l)\,|| $$
and with pointwise defined algebraic operations,
where the functions are also assumed to be continuous with respect to 
the standard lift of the
gauge group action, meaning that
$$
||\,\underline{\b}_g({\sf F}) - {\sf F}\,|| = \sup_{\l > 
0}\,||\,\b_{g}({\sf F}(\l)) - {\sf F}(\l)\,|| \to 0 \quad {\rm for}
 \quad g \to 1_G\,.$$
 Then we define $C_{\a}(\bR^+,B(\cH))$ as the $C^*$-subalgebra
of $B_{\b}(\bR^+,B(\cH))$ whose elements fulfill
$$
||\,\underline{\a}_a({\sf F}) - {\sf F}\,|| = \sup_{\l > 
0}\,||\,\a_{\l a}({\sf F}(\l)) - {\sf F}(\l)\,|| \to 0 \quad {\rm for}
 \quad a \to 0\,.$$
With this notation, we are ready to define scaling algebras for field 
operators localizable in
spacelike cones.
\begin{Definition}\label{def:scalingalgebracone}{\rm
Let $O$ be a double cone and $C \supset O$ a spacelike cone. We 
define the scaling algebra $\underline{\fA}(O)$ as in
Def.\ 2.2, i.e.\ consisting of all $\underline{A} \in 
C_{\a}(\bR^+,B(\cH))$ where $\underline{A}_{\l} \in \cA(\l O)$, $\l > 
0$.
Then we denote by $\underline{\fA}(O)_1$ the subset of all elements 
of $\underline{\fA}(O)$ whose norm
is bounded by unity, and we define:
\begin{itemize}
\item[(I)] $\underset{\sim}{\fF}(C,O)$ is the $C^*$-subalgebra of all 
${\sf F}$ in $B_{\b}(\bR^+,B(\cH))$ having the
properties
\begin{equation}\label{eq:asintloc}
{\sf F}(\l) \in \cF(\l C) \quad {\rm and} \quad \limsup_{\l \to 0} 
\,\left( \sup \{ \nnorm{\,[{\sf F}(\l),\underline{A}_{\l}]\,}\,:\, 
\underline{A}
\in \underline{\fA}(O')_1\} \right) = 0\,,
\end{equation}
where $[A,B] = AB-BA$ denotes the commutator.
\item[(II)] $\underline{\fF}(C,O) := \underset{\sim}{\fF}(C,O) \cap 
C_{\a}(\bR^+,B(\cH))$.
\end{itemize}}
\end{Definition}   

%%% ADDED %%%%%%%%%%%%%%%%%%%%%%%%%%%%%%%%%%%%%%%%%%%%%%
Some remarks about this definition are in order. The second 
condition in~\ref{eq:asintloc} expresses the fact that the field 
operators $\sFl$ that we are considering are \emph{asymptotically 
localized}, as $\l \to 0$, in the double cone $\l O$, in the sense 
that their effect on measurements performed in the spacelike 
complement of this bounded region vanishes in the limit. Through this 
requirement, then, we implement in this more general case the basic 
idea of the scaling algebra approach. As a physical motivation for 
such a condition, we may note that a behaviour of this kind is 
expected to show up at least in nonabelian asymptotically free gauge 
theories: as remarked by Buchholz and Fredenhagen, the spacelike cone 
in which BF charges are localized has to be thought of as a fattened 
version of a gauge flux string between two opposite charges, one of 
which has been shifted at spacelike infinity, and it is then natural 
to expect that such string should become weaker and weaker at small 
scales if the theory is asymptotically free, leaving in the limit a 
compactly localizable charge. 

We define the auxiliary C$^*$-algebra $\uFx$ as the C$^*$-subalgebra 
of $C_\a(\bR^+,B(\cH))$ generated by all the algebras $\uFCO$, and we 
note that for this system of algebras we have the obvious covariance 
properties
\begin{equation*}
\ua_a(\uFCO) = \uFF(C+a,O+a), \qquad \ub_g(\uFCO) = \uFCO,
\end{equation*}
for the actions $\ua_{\bR^n}$ and $\ub_G$ of the translations and 
gauge group defined above, so that these restrict to automorphic 
actions on $\uFx$, denoted by the same symbols. For each normal state 
$\o'$ on $B(\cH)$ we define the family of states $(\uo'_\l)_{\l > 0}$ 
on $\uFx$ in analogy to the case of localized fields,
\begin{equation*}
\uo'_\l(\uF) := \o'(\uFl), \qquad \uF \in \uFx,
\end{equation*}
and ${\rm SL}^{\uFx}(\o') = \{\o'_{0,\iota} \, : \, \iota \in \bI\}$
will be the set of weak* limit points of the net $(\uo'_\l)_{\l>0}$
(this is non-void, as in the localizable case), and will be called the
set of \emph{scaling limit states of} $\o'$.

\begin{Lemma}\label{lem:unicity}
Assume that the net of observable algebras in the vacuum 
representation satisfies the following condition: for each double 
cone $O$ containing the origin, there holds
\begin{equation}\label{eq:AvacO}
\Avac(O) = \bigvee_{O_0 \ni 0} \Avac(O\cap O'_0),
\end{equation}
where $O_0$ runs through all double cones containing the origin. Then 
${\rm SL}^{\uFx}(\o')$ is independent of the normal state $\o'$.
\end{Lemma}

\noindent \emph{Remark.} The above condition~\eqref{eq:AvacO} is 
suggested by the fact that $\bigcap_{O_0 \ni 0} \Avac(O_0) = \bC 
\Id$, by Haag duality and by the time-slice axiom. Its validity can 
also be proven in free field models.

\begin{proof}[Proof of Lemma~\ref{lem:unicity}]
As the union of all C$^*$-algebras $\uFCO$ is norm dense in $\uFx$, 
it is sufficient to show that, for any two normal states $\o^1$, 
$\o^2$ on $B(\cH)$, and for any choice of $C$, $O$ with $O \subset C$ 
and any $\uF \in \uFCO$, there holds
\begin{equation*}
\lim_{\l \to 0}\ \uol^1(\uF) - \uol^2(\uF) = 0.
\end{equation*}
To this end, we adapt Roberts' argument~\cite{Rob.dil} and assume that
this is not true. Then we can find a subnet $(\uF_{\l_\n})_{\n}$ of
$(\uFl)_{\l>0}$ weakly convergent to some $F_0 \in B(\cH)$ and such 
that
\begin{equation*}
\nabs{\o^1(F_0) - \o^2(F_0)} = \lim_\n \nabs{\uo^1_{\l_\n}(\uF) - 
\uo^2_{\l_\n}(\uF)}>0.
\end{equation*}
Now we intend to show that $F_0$ is a multiple of $\Id$ which leads to a
contradiction, and hence shows validity of the statement of the lemma. We first observe that,
if $D$ is the double cone at spacelike infinity defined by the
spacelike cone $C$, then from $\uF_{\l_\n} \in \cF(\l_\n C) \subset
\cF[D]$ for each $\n$, $F_0 \in \cF[D]$ follows. Let then $O_1$ be a 
double cone containing the origin, take $A \in \fA(O_1')$ such that 
$\nnorm{A}\leq 1$ and $x \mapsto \a_x(A)$ is norm continuous, and 
define for each $\m > 0$ an element $\uA^{(\m)} \in 
\uAA(\frac{1}{\m}O'_1)$ by 
\begin{equation*}
\uA^{(\m)}_\l := \begin{cases}A& \text{if $\l = \m$}, \\
                              0& \text{if $\l \neq \m$}.
                 \end{cases}
\end{equation*}
Then since $\uA^{(\m)} \in \uAA(O')_1$ for $\m$ sufficiently small, 
from the asymptotic localizability of $\uF \in \uFCO$ it follows that
\begin{equation*}
\limsup_{\l \to 0} \nnorm{\,[\uFl,A]\,} = \limsup_{\l \to 
0}\nnorm{\,[\uFl,\uA^{(\l)}_\l]\,} \leq \lim_{\l\to 0}\sup_{\uA \in 
\uAA(O')_1} \nnorm{\,[\uFl,\uAl]\,}=0,
\end{equation*}
which implies $[F_0,A]=0$, so that, as the multiples of the $A$'s
satisfying the stated requirements form a weakly dense set in each
algebra $\cA(O_0') := \fA(O_0')''$ with $O_0 \supset \overline{O}_1$, 
we get
\begin{equation*}
F_0 \in \bigg[ \bigvee_{O_0 \ni 0} \cA(O_0')\bigg]'.
\end{equation*}
But from the assumption~\eqref{eq:AvacO}, together with local 
normality of the representation $\pi$, we have $\pi(\fA_{\text{vac}}) 
\subseteq \bigvee_{O_0\ni 0} \cA(O_0')$, so that, 
by~$(^{{}\,*}_{**})$ of Definition~\ref{def:coneQFSGS}, $F_0 \in 
\pi(\fA_{\text{vac}})' \cap \cF[D] = \bC \Id$.
\end{proof}

In view of the above lemma, from now on we will only consider the 
scaling limit states ${\rm SL}^{\uFx}(\o)$, with $\o := 
\angscalar{\O}{(\cdot)\O}$ the underlying vacuum state. For $\ooi \in 
{\rm SL}^{\uFx}(\o)$, let $(\poix,\Hoix,\Ooi)$ be the associated GNS 
representation and let $\Uscoix(a)$, $a \in \bR^n$, $\Uoix(g)$, $g 
\in G$, be respectively the translations group and gauge group 
representations, obtained as in the case of localized fields (Part 2
of Proposition~\ref{SLQFTGA}), thanks to $\ua_{\bR^n}$- and 
$\ub_G$-invariance of $\ooi$. We define then for each double cone $O$ 
the von Neumann algebra
\begin{equation}
\Foix(O) := \bigcap_{C \supset O}\poix(\uFCO)'',
\end{equation}
and correspondingly a cyclic Hilbertspace $\Hoi :=
\overline{\bigcup_{O}\Foix(O)\Ooi}$, a net of von Neumann algebras
over it given by $\Foi(O):=\Foix(O)\rest\Hoi$, a translation group
representation $\Uscoi(a) := \Uscoix(a)\rest\Hoi$, $a \in \bR^n$, and 
a
gauge group representation $\Uoib(\gb):=\Uoix(g)\rest\Hoi$, $\gb \in
\Goib$, where, in analogy to the localizable case, $\Goib := G/\Noi$
with $\Noi$ the closed normal subgroup of $G$ of the elements $g \in 
G$ such
that $\Uoix(g)\rest\Hoi = \Id_{\Hoi}$, and $g \in G \mapsto \gb \in
\Goib$ the quotient map, so that $\Uoi$ is a faithful unitary
representation of $\Goib$ on $\Hoi$. We also denote by $\poi$ the
subrepresentation of $\poix$ determined by $\Hoi$.

\begin{Proposition}\label{prop:SLQFTGAcone}
The quintuple $(\Foi,\Uscoi(\bR^n),\Uoi(\Goib),\Ooi,\koib)$ defined
above is a
normal, covariant quantum field theory with gauge group action, which
will be called a \emph{scaling limit QFTGA}, corresponding to $\ooi$, 
of
the QFSGS determined by $(\cA_{\rm vac},\mathscr{U}_{\rm
  vac}(\bR^n),\O_{\rm vac})$ and $\D_{\rm fin}^{\rm BF,cov}$.
\end{Proposition}

\begin{proof}
The proof is completely analogous to the one of
Proposition~\ref{SLQFTGA}, so we don't repeat it here. The only thing
that deserves a comment is the normality of commutation relations
within $\Foi$. We first note that they hold for the system of algebras
$\uFCO$, in the sense that, by defining $\uF_\pm :=
\frac{1}{2}(\uF\pm\ub_k(\uF))$, relations analogous to
~\eqref{comrel} are satisfied for $\uF_i \in \uFF(C_i,O_i)$,
with spacelike separated $C_i$, $i=1,2$. Clearly this also carries over
to the system of algebras $\poix(\uFCO)''$, with respect to the
grading defined by $\boi_k$. If then $\fbs_i \in \Foix(O_i)$ with
spacelike separated $O_i$, $i=1,2$, we can assume that
$\fbs_i\in\poix(\uF(C_i,O_i))''$ with $C_i \supset O_i$, $i=1,2$,
spacelike separated to each other, so that normal commutation
relations also hold for the net $\Foix$, and hence for $\Foi$.
\end{proof}

The classification of the underlying theory in terms of the resulting
structure of the scaling limit theories given in 
Section~\ref{sec:QFTGA} can be
clearly applied also here, the isomorphism notion being again the one 
of
Definition~\ref{def:netisomorphism}.

We can also show a result analogous to 
Proposition~\ref{prop:dilation}, stating that if all the scaling limit
QFTGAs are isomorphic, then they are dilation covariant. Since the
formulation of this result and its proof are straightforward, we omit 
them.

It is also straightforward to show that if the underlying QFSGS is
also Lorentz covariant, where Lorentz covariance is defined as in
(QFTGA.6), understanding that spacelike cones substitute double
cones,\footnote{This happens for instance if one considers the QFSGS
  determined through the Doplicher-Roberts reconstruction theorem from
  the set of Poincar\'e covariant BF sectors.} then the same is true
for each scaling limit QFTGA, provided that in this case one considers
the scaling algebras $\uFCO$ obtained by redefining the
C$^*$-algebra $C_\a(\bR^+,B(\cH))$ appearing in 
Definition~\ref{def:scalingalgebracone}
as the C$^*$-subalgebra of $B_\b(\bR^+,B(\cH))$ whose elements ${\sf 
F}$ fulfill
\begin{equation*}
\nnorm{\,\underline{\a}_s({\sf F}) - {\sf F}\,} = \sup_{\l >
  0}\,\nnorm{\,\a_{s_\l}({\sf F}(\l)) - {\sf F}(\l)\,} \to 0 \quad
{\rm for} \quad s \to 1_{\rPport}\,,
\end{equation*}
where $s := (L,a)\in \rPport$ is a generic element of the covering of
the (proper orthochronous) Poincar\'e group $\rPport = \Lpocov \ltimes
\bR^n$, $s_\l := (L,\l a)$, and $\a_s := {\rm Ad}\Usc(s)$, $\Usc(L,a)
:= \Usc(a)\tU(L)$ (slightly abusing notation). We will denote by
$\Uscoi(s)$, $s \in \rPport$, the corresponding unitary 
representation of the
Poincar\'e group on $\Hoi$, with respect to which $\Foi$ is
covariant. We also note, for future use, that $\Uscoi(s) =
\Uscoix(s)\rest\Hoi$, where $\Uscoix(s)$ is the unitary representation
of $\rPport$ on $\Hoix$ defined by $\Uscoix(s)\poix(\uF)\Ooi =
\poix(\a_s(\uF))\Ooi$. 

We introduce here the standard notations
\begin{equation*}
\aoix_s := {\rm Ad}\Uscoix(s),\qquad \aoi_s := {\rm Ad}\Uscoi(s),
\qquad s\in \rPport,
\end{equation*}
and, for simplicity, we will
identify $L\in \Lpocov$ and $a\in \bR^n$ with
$(L,0),(1_{\Lpocov},a)\in\rPport$ respectively.

We close this section by putting on record a result generalizing Part
4 of Proposition~\ref{prop:QFTGAadd}.

\begin{Proposition}\label{prop:geometricmodular}
Assume that the underlying QFSGS determined by BF sectors is
Poin\-ca\-r\'e covariant and satisfies
the condition of geometric modular action (QFTGA.9), where in this
case $\cF(W) := \bigvee_{\overline{C}\subset W}\cF(C)$, and where
equation~\eqref{eq:geometricmodularJO} is substituted by
\begin{equation}
J_W\cF(C)J_W = \cF^t(r_W C), \qquad C \in \cS.
\end{equation}
Then each scaling limit QFTGA satisfies the condition (QFTGA.9) of geometric
modular action.
\end{Proposition}

\begin{proof}
Let $\fFtoix(W)$ be the C$^*$-algebra generated by all $\poix(\uFCO)$ with
$O \subset C$ and $\overline{C} \subset W$, and let $\cFtoix(W) :=
\fFtoix(W)''$. In Appendix~\ref{app:reehschlieder} we prove that the 
scaling limit vacuum $\Ooi$ is cyclic and
separating for $\cFtoix(W)$ and that $\ooi$ is a $(-2\pi)$-KMS state 
for
the C$^*$-dynamical system $(\fFtoix(W),\aoix_{\tLWd})$, and then also
for the W$^*$-dynamical system $(\cFtoix(W),\aoix_{\tLWd})$. Then, 
arguing as in the proof
of Lemma 6.2 in~\cite{BV1} and denoting by
$(\Doix,\Joix)$ the modular objects
determined by $(\cFtoix(W),\Ooi)$, it is easy to verify the relations
\begin{align}
\Joix \Uscoix(L,a)\Joix &= \Uscoix(\widetilde{{\rm Ad}r_W}L,r_W a), 
\label{eq:Joixpoincare}\\
(\Doix)^{it}&=\Uscoix(\tLWt),\label{eq:Doixt}\\
\Joix\poix(\uFCO)''\Joix &= \poix(\uFF(r_W C, r_W O))''. 
\label{eq:JoixCO}
\end{align}
But since $\Foix(W) := \bigvee_{\overline{O}\subset W}\Foix(O)$ is
contained in $\cFtoix(W)$, and $\Foi(W) = \Foix(W)\rest\Hoi$, it 
follows easily that $\ooi$ is also
$(-2\pi)$-KMS for $(\Foi(W),\aoi_{\tLWd})$, so that we also get
\begin{equation}
\Doi^{it} = \Uscoi(\tLWt).\label{eq:Doit}
\end{equation}
This, together with~\eqref{eq:Doixt} and standard arguments of
Tomita-Takesaki theory, implies that if $\fbs \in 
\Foi(W)$ is analytic for $\aoi_{\tLWd}$ there holds
\begin{equation*}
\Joi\fbs\Ooi = \Uscoi(\widetilde{\L_W(-i\pi)})\fbs^*\Ooi =
\Uscoix(\widetilde{\L_W(-i\pi)})\fbs^*\Ooi = \Joix \fbs\Ooi,
\end{equation*}
so that one obtains  $\Joix \rest \Hoi = \Joi$
(since the analytic elements for  $\aoi_{\tLWd}$ weakly dense
in $\Foi(W)$). The
equations~\eqref{eq:Joixpoincare}, \eqref{eq:JoixCO} then imply geometric modular
action for the scaling limit QFTGA.
\end{proof}

\section{Preservance of BF-type Charges in the Scaling
  Limit}\label{sec:BFpreservance}
\setcounter{equation}{0}

We will now generalize the notion of preservance of charges given in
Section~\ref{sec:DHRpreservance} so as to encompass the more general
situation of BF-type charges considered in the previous section. In
particular, in view of the physical picture of asymptotically free
theories discussed after 
Definition~\ref{def:scalingalgebracone}, and
of the ensuing construction of the scaling algebras and scaling limit,
we will formulate a criterion implying that a given BF charge of the
underlying theory gives rise to a localizable charge in the scaling
limit.

We consider then a Poincar\'e covariant observable net $(\Avac,
\Usc_{\rm vac}(\rPport),
\O_{\rm vac})$ and the corresponding Poincar\'e covariant QFSGS
$(\cF,\Usc(\rPport),U(G),\O,k)$ determined by the set $\D_{\rm
  fin}^{\rm BF,cov}$ of finite statistics, Poincar\'e covariant BF
sectors. In addition, we assume throughout this section that 
$(\cF,\Usc(\rPport),U(G),\O,k)$
satisfies the condition of geometric modular action (as formulated in
Proposition~\ref{prop:geometricmodular}). We note that also in this
case, as for localizable charges, this can be deduced from geometric
modular action of the observable net, supplemented by mild additional
assumptions~\cite{GLspst, Kuckert}.

The considerations about the possible phase space behaviours of 
localizable charges
discussed in Section~\ref{sec:DHRpreservance} as a motivation for the 
preservance
criterion for DHR-charges clearly apply also to the present case, as we
are considering asymptotically localized field operators, for which an
asymptotic phase space notion can be recovered. By this, we mean that
if $\psi_1(\l),\dots,\psi_d(\l)\in\cF(\l C)$ is some scaled multiplet
associated to a fixed BF sector $[\rho]$ of the underlying theory, and
if this multiplet is asymptotically localized in some $O \in \cK$,
then we can still think of the states $\psi_j(\l)\O$ as describing
a charge $[\rho]$ which is, for small $\l$, essentially localized
in $\l O$ so that, by looking at the energy content of these states,
we can define their phase-space occupation. Furthermore, the
direction of the cone $C$ in which this multiplet is localized is
irrelevant, in the sense that if $\hat{C}\supset O$ is another
spacelike cone, we can find another multiplet
$\hat{\psi}_1(\l),\dots,\hat{\psi}_d(\l) \in \cF(\l\hat{C})$ still
implementing the sector $[\rho]$. But from the picture of spacelike
cones as strings which tend to vanish at small scales, we expect that,
if also the multiplet $\hat{\psi}_j(\l)$ is asymptotically localized
in $O$, then t should be possible to choose it in such a way that the charged states $\psi_j(\l)^*\O$, $\hat{\psi}_j(\l)^*\O$
should become close to each other as $\l \to 0$. This motivates the
following generalization to the present setting of the notion of
asymptotic containment, for whose formulation we introduce the notation $R_1
\Subset R_2$ for two arbitrary spacetime regions $R_1$, $R_2$, to mean
that there exists some neighbourhood of the identity $\cN \subset
\Pport$ such that $\cN\cdot R_1\subseteq R_2$.

\begin{Definition}\label{def:asympcontcone}{\rm Let $O_1 \in \cK$, 
$C_1 \in
  \cS$ be such that $O_1 \subset C_1$, and let $\sF \in
  \utF(C_1,O_1)$. $\sF$ is said to be \emph{asymptotically contained
    in} $\Foi(O)$ with $O\supset\overline{O_1}$,
if for each spacelike cone $\hat{C}_1\supset O_1$ there exist some
$\hsF \in \utF(\hat{C}_1,O_1)$, with $\hsF = \sF$ for $\hat{C}_1 =
C_1$, fulfilling the following properties:
\begin{itemize}
\item[(A)] $\lim_{\k}\,\left(\nnorm{(\hsF(\l_{\k}) -\sF(\l_{\k}))\O} 
+\nnorm{(\hsF(\l_{\k}) -\sF(\l_{\k}))^*\O} \right) = 0$;
\item[(B)] for any $C \supset O$ such that $\hat{C}_1 \Subset C$, and 
for any $\epsilon > 0$, there exist elements $\uF,
  \uF' \in \underline{\fF}(C,O)$ (depending on $C$ and $\epsilon$) so 
that 
$$ \limsup_{\k}\,\left(\nnorm{(\hsF(\l_{\k}) - \uF_{\l_{\k}})\O} 
+\nnorm{(\hsF(\l_{\k}) - \uF'_{\l_{\k}})^*\O}\right) < \epsilon \,;$$
\end{itemize}
where $\{\l_\k\}_{\k\in K} \subset \bR^+$ is such that $\ooi = \lim_\k
\uo_{\l_\k}$.}
\end{Definition}

\begin{Lemma}\label{lem:alliseqcone}
Let $\sF \in \utF(C_1,O_1)$ and let $O \supset \overline{O_1}$. The 
following statements are equivalent:
\begin{itemize}
\item[$(a)$]  $\sF$ is  asymptotically
  contained in $\Foi(O)$;
\item[$(b)$] for each $\hat{C}_1 \supset O_1$ there exist some $\hsF 
\in \utF(\hat{C}_1,O_1)$, with $\hsF = \sF$ for $\hat{C}_1 =
C_1$, fulfilling 
\\[4pt]
property (A) of
  Defintion~\ref{def:asympcontcone} and the following property
\begin{itemize}
\item[$(B')$] for any $C\supset O$ such that $\hat{C}_1 \Subset C$,
  and for any given $\epsilon > 0$
  and finitely many $\uF^{(1)},\ldots,\uF^{(N)} \in \uFF^\times$, 
there is
  $\uF \in \uFF(C,O)$ such that for $j = 1,\dots,N\,;$
$$ \limsup_{\k}\,\left(\nnorm{(\hsF(\l_{\k}) - 
\uF_{\l_{\k}})\uF^{(j)}_{\l_{\k}}\O} +\nnorm{(\hsF(\l_{\k}) - 
\uF_{\l_{\k}})^*\uF^{(j)}_{\l_{\k}}\O}\right) < \epsilon \,;$$
\end{itemize}
\item[$(c)$] for each $\hat{C}_1 \supset O_1$ there exist 
some $\hsF \in \utF(\hat{C}_1,O_1)$, with $\hsF = \sF$ for $\hat{C}_1 
=
C_1$, such that
\begin{equation} \label{asapproxcone}
  \lim_{(\k,\n)}\,\left(\nnorm{\,(\, (\ua_{h_{\nu}}\hsF)(\l_{\k})
    - \sF(\l_\k)\,)\O\,} + \nnorm{\,(\, (\ua_{h_{\nu}}\hsF)(\l_{\k})
    - \sF(\l_\k)\,)^*\O\,} \right) = 0 
\end{equation}
whenever $\{h_{\nu}\}$ is a $\d$-sequence on $\rPport$,
where the limit is taken with respect to the product partial ordering
on $K \times \bN$, and where
$$ (\ua_h\hsF)(\l) := \int_{\rPport} ds\, h(s) \alpha_{s_\l}(\hsF(\l))\,, \quad 
\l > 0\,, \ h \in L^1(\rPport)$$
(integral in the weak sense, using the standard invariant measure on 
$\rPport$).
\end{itemize}
\end{Lemma}

\begin{proof} $(a)\Rightarrow(c)$. The proof proceeds analogously to the
  proof of the corresponding implication in Lemma~\ref{alliseq} using
  the estimate (with the notation introduced there)
\begin{equation*}\begin{split}
\nnorm{\,((\ua_{h_\n}\hsF)(\l_\k)-\sF(\l_\k))^\sharp\O\,} &\leq 
2\nnorm{\,(\uF^\sharp_{\l_\k}-\hsF(\l_\k))^\sharp\O\,}
\\
&+\sup_{s \in {\rm supp}\,h_\n}\nnorm{\,\ua_s(\uF^\sharp)-\uF^\sharp\,}+ 
\nnorm{\,(\hsF(\l_\k)-\sF(\l_\k))^\sharp\O}.
\end{split}\end{equation*}

$(c)\Rightarrow(b)$. From the estimate
\begin{equation*}\begin{split}
\nnorm{\,(\hsF(\l_\k)-\sF(\l_\k))^\sharp\O\,}&\leq\sup_{s\in{\rm
    supp}\,h_\n}\nnorm{\,[(\Usc(s_{\l_\k})-\Id)\hsF(\l_\k)]^\sharp\O\,}\\
&+\nnorm{\,((\ua_{h_\n}\hsF)(\l_\k)-\sF(\l_\k))^\sharp\O,}
\end{split}\end{equation*}
and from strong continuity of $s \mapsto \Usc(s)$ property (A) for
$\hsF$ follows at once, by first choosing $\k$ and
then $\n$, depending on $\k$, sufficiently large. Property 
(B')
is proven by the same argument as the one in the proof of the
corresponding implication in Lemma~\ref{alliseq}, using here the fact
that $\tilde{\cF}_{0,\iota}^{\times\,t}(W)\Ooi = 
V_{0,\iota}^\times\cFtoix(W)\Ooi$, with $V_{0,\iota}^\times$
the natural twisting operator on $\Hoix$, is dense in $\Hoix$,
Theorem~\ref{thm:reehschliederwedge}.

$(b)\Rightarrow(a)$. Obvious.
\end{proof}

\noindent \emph{Remark.} The field net on double cones defined by
$\hat{\cF}(O) := \bigcap_{C \supset O}\cF(C)$ is essentially the net
of the QFSGS determined by the localizable sectors of $\Avac$
(see~\cite{DRwhy} for precise statements), and we can associate to it
scaling algebras $\hat{\uFF}(O)$ and a scaling limit net
$\hat{\cF}_{0,\iota}(O)$ in the way discussed in previous sections. 
It is then
clear that $\hat{\uFF}(O) \subset \uFCO$ for each $C \supset O$ and
that a function $\sF:\bR^+\to\hat{\fF}$ complying with properties
(i)-(iii) in definition~\ref{asympcont} is an element of
$\utF(\hat{C}_1,O_1)$ for each $\hat{C}_1 \supset O_1$, so that if 
$\sF$ is
asymptotically contained in $\hat{\cF}_{0,\iota}(O)$, it is also
asymptotically contained in $\Foi(O)$ as it suffices to take $\hsF
= \sF$ for each $\hat{C}_1$ in Definition~\ref{def:asympcontcone}. Conversely, if
$\sF$ with $\sF(\l)\in\hat{\cF}(\l O_1)$ is asymptotically contained 
in
$\Foi(O)$, then thanks to the characterizations (c) in
Lemmas~\ref{alliseq} and~\ref{lem:alliseqcone}, it is also
asymptotically contained in $\hat{\cF}_{0,\iota}(O)$. 

\vspace{\baselineskip}
As in the case of localizable charges, a collection of multiplets
$\{\psi_1(\l),\ldots,\psi_d(\l)\}_{\l > 0}$ inducing a fixed BF sector
$[\r]$ and with $\psi_j(\l)\in\cF(\l C)$ for some $C \in \cS$, will 
be called a \textbf{scaled multiplet} for $[\r]$.

\begin{Definition}\label{def:BFpreservance} 
{\rm Let $\ooi \in {\rm SL}^{\uFx}(\o)$ be a scaling limit state of 
the underlying
QFSGS, and let $[\rho] \in {\rm Sect}^{\rm BF, cov}_{\rm fin}$ be a
BF superselection sector. Then we say that the charge $[\rho]$ is {\em
  preserved in the scaling limit QFTGA of $\ooi$} if, for each $O_1
\in \cK$, $C_1 \in \cS$ with $C_1 \supset O_1$, there is some scaled 
multiplet
$\{\psi_1(\l),\ldots,\psi_d(\l)\}_{\l > 0}$ for $[\rho]$ such that 
all functions
$\l \mapsto \psi_j(\l)$, $j = 1,\ldots,d$, are elements of
$\utF(C_1,O_1)$ and are asymptotically contained
in $\Foi(O)$ if $O \supset \overline{O_1}$. }
\end{Definition}

\begin{Proposition}\label{prop:limitmultipletcone}
Suppose that the charge $[\rho]$ is preserved in the scaling limit
QFTGA of $\ooi$. Let $\{\psi_1(\l),\ldots,\psi_d(\l)\}_{\l > 0}$ be a
scaled multiplet for $[\rho]$ such that
$\psi_j(\cdot)\in\utF(C_1,O_1)$
\\[4pt]
 is asymptotically contained in
$\Foi(O)$. Let, for $j=1,\dots,d$ and $\hat{C}_1 \supset O_1$, 
$\psC_j
\in \utF(\hat{C}_1,O_1)$ be as in Lemma~\ref{lem:alliseqcone}(c). Then the limit
operators 
\begin{equation} \label{SLmulticone}
 \boldsymbol{\psi}_j = s\mbox{-}\lim_{\n \to +\infty}\,
\poi(\ua_{h_\n}\psC_j) \quad {\it and}\quad
 \boldsymbol{\psi}_j^* = s\mbox{-}\lim_{\n \to +\infty}\,
\poi(\ua_{h_\n}\psC_j)^* 
\end{equation}
exist for any $\d$-sequence $\{h_\n\}$, are independent of $\hat{C}_1$ and of the chosen $\d$-sequence, and are contained in $\Foi(O)$
whenever $O \supset \overline{O_1}$. Furthermore,
$\boldsymbol{\psi}_1,\ldots, \boldsymbol{\psi}_d$ forms a multiplet
transforming under the adjoint action of $\Uoib(\Goib)$ according to
the irreducible, unitary representation $v_{[\rho]}$. More precisely,
denoting by $G \owns g \mapsto g^{\bullet} \in \Goib$ the quotient
map, there is a finite-dimensional, irreducible, unitary
representation $v_{[\rho]}^{\bullet}$ of $\Goib$ so that
$v_{[\rho]}^{\bullet}(g^{\bullet}) = v_{[\rho]}(g)$ for all $g\in G$
and 
$$ \Uoib(g^{\bullet})\boldsymbol{\psi}_j \Uoib(g^{\bullet})^* =
\sum_{i=1}^d \boldsymbol{\psi}_i
v_{[\rho]ij}^{\bullet}(g^{\bullet})\,, \quad g^{\bullet} \in
\Goib\,.$$
\end{Proposition}

\begin{proof}
The proof is similar to the proof of Proposition~\ref{prop:limitmultiplet},
so we will only indicate the major differences. From the inequality
\begin{multline*}
\nnorm{\,(\poix(\ua_{h_\n}\psC_j)-\poix(\ua_{\tilde{h}_{\tilde{\n}}}\psi^{\tilde{C}_1}_j))\Ooi\,}\\
\leq \limsup_\k 
\left(\nnorm{\,((\ua_{h_\n}\psC_j)(\l_\k)-\psi_j(\l_\k))\O\,}+
  \nnorm{\,((\ua_{\tilde{h}_{\tilde{\n}}}\psi^{\tilde{C}_1}_j)(\l_\k)-\psi_j(\l_\k))\O\,}\right),
\end{multline*}
valid for arbitrary $\d$-sequences $\{h_\n\}$, $\{\tilde{h}_{\tilde{\n}}\}$ and spacelike cones
$\hat{C}_1,\tilde{C}_1\supset O_1$, together with
Lemma~\ref{lem:alliseqcone}(c), it follows that $\lim_{\n \to +\infty}\poix(\ua_{h_\n}\psC_j)\Ooi =:
\Phi_j$ exists and is independent of $\hat{C}_1$ and of the chosen $\d$-sequence. As
$\nnorm{\ua_{h_\n}\psC_j}$ is uniformly bounded in $\n$, and since $\Ooi$ is separating
for $\cFtoix(W)$ with $W\Supset \hat{C}_1$, this implies that
$s\mbox{-}\lim_{\n \to +\infty} \poix(\ua_h\psC_j) =: \psbs^\times_j$ 
exists and
is defined by $\psbs^\times_j \fbs\Ooi = \fbs\Phi_j$ for $\fbs \in
\cFtoix(W)'$. Then, since for any two spacelike cones $\hat{C}_1$,
$\tilde{C}_1$ we can find spacelike cones $\hat{C}_2, \dots,
\hat{C}_n$ with $\hat{C}_n = \tilde{C}_1$, and wedges
$W_1,\dots,W_{n-1}$ such that $\hat{C}_i \cup \hat{C}_{i+1} \Subset
W_i$, $i=1,\dots,n-1$, we conclude that $\psbs^\times_j$ is
independent of $\hat{C}_1$, as well as of $\{h_\n\}$. Thus, since for any spacelike cone
$C \supset O$ there is a $\hat{C}_1 \supset O_1$ such that $\hat{C}_1 
\Subset C$, we have
$\psbs^\times_j\in\poix(\uFF(C,O))''$, and then $\psbs_j :=
\psbs_j^\times \rest \Hoi\in \Foi(O)$. The same argument shows that
$\poi(\ua_{h_\n}\psC_j)^*$ converges strongly to $\psbs_j^*$.

The rest of the proof is essentially identical to the corresponding
part of the proof of Prop.~\ref{prop:limitmultiplet}.
\end{proof}

If, for $\sF \in \utF(C_1,O_1)$, we define $\overline{\sF}(\l) := J_W
V \sF(\l)V^* J_W$, where $W \supset C_1$ is a wedge and $J_W$ is the
associated modular conjugation, and recalling that we assume that the
underlying QFSGS satisfies geometric modular action, it is easily 
checked that
$\overline{\sF} \in \utF(r_W C_1, r_W O_1)$
\\[4pt]
 and that
$(\ua_h\overline{\sF})(\l) = J_W V (\ua_{h\circ \widetilde{{\rm 
Ad}r_W}}
  \sF)(\l)J_W V^*$ (where $\widetilde{{\rm Ad}r_W}(L,a) := 
(\widetilde{{\rm
      Ad}r_W}L,r_W a)$), so that if $\sF$ is
asymptotically contained in $\Foi(O)$ then $\overline{\sF}$ is
asymptotically contained in $\Foi(r_W O)$, and it is then
straightforward to verify that the following generalization of
Theorem~\ref{PCT} holds.

\begin{Theorem}\label{thm:PCTcone}
Let $\ooi \in {\rm SL}^{\uFx}(\o)$ be a scaling limit state. Then a BF
charge $[\rho] \in {\rm Sect}^{\rm BF,
  cov}_{\rm fin}$ is preserved in the scaling limit state $\ooi$
 if and only if also the conjugate
charge $[\overline{\rho}]$ is preserved.
\end{Theorem}

We would then like to obtain a result corresponding to
Proposition~\ref{prop:SLendo}. However, at the present stage of our
work, this can be achieved only at the price of some additional
assumptions on the net $\Foi$, namely that $\Aoi(O) =
\Foi(O)^{\Goib}$. We will comment on this assumption below. Nevertheless, without
making this assumption, we can
at any rate show that the fields $\psbs_j$ constructed above give
rise, in a sense made precise in the following theorem, to positive
energy representations of $\AAoi$.

\begin{Theorem}\label{thm:covariantrep}
Let $\xi = [\rho]$ be a BF charge which is preserved in the scaling
limit QFTGA of a given $\ooi$, and let $\psbs_j \in \Foi(O)$,
$j=1,\dots,d$ a multiplet for $\xi$ arising as in
proposition~\ref{prop:limitmultipletcone}. Then the state $\o_\xi$ on
$\AAoi$ defined by
\begin{equation}
\o_\xi(\aabs) := \sum_{j=1}^d
\angscalar{\Ooi}{\psbs_j\aabs\psbs_j^*\Ooi}, \qquad \aabs \in \AAoi,
\end{equation}
obeys $\o_\xi \rest \AAoi(O') = \ooi \rest \AAoi(O')$ and
induces, via the GNS construction, a representation $\pi_\xi$ of
$\AAoi$ which is locally normal and translation covariant.
\end{Theorem}

\begin{proof}
It is evident that $\o_\xi\rest \Aoi(O)$ is a normal state, so that
$\pi_\xi$ is a locally normal representation of $\AAoi$. According to
a theorem of Borchers~\cite[Thm. II.6.6]{Borchers:1996}, in order to
show that $\pi_\xi$ is translation covariant, it
is necessary and sufficient to show that the set of vector states of
$\pi_\xi$ is contained in the norm closure of the set $\AAoi^*(V_+)$
of functionals $\phi \in \AAoi^*$ 
with the following property.
For each pair
$\aabs,\bbs\in\AAoi$, the function $x \in \bR^4
\mapsto \phi(\aabs\aoi_x(\bbs))$ is continuous and is the boundary
value of a function $W$ which is analytic in the forward tube $\cT :=
\bR^4 + i V_+$ and satisfies the bound
\begin{equation*}
\abs{W(z)}\leq \nnorm{\aabs}\nnorm{\bbs}e^{m\abs{{\rm Im}z}}, \qquad z
\in \cT,
\end{equation*}
for some constant $m > 0$ which may depend on $\phi$ but not on
$\aabs$, $\bbs$; furthermore the same conditions must be satisfied by
$\phi^*$ ($\phi^*(\aabs) := \overline{\phi(\aabs^*)}$).

Now, the set of operators $\cbs \in \AAoi$ with compact support in
momentum space, i.e. where there exists a compact $\D \subset
\bR^4$ such that $\aoi_h(\cbs) = 0$ for each $h \in L^1(\bR^4)$ with
${\rm supp}\,\hat{h} \subset \bR^4\setminus\D$, is strongly dense in
$\AAoi$.
To see this, take $\cbs = \aoi_f(\cbs_1)$ with $\cbs_1 \in \AAoi$ and compact
${\rm supp}\,\hat{f}$. Then $\cbs$ has compact momentum space support.
The set of $L^1$ functions $f$ with compact ${\rm supp}\,\hat{f}$ is dense in $L^1$.
Owing to the fact that the action of $\aoi$ is strongly continuous on $\AAoi$,
this implies that there exists a sequence of $L^1$-functions $f_n$ with
compact ${\rm supp}\, \hat{f}_n$ so that $\aoi_{f_n}(\cbs_1)$ approaches
$c_1$ in norm.
 
Then, using
\begin{equation*}
\nnorm{\pi_\xi(\cbs)\O_\xi-\pi_\xi(\boldsymbol{d})\O_\xi}^2=
\sum_{j=1}^d \nnorm{(\cbs - \boldsymbol{d})\psbs_j^*\Ooi}^2,
\end{equation*}
where $\O_\xi$ is a cyclic vector for $\pi_\xi$, it is sufficient to
show that the functionals $\phi_{\cbs}(\aabs) :=
\o_\xi(\cbs^*\aabs\cbs)$, with $\cbs \in \AAoi$ having compact
momentum space support, are contained in the norm closure of
$\AAoi^*(V_+)$.

To this end let $\cbs \in \AAoi$ have momentum support in a compact
set $\D$, and let $\D_n$ be the closed double cone in momentum space
with vertices $0$ and $(n,\boldsymbol{0})$, $n>0$. Then the functional
$\phi_{\cbs,n} \in \AAoi^*$ defined by
\begin{equation*}
\phi_{\cbs,n}(\aabs) := \sum_{j=1}^d\angscalar{\Ooi}{\psbs_j
  E(\D_n)\cbs^*\aabs\cbs E(\D_n)\psbs_j^*\Ooi},
\end{equation*}
where by $E$ we denote the spectral measure associated to
translations of $\Foi$, is such that the function $x \mapsto 
\phi_{\cbs,n}(\aabs\aoi_x(\bbs))$
is continuous and has distributional Fourier transform with support in
$-(\D+\D_n)+\overline{V}_+$. Then, if $p = (-m,\boldsymbol{0})$ is
such that $-(\D+\D_n)+\overline{V}_+ \subset p + \overline{V}_+$,
one concludes by \cite[Thm. II.1.7]{Borchers:1996} that this function is
a boundary value of a function $W$ analytic in $\cT$ and satisfies,
for suitable constants $k$, $M$, $N > 0$, the bound
\begin{equation*}
\abs{W(z)}\leq k (1+\abs{x})^N(1+{\rm dist}(y,\partial V_+)^{-1})^M
e^{m\abs{y}},\qquad z=x+iy \in \cT.
\end{equation*}
But there also holds
$\abs{W(x)}\leq\nnorm{\phi_{\cbs,n}}\nnorm{\aabs}\nnorm{\bbs}$ for $x
\in \bR^4$, and
this implies, by a standard analytic function argument,\footnote{We
  are indebted to J. Bros for helpful remarks on this point.} the
desired estimate
$\abs{W(z)}\leq\nnorm{\phi_{\cbs,n}}\nnorm{\aabs}\nnorm{\bbs}e^{m\abs{y}}$,
showing that $\phi_{\cbs,n}\in\AAoi^*(V_+)$. Then by the inequality
\begin{equation*}
\abs{\phi_{\cbs}(\aabs)-\phi_{\cbs,n}(\aabs)}\leq \sum_{j=1}^d 2
\nnorm{\cbs}^2\nnorm{\aabs}\nnorm{[E(\D_n)-\Id]\psbs_j^*\Ooi},
\end{equation*}
we get the statement.
\end{proof}

We denote by $\poi^{\rm vac}$ the vacuum
representation of $\AAoi$, defined by $\poi^{\rm vac}(\aabs) := \aabs
\rest \Hoi^{\rm vac}$, where $\Hoi^{\rm vac} = \overline{\AAoi\Ooi}$
is the scaling limit vacuum Hilbert space. Thanks to the separating
property of $\Ooi$ for local algebras, $\poi^{\rm vac}\rest\Aoi(O)$ is
an isomorphism of von Neumann algebras.

\begin{Corollary}
If the scaling limit vacuum Hilbert space $\Hoi^{\rm vac}$ is 
separable, then for each $x \in \bR^4$
\begin{equation}\label{eq:DHR}
\pi_\xi \rest \AAoi(O'+x) \cong \poi^{\rm vac}\rest \AAoi(O'+x),
\end{equation}
i.e. $\pi_\xi$ has the DHR property for the class of all translates of
the given double cone $O$.
\end{Corollary}

\begin{proof}
By the argument in the appendix of~\cite{DHR:1971},
the fact that $\o_\xi \rest \AAoi(O') = \ooi \rest \AAoi(O')$,
together with translation covariance of $\pi_\xi$,
imply~\eqref{eq:DHR} if it is known that property B holds in the
representation $\pi_\xi$. But if $\Hoi^{\rm vac}$ is separable, then
each local algebra $\poi^{\rm vac}(\Aoi(O))$ has a separable predual,
and being $\o_\xi$ locally normal, from~\cite[corollary
3.2]{Takesaki:1973} it follows that the Hilbert space $\cH_\xi$ of
$\pi_\xi$ is separable, and then, by the already recalled argument of
Roberts~\cite{Rob82}, property B holds in the representation
$\pi_\xi$.
\end{proof}

We recall that separability of $\Hoi^{\rm vac}$ follows from suitable
nuclearity properties of the underlying observable net~\cite{Bu.phsp}.

We now turn to discussing the conditions under which it is possible to
generalize Proposition~\ref{prop:SLendo}. At the technical level, the 
main
obstruction is represented by the fact that, in general, $\Aoi(O)
\subsetneq \Foi(O)^{\Goib}$, as it is easy to construct gauge
invariant combinations of the $\ua_h \psi_j$'s which 
need
not belong to some scaling algebra $\uAA(O)$ but are only
localized in spacelike cones. However, thanks to the fact that these
functions are asymptotically localizable in $O$, it may well happen
that, at least in favourable cases, their scaling limits do belong to
$\Aoi(O)$. Adding the simple hypotesis that this is indeed the case,
yields a quite satisfactory picture of the scaling limit of
morphisms.

\begin{Proposition}\label{prop:SLendocone}
Let $\ooi \in {\rm SL}^{\uFx}(\o)$ and assume that $\Aoi(O) =
\Foi(O)^{\Goib}$ and that $\fF_{0,\iota}$ acts irreducibly on
$\Hoi$. Moreover, let $[\rho] \in {\rm Sect}^{\rm
BF, cov}_{\rm fin}$ be a charge of the underlying QFSGS which is
preserved in the scaling limit QFTGA of $\ooi$, let
$\{\psi_1(\l),\ldots,\psi_d(\l)\}_{\l > 0}$ be a scaled multiplet for
$[\rho]$ asymptotically contained in $\Foi(O)$ and let, with respect
to this scaled multiplet,
$\boldsymbol{\psi}_1,\ldots,\boldsymbol{\psi}_d$ be defined as in
\eqref{SLmulticone}.

If we define, for each $\uA \in \underline{\fA}$, the family
$\{\rho(\uA)(\l)\}_{\l > 0}$ as
$$ \rho(\uA)(\l) = \sum_{j=1}^d \psi_j(\l)\uA_{\l}\psi_j(\l)^*, $$
then there holds
\begin{equation} 
s\mbox{-}\lim_{\n \to +\infty}\,\poi(\underline{\alpha}_{h_\n}\rho(\uA)) = 
 \sum_{j=1}^d \boldsymbol{\psi}_j\poi(\uA)\boldsymbol{\psi}_j^*\,,
 \quad \underline{A} \in \uAA\,;
\end{equation}
and $\boldsymbol{\rho}$ defined by
\begin{equation} \label{SLendocone}
 \boldsymbol{\rho}(\boldsymbol{a}) = \sum_{j=1}^d
\boldsymbol{\psi}_j
 \boldsymbol{a} \boldsymbol{\psi}_j^*\,, \quad \boldsymbol{a} \in
 \AAoi\,,
\end{equation}
is a localized, transportable, irreducible endomorphism of $\AAoi$
which is moreover covariant and has finite 
statistics.
\end{Proposition}

The proof of these statements is completely parallel to that of
Proposition~\ref{prop:SLendo}.

As a final comment, we would like to remark that the condition
$\Aoi(O) = \Foi(O)^{\Goib}$, introduced here as a technical assumption
in order to get a well defined scaling limit of morphisms, may
turn out to have a sensibile physical interpretation. By the above
remarks, we see that $\Foi(O)^{\Goib}$ contains, apart from the
scaling limit observables localized in $O$, the scaling limit of
functions $\l \mapsto \uA_\l \in \cA(\l C)$, for every spacelike cone
$C \supset O$, i.e. there are gauge invariant families of operators,
with localization regions extending to spacelike infinity, which give
rise to objects in the scaling limit which are charged with respect to
the intrinsic gauge group of $\Aoi$, so that new charges appear at
small scales. This situation, which must not be confused with
confinement where the fields carrying the new charges
cannot be approximated at all at finite scales, is instead reminiscent
of the phenomenon of \emph{charge screening},\footnote{This connection
  was pointed out to us by Detlev Buchholz.} much discussed in the
physical literature (cf.\ for instance~\cite{Rothe:1979fi, Swieca:1976xc} and references quoted). In this scenario, a charge
which is described by an asymptotically free theory at small scales
disappears at finite scales because, due to nonvanishing interactions,
it is always accompanied by a cloud, extending to spacelike
infinity, of charge-anticharge pairs, so that one can expect that the
corresponding ``charge carrying fields'' are neutral and non-compactly
localized at finite scales, and become instead charged and localized
in the scaling limit. Then the condition  $\Aoi(O) = \Foi(O)^{\Goib}$ 
could be interpreted as the requirement that in the theory under 
consideration, no charges are screened.
%%%%%%%%%%%%%%%%%%%%%%%%%%%%%%%%%%%%%%%%%%%%%%%%%%%%%%%%%%
${}$\\[32pt]
{\Large {\bf Concluding Remarks}}\\[20pt]
A generalization of the scaling algebra framework to the situation where the operator
algebras describing the underlying quantum field theory contain charge-carrying fields
has been developed in this work, 
together with a proposal as to what it means that a 
charge present in the underlying theory is preserved in the scaling limit.
A natural concept of confined charge arises as a charge in the scaling limit
theory which is not obtainable as a charge of the underlying theory
which is preserved in the scaling limit process \cite{Bu.Conf,Bu.QGC}. 
 We have
indicated two basic physical mechanisms for the disappearance of charges in the scaling limit.
Moreover, we have seen that the preservance of all charges in the scaling limit leads
to the equivalence of local and global intertwiners for the superselection sectors in the
underlying theory.

We hope that in the future it will be possible to illustrate the mechanisms for charge disappearance in the scaling limit
by instructive examples, possibly in lower spacetime dimensions. This should also shed light
on the very important issue if the lifted action of the gauge transformations shouldn't be defined
differently than in \eqref{liftgauge} in the case where the physical dimension of charge is
related to the dimension of length. Furthermore, it would also appear desirable to
develop, based on the method of scaling algebras for the observables of the underlying theory,
an abstract renormalization group analysis for superselection charges without using the Doplicher-Roberts
reconstruction theorem. This would eventually make superselection charges with 
braid group statistics and with infinite statistics
accessible to short-distance analysis.
%%%%%%%%%%%%%%%%%%%%%%%%%%%%%%%%%%%%%%%%%%%%%%%%%%%%%%%%%%%%
${}$\\[32pt]
{\large {\bf Acknowledgements}}\quad  It is a pleasure to thank D.\ 
Buchholz
for many helpful discussions and instructive comments. Similar thanks
are also extended to S. Doplicher, who was PhD supervisor of one of us
(G.M.), and to J.E.\ Roberts and K.-H.\ Rehren. 
  R.V. gratefully acknowledges financial support by
INDAM-GNAMPA. G.M. gratefully acknowledges financial support by DAAD
and MPI-MIS.

\renewcommand{\thesection}{\Alph{section}}
\setcounter{section}{0}
\vspace{2 \baselineskip}
\section{An example of a preserved localizable charge}\label{app:example}
\setcounter{equation}{0}

In this Appendix we shall show that the localizable charge described 
by the Majorana field in $n=1+3$ spacetime dimensions with $\bZ_2$ 
gauge group satisfies the preservance condition, 
Definition~\ref{def:DHRpreservance}, in all scaling limit states.

For the definition of the Majorana field, we will mainly 
follow~\cite{Fredenhagen:1995a}, where also a discussion of the 
superselection structure is given.

We begin with some notational conventions. Let 
\begin{equation}
\gamze = \begin{pmatrix} 0 & {\bf 1} \\ {\bf 1} & 0\end{pmatrix}, 
\qquad \gamma^{\, j} = \begin{pmatrix} 0& \sigma_j \\ -\sigma_j &0 
\end{pmatrix}, \qquad j=1,2,3,
\end{equation}
be the Dirac matrices in chiral representation, where $\sigma_j$ are 
the Pauli matrices. A vector $u \in \bC^4$ (also called a spinor) 
will be thought as a column matrix and correspondingly its adjoint 
$u^\dagger$ will be a row matrix, so that the standard scalar product 
on $\bC^4$ is given by $(u,v) \mapsto u^\dagger v$ (rows by columns 
product of matrices). We adopt the notation $\slash{v}:= v_\mu 
\gammu$ for any (covariant) vector $v \in \bR^4$.  By $\O_m^\pm$ we 
shall indicate the upper and lower mass $m>0$ hyperboloid, $\O_m^\pm 
:= \{p \in \bR^4:p^2=m^2,\pm p_0 > 0\}$. 

For a given mass $m>0$, the Dirac operator is $D := \gammu \de_\mu + 
i m$, and, denoting as usual by $\cD(\bR^4;\bC^4)$ the space of 
spinor valued, compactly supported smooth functions on Minkowski 
space, we endow the space $H_{0,m} := \cD(\bR^4;\bC^4)/\Imag D$ with 
the scalar product
\begin{equation}\label{eq:scalarprod}
\angscalar{f}{g}_m := \int_{\bR^3} d^3\spp\; 
\sum_\pm\hat{f}(\pm\omp,\spp)^\dagger 
P_\pm(\spp)\hat{g}(\pm\omp,\spp),
\end{equation} 
where
\begin{equation}
P_\pm(\spp) = \frac{\gamze(\ps+ m)}{2 p_0}\bigg|_{p_0 = \pm\omp}, 
\qquad \omp = \sqrt{\abs{\spp}^2+m^2},
\end{equation}
and where we made no notational distinction between elements in 
$H_{0,m}$ and their representatives in $\cD(\bR^4;\bC^4)$. Let $H_m$ 
be the completion of $H_{0,m}$ in this scalar product.

The action of the universal covering of the Poincar\'e group on $H_m$ 
is defined, for $(A,a) \in \rPport$, by
\begin{equation}
\big(u(A,a)f\big)(x) := S(A) f\big(\L(A)^{-1}(x-a)\big), \qquad S(A) 
:= \begin{pmatrix} A & 0 \\ 0 & (A^\dagger)^{-1} \end{pmatrix},
\end{equation}
$A \in SL(2,\bC) \mapsto \L(A) \in \SOo$ being the covering 
homomorphism. 

Let $C$ be the antilinear operator on $\bC^4$ defined by $C u := 
i\gamtwo \overline{u}$, where the bar denotes complex conjugation, 
which satisfies $C^2 = {\bf 1}$, $C^\dagger = C$ and $C \gammu C = 
-\gammu$, and define then an antilinear involution $\Gamma$ on 
$H_{0,m}$ by $(\Gamma f)(x) := Cf(x)$,  which is antiunitary, 
$\angscalar{\Gamma f}{\Gamma g}_m = \angscalar{g}{f}_m$ (so that it 
extends to $H_m$) and commutes with the action of the Poincar\'e 
group.

Let $\fB(H_m)$ be the self-dual CAR algebra over 
$H_m$~\cite{Araki:1968}, generated as a C$^*$-algebra by elements 
$B(f)$, $f \in H_m$, such that $f \mapsto B(f)$ is antilinear, and
\begin{equation}
\{B(f),B(g)\} = \angscalar{g}{\Gamma f}_m {\bf 1}, \qquad B(f)^* =
B(\Gamma f).
\end{equation}
By CAR unicity, the representation $u$ of $\rPport$ on $H_m$ induces 
an automorphic action $\a$ of $\rPport$ on $\fB(H_m)$, defined by
\begin{equation*}
\a_{(A,a)}(B(f)) := B(u(A,a)f), \qquad (A,a) \in \rPport, f \in H_m,
\end{equation*}
and, by the fact that $\norm{B(f)} \leq 2\norm{f}_m$ and strong 
continuity of $u$, it follows that this action is strongly 
continuous, i.e. $(A,a) \mapsto \a_{(A,a)}(B)$ is norm continuous for 
each $B \in \fB(H_m)$.

We consider on $\fB(H_m)$ the quasifree state $\o$ defined, according to~\cite{Araki:1970}, by the 2-point function
\begin{equation}
\o\big(B(f)B(g)\big) := \angscalar{\Gamma f}{P_+g}_m,
\end{equation}
where $P_+$ is the projection on the positive
energy states in $H_m$, defined by $\widehat{P_+ f}(p_0,\spp) = 
P_+(\spp)\hat{f}(p_0,\spp)$.

The action $\a$ of $\rPport$ leaves $\o$ invariant, so that if we 
consider the GNS representation $(\pi,\cH,\O)$ induced by $\o$, we 
get on $\cH$ a unitary strongly continuous representation 
$\mathscr{U}$ of $\rPport$ leaving $\O$ invariant and such that 
$(\pi,\mathscr{U})$ is a covariant representation of $(\fB(H_m),\a)$. 

\begin{Definition}
The \emph{free Majorana field of mass $m > 0$} is the operator valued 
distribution $f \in \cD(\bR^4;\bC^4) \mapsto \psi(f) \in 
\mathscr{B}(\cH)$ given by $\psi(f) := \pi\big(B(f)\big)$, $f \in 
\cD(\bR^4;\bC^4)$, where on the right hand side $f$ is identified 
with its image in $H_m$.
\end{Definition}

It is straightforward to verify that 
$\psi$ is covariant with respect to $\mathscr{U}$,
\begin{equation*}
\mathscr{U}(A,a) \psi(f) \mathscr{U}(A,a)^* = \psi(u(A,a)f),
\end{equation*}
that the translations $a \mapsto \mathscr{U}({\bf 1},a)$ satisfy the 
spectrum condition with $\O$ as the unique (up to a phase) 
translation invariant unit vector in $\cH$, and that $\psi(f)$, 
$\psi(g)$ anticommute for spacelike separated $\supp f$, $\supp g$.

We now turn to the consideration of the net of local von~Neumann 
algebras associated to the free Majorana field,  and defined by
\begin{equation}
\FcO := \{ \psi(f) \,:\, \supp f \cont O\}'',
\end{equation}
for $O \subset \bR^4$ open and bounded. On this net the group $\bZ_2$ 
acts by an automorphism $\beta_k$ induced by the automorphism of 
$\fB(H_m)$ defined by $B(f) \mapsto -B(f)$, which leaves the vacuum 
state $\o$ invariant, and is therefore implemented by a unitary 
operator $U(k)$ on $\cH$ such that $U(k)^2 = U(k^2) = {\bf 1}$, so 
that it induces a direct sum decompostion $\cH = \cH_+\oplus\cH_-$ 
according to its eigenspaces, i.e. $U(k)\rest\cH_\pm = \pm{\bf 
1}_{\cH_\pm}$, which is Poincar\'e and gauge invariant. Define then 
$\mathscr{U}_{\text{vac}}(A,a) := \mathscr{U}(A,a)\rest\cH_+$, and 
the net of observable von~Neumann algebras associated to the free 
Majorana field as
\begin{equation}
\AvacO := \FcO^{\bZ_2}\rest\cH_+.
\end{equation}
That in this way we get an example satisfying the assumptions made in 
Sections~\ref{sec:QFTGA} and~\ref{sec:QFSGS} is the content of the 
following proposition, the proof of which, being standard, is omitted.

\begin{Proposition}
With the above notations, and with $\Hvac := \cH_+$, let $\pi$ be the 
representation of the quasi-local algebra $\fAvac$ defined by 
$\pi(A\rest\Hvac) := A$.  Then $(\cA, \Uvac(\Pport), \O)$ is a 
Poincar\'e covariant observable net, and $(\cF, 
\mathscr{U}(\rPport),U(\bZ_2),\O,k)$ is a QFSGS on it. 
\end{Proposition}

In the next proposition, the very simple superselection structure of
$\fAvac$ described by the field net $\cF$ is analysed, 
cf.~\cite{Fredenhagen:1995a}.

\begin{Proposition}
The representation $\pi_-$ of $\fAvac$ given by $\pi_- := 
\pi(\cdot)\rest\cH_-$ satisfies the DHR criterion, is covariant, 
irreducible and with finite statistics, and any irreducible 
representation of $\fAvac$ appearing in $\cH$ is equivalent either to 
$\iota$, the identity representation, or to $\pi_-$. Moreover, if for 
$f \in \cD(O,\bC^4)$ with $\supp f \cont O$, $\Gamma f = f$ and 
$\norm{f}_m = \sqrt{2}$, $\rho_f$ is the automorphism of $\fAvac$ 
induced by the unitary operator $\psi(f) \in \FcO$, then $\rho_f \in 
\D_{\text{\emph{fin}}}^{\text{\emph{cov}}}(O)$ and $\rho_f \unitequiv 
\pi_-$.
\end{Proposition}

\begin{proof}
The irreducibility of $\pi_-$ follows from the arguments 
in~\cite{Doplicher:1969tk}, taking into account that $\cH_-$ is the 
subspace associated to the irreducible representation $k \mapsto -1$ 
of $\bZ_2$ in the factorial decomposition of $U$. This also implies 
that any other irreducible representation of $\fAvac$ in $\cH$ is 
equivalent to $\iota$ or $\pi_-$. To show that $\pi_-$ satisfies the 
DHR criterion, fix a double cone $O$ and an $f \in \cD(O;\bC^4)$ with 
$\Gamma f = f$ and $\norm{f}_m = \sqrt{2}$. Then $\psi(f)$ is unitary 
by the CARs, and $\psi(f)\cH_\pm = \cH_\mp$. Let then $V_f := 
\psi(f)\rest\cH_-$, and $g,h \in \cD(O',\bC^4)$. We have 
$V_f\pi_-\big(\psi(g)\psi(h)\big) = \psi(g)\psi(h)V_f$, so that $V_f$ 
intertwines between $\pi_-\rest\fAvac(O')$ and 
$\iota\rest\fAvac(O')$. Covariance of $\pi_-$ follows by 
$\mathscr{U}_- := \mathscr{U}(\cdot)\rest\cH_-$. Finally if 
$\pi\rho_f(A) = \psi(f) \pi(A) \psi(f)^*$ then $\rho_f$ is localized 
in $O$ and has finite statistics, and $V_f$ intertwines between 
$\pi_-$ and $\rho_f$, and if $\supp f_1 \cont O_1$, $V := 
\psi(f)\psi(f_1)^* \rest \cH_+$ intertwines between $\rho_{f_1}$ and 
$\rho_f$, which is therefore transportable.
\end{proof}

Finally we come to the proof of the fact that the charge $\xi := 
[\pi_-]$ in the above proposition is preserved in any scaling limit 
theory, in the sense of Definition~\ref{def:DHRpreservance}. To this 
end it is sufficient to find, for every double cone $O_1$, a family 
$(f_\l)_{\l \in (0,1]}$ of functions such that $\supp f_\l \cont \l 
O_1$, $\norm{f_\l}_m = \sqrt{2}$, and such that 
condition~\eqref{def:DHRpreservance} is satisfied for $\psi(f_\l)$.

\begin{Proposition}\label{prop:majoranastable}
For every double cone $O_1$, there exists $f \in \cD(O_1;\bC^4)$ such 
that, if \mbox{$f_\l \in \cD(\l O_1;\bC^4)$} is defined by
\begin{equation*}
f_\l(x) := \l^{\frac{3}{2}-4} f(\l^{-1}x), \qquad \l \in (0,1],
\end{equation*}
then $\norm{f_\l}_m = \sqrt{2}$ and $\Gamma f_\l = f_\l$ for $\l \in 
(0,1]$, and $\l \mapsto \psi(\l) := \psi(f_\l)$ is asymptotically 
contained in $\FFoi(O)$ for each $O \supset \overline{O}_1$ and each 
scaling limit state $\ooi$. 
\end{Proposition}

In the course of the proof of this proposition, we will need the 
following simple result concerning the action of the Lorentz group on 
Minkowski space.

\begin{Lemma}
Fix a mass $m>0$. For any sufficiently large $R > 0$, there exists a 
neighbourhood of the identity $\cN$ in $\SOo$ such that, for any $p 
\in \Vpc$ with $0 \leq p^2 \leq m^2$ and $\abs{\spp} > R$, and for 
any $\L \in \cN$, it holds, for $p' := \L p$, $\abs{\spp'} > 
\abs{\spp}/\sqrt{2}$.
\end{Lemma}

\begin{proof}
To simplify the notation, we will write $\L\cdot\spp$ for the spatial 
part, in a given Lorentz frame, of the 4-vector $\L p$, $\L \in 
\SOo$, $p \in \bR^4$. Let $\L_1(s)$, $s \in \bR$, denote the 
1-parameter group of boosts in the $p_1$ direction. 

If $p \in \bR^4$ is such that $\abs{p_1}^2 \leq 
\abs{p_2}^2+\abs{p_3}^2$, since $\L_1(s)$ leaves the components 
$p_2$, $p_3$ unaffected, we have, for any $s \in \bR$, 
$\abs{\L_1(s)\cdot\spp}^2 \geq \abs{p_2}^2 + \abs{p_3}^2 \geq 
\abs{\spp}^2/2$.

Assume now that $\abs{p_1}^2 > \abs{p_2}^2+\abs{p_3}^2$. This implies 
$\abs{p_1} \geq \abs{\spp}/\sqrt{2}$ and, since for any sufficiently 
large $R > 0$,
\begin{equation*}
\inf_{\substack{0\leq p^2 \leq m^2 \\ \abs{\spp} > R}} 
\frac{\abs{\spp}}{\abs{p_0}} \geq \inf_{\abs{\spp}>R} 
\frac{\abs{\spp}}{\sqrt{\abs{\spp}^2 + m^2}} > 0,
\end{equation*}
we can find a $\delta > 0$ such that, if $\abs{s} < \delta$, 
$\abs{(\L_1(s)p)_1} = \abs{\sinh s \,p_0 + \cosh s \,p_1} \geq 
\abs{p_1}/\sqrt{2}$ for any $p \in \Vpc$ with $0 \leq p^2 \leq m^2$ 
and $\abs{\spp} > R$, so that $\abs{\L_1(s)\cdot \spp}^2 \geq p_1^2/2 
+ p_2^2 + p_3^2 \geq \abs{\spp}^2/2$.

Then, if we identify in the canonical way $SO(3)$ with a subgroup of 
$\SOo$, we conclude with $\cN := \{R_1 \L_1(s) R_2 \; : \; \abs{s} < 
\delta, \;R_1,R_2 \in SO(3)\}$.
\end{proof}

\begin{proof}[Proof of Proposition~\ref{prop:majoranastable}]
In order to shorten formulae, we will use the notation $p_{\l,\pm} := 
(\pm\o_{\l m}(\spp),\spp)$, as well as the notation $\L \cdot \spp$ 
introduced in the proof of the above lemma. Also, $\abs{\cdot}$ will 
denote the norm of a vector both in $\bR^3$ and in $\bC^4$. A 
calculation shows
\begin{equation*}
\norm{f_\l}_m^2 = \norm{f}_{\l m}^2 = \int_{\bR^3} \frac{d^3\spp}{4 
\o_{\l m}(\spp)^2} \; \sum_\pm\bigabs{\gamze(\slash{p_{\l,\pm}}+\l 
m)\hat{f}(p_{\l,\pm})}^2,
\end{equation*}
and then, in order to show that there is an $f \in \cD(O,\bC^4)$ such 
that $\norm{f_\l}_m = \sqrt{2}$ and $\Gamma f_\l = f_\l$ for each $\l 
\in (0,1]$, it is sufficient to exhibit an $f \in \cD(O,\bC^4)$ such 
that $\Gamma f = f$, and for which $(\ps + \mu)\hat{f}(p)$ is not 
identically zero on each hyperboloid $\O_\mu := \O_\mu^+ \cup 
\O_\mu^-$, $\mu > 0$. A direct check shows that these conditions are 
met by $f(x) := g(x)({\bf 1} + i\gamtwo)\begin{pmatrix} 1 &0 &0 &0 
\end{pmatrix}^t$ where $g \in \cD(O;\bR)$.

We now show that actually $\l \mapsto \psi(\l)$ is an element of the 
scaling algbera itself, i.e. that
\begin{equation}
\lim_{(A,a) \to ({\bf 1},0)} \sup_{\l\in(0,1]} \bignorm{\a_{(A,\l 
a)}\big(\psi(f_\l)\big) - \psi(f_\l)} = 0,
\end{equation}
which clearly implies the statement. We have
\begin{multline*}
\bignorm{\a_{(A,\l a)}\big(\psi(f_\l)\big) - \psi(f_\l)}^2 \leq 
4\norm{u(A,a)f - f}^2_{\l m} \\
= 4\int_{\bR^3}\frac{d^3\spp}{4 \o_{\l m}(\spp)^2} \sum_\pm 
\Bigabs{\gamze(\slash{p_{\l,\pm}}+\l m)\Big(e^{ip_{\l,\pm}\cdot 
a}S(A)\hat{f}(\L(A)^{-1}p_{\l,\pm})- \hat{f}(p_{\l,\pm})\Big)}^2,
\end{multline*}
and, considering only the $+$ term in the sum inside the integral 
(the other one is estimated in the same way), and writing $p_\l := 
p_{\l,+}$,
\begin{equation}\begin{split}\label{eq:threeterms}
\int_{\bR^3}\frac{d^3\spp}{4 \o_{\l 
m}(\spp)^2}&\Bigabs{\gamze(\slash{p_\l}+\l m)\Big(e^{ip_\l\cdot 
a}S(A)\hat{f}(\L(A)^{-1}p_\l)- \hat{f}(p_\l)\Big)}^2 \\
&\leq 
\frac{5}{4}\int_{\bR^3}\frac{d^3\spp}{\abs{\spp}}\bigabs{e^{ip_\l\cdot 
a}S(A)\hat{f}(\L(A)^{-1}p_\l)- \hat{f}(p_\l)}^2 \\
&\leq \frac{5}{4}\Big\{ 
\norm{S(A)}\Big[\bignorm{\hat{f}(\L(A)^{-1}p_\l) - \hat{f}(p_\l)}_2 + 
\bignorm{(e^{ip_\l\cdot a} -1)\hat{f}(p_\l)}_2 \Big]\\
&\phantom{\sqrt{\frac{5}{4}}\Big\{}+\norm{S(A)-{\bf 
1}}\lVert\hat{f}(p_\l)\rVert_2\Big\}^2,
\end{split}\end{equation}
where $\norm{\cdot}_2$ denotes the standard norm in 
$L^2(\bR^3,d^3\spp/\abs{\spp})\otimes\bC^4$, and where, for more 
clarity, we indicated explicitly the variable of integration inside 
the norms. The last term of the last line in this equation can be 
estimated uniformly in $\l$ by the fact that, being \mbox{$\hat{f} 
\in \cS(\bR^4;\bC^4)$}, there are constants $C> 0$, $n>1$, such that
\begin{equation*}
\int_{\bR^3} \frac{d^3\spp}{\abs{\spp}}\bigabs{\hat{f}(p_\l)}^2 \leq 
C\int_{\bR^3} \frac{d^3\spp}{\abs{\spp}}\frac{1}{(1+\o_{\l m}(\spp)^2 
+ \abs{\spp}^2)^n} \leq C\int_{\bR^3} 
\frac{d^3\spp}{\abs{\spp}}\frac{1}{(1+2\abs{\spp}^2)^n},
\end{equation*}
so that it can be made arbitrarily small, as $A \to {\bf 1}$, 
uniformly in $\l \in (0,1]$. For the second term in square brackets 
at the end of~\eqref{eq:threeterms}, we have, by an application of 
Lagrange's theorem to the exponential,
\begin{equation*}
\int_{\bR^3}\frac{d^3\spp}{\abs{\spp}} \bigabs{(e^{ip_\l\cdot a} 
-1)\hat{f}(p_\l)}^2 \leq C(\lvert 
a^0\rvert^2+\abs{\boldsymbol{a}}^2)\int_{\bR^3}\frac{d^3\spp}{\abs{\spp}}\frac{\abs{\omp}^2 
+ \abs{\spp}^2}{(1+2\abs{\spp}^2)^n},
\end{equation*}
and then, if $n > 2$, this term is also uniformly small in the 
relevant limit. Finally, we use the above lemma to estimate the first 
term in square bracket at the end of~\eqref{eq:threeterms}. For each 
sufficiently large $R > 0$ let $\cN_R$ be a neighbourhood of the 
identity in $SL(2,\bC)$, such that $\L(\cN_R) \cont \SOo$ is as in 
the lemma. Then, for $\abs{\spp} > R$, $A \in \cN_R$,
\begin{equation*}\begin{split}
\bigabs{\hat{f}(\L(A)^{-1}p_\l) - \hat{f}(p_\l)} &\leq 
C\bigg[\frac{1}{(1+2\abs{\L(A)^{-1}\cdot\spp_\l}^2)^n} 
+\frac{1}{(1+2\abs{\spp_\l}^2)^n} \bigg]\\
&\leq C\bigg[\frac{1}{(1+\abs{\spp}^2)^n} 
+\frac{1}{(1+2\abs{\spp}^2)^n}\bigg].
\end{split}\end{equation*}
Thus, again by Lagrange theorem, we have, for $A \in \cN_R$,
\begin{equation*}\begin{split}
\int_{\bR^3}\frac{d^3\spp}{\abs{\spp}} 
\bigabs{\hat{f}(\L(A)^{-1}p_\l) - \hat{f}(p_\l)}^2 &\leq 
C\int_{\abs{\spp}>R}\frac{d^3\spp}{\abs{\spp}}\bigg[\frac{1}{(1+\abs{\spp}^2)^n} 
+\frac{1}{(1+2\abs{\spp}^2)^n}\bigg]^2 \\
&+ \lVert\de \hat{f}\rVert_\infty^2 \norm{\L(A)^{-1}-{\bf 
1}}^2\int_{\abs{\spp}<R}\frac{d^3\spp}{\abs{\spp}}(\abs{\omp}^2 + 
\abs{\spp}^2).
\end{split}\end{equation*}
and the $\l$ independent right hand side can be made arbitrarily 
small by taking $R$ sufficiently large, and $A$ in a corresponding 
neighbourhood $\tilde{\cN}_R \cont \cN_R$.
\end{proof}

\section{Reeh-Schlieder property for 
$\cFtoix(W)$}\label{app:reehschlieder}
We employ the notations introduced in Section~\ref{sec:scalingcone}. 
Let $\fFtoix(C)$ be the C$^*$-algebra generated by $\poix(\uFCO)$ as 
$O \subset C$.

\begin{Theorem}\label{thm:reehschliederwedge}
The vacuum $\Ooi$ is a cyclic and separating vector for the algebras 
$\cFtoix(W)$.
\end{Theorem}Let $\fFtoix(C)$ be the C$^*$-algebra generated by 
$\poix(\uFCO)$ as $O \subset C$.

We will give a sketch of the proof of this theorem, which uses in an 
essential way analyticity of both translations and Lorentz boosts, 
consequence of geometric modular action and Tomita-Takesaki theory. 
Similar results can be found in~\cite{Borchers:1999a, 
Driessler:1986a}, to which we refer the interested readers for the 
details, which can also be found in~\cite{Morsella2}.

We need some preparations. We recall that we denote by $\L_W(t) \in 
\Pport$, $t \in \bR$, the one parameter group of Poincar\'e 
transformations leaving the wedge $W$ invariant. In order to simplify 
notations, we will identify $\L_W(t)$ with its unique smooth lift to 
$\rPport$ which is the identity for $t=0$.

\begin{Lemma}\label{lem:poincareboost}
Let $\Uscoix$ be a strongly continuous unitary representation of 
$\rPport$, and $\cN \subseteq \rPport$ an open neighbourhood of the 
identity. Then $\Uscoix(\rPport)$ is the strong closure of the group 
$\mathscr{U}_{\cN}$ generated by the elements 
$\Uscoix(s\L_W(t)s^{-1})$, $t \in \bR$, $s \in \cN$.
\end{Lemma}

\begin{proof}
We can assume that $W = W_R$, and that $\cN =\cN_1\times\cN_2 \cont 
SL(2,\bC)\times\bR^4$. Then by~\cite[lemma 2.1]{Borchers:1999a}, 
$\Uscoix(SL(2,\bC))$ is the strong closure of the subgroup of 
$\mathscr{U}_{\cN}$ generated by $\Uscoix(\L\LaWo(t)\L^{-1})$, $t \in 
\bR$, $\L \in \cN_1$,\footnote{the cited results refers actually to 
representations of $\SOo$, but since the proof uses only properties 
of its Lie algebra, it can be also applied to the present case} so 
that it is sufficient to show that $\Uscoix(x) \in 
\mathscr{U}_{\cN}^-$. Furthermore, as $\Uscoix(\Lx\LaWo(t)\Lx^{-1}) 
\Uscoix(\L\LaWo(t)\L^{-1})^* = \Uscoix(\L(\Id-\LaWo(t))\L^{-1}x)$, 
and $\Uscoix(x) = \Uscoix(x/n)^n$, $x \in \bR^4$, $n \in \bN$, we 
reduce the problem to showing that the set $E 
:=\{\sum_i\L_i(\Id-\LaWo(t_i))\L_i^{-1}x_i\;:\; (\L_i,x_i) \in \cN,\; 
t_i \in \bR\}$ is a neighbourhood of zero in $\bR^4$. Then with 
$e_\mu$, $\mu = 0,\dots,3$ the canonical basis of $\bR^4$ and $e_\pm 
:= e_1 \pm e_0$, it is easily verfied, by first choosing the $\L_i$ 
in the definition of $E$ to be $\Id$ and then to be a small rotation 
around the $e_3$ axis, that $s e_\a \in E$ for $\abs{s}$ sufficiently 
small and $\a = +,-,2,3$, so that $E$ contains a neighbourhood of 0.
\end{proof}

\begin{Lemma}\label{lem:ozkms}
The state $\ooi = \angscalar{\Ooi}{(\cdot)\Ooi}$ is a $(-2\pi)$-KMS 
state for the C$^*$-dynamical system $(\fFtoix(W), \aoix_{\L_W})$.
\end{Lemma}

The proof is completely analogous to the one of the first part of 
Lemma 6.2 in~\cite{BV1}.
For any finite set of spacelike cones $C_1,\dots,C_n$, we introduce 
the C$^*$-algebra $\FtzCC$ as the one generated by the algebras 
$\fFtoix(C_1),\dots,\fFtoix(C_n)$. We also define $\GzCC$ to be the 
set of operators $G \in \FtzCC$ for which there exists a 
neighbourhood $\cN$ of the identity in $\rPport$ such that 
$\aoix_s(G) \in \FtzCC$ for any $s \in \cN$. It is clear that $\GzCC$ 
is a $*$-algebra and that for any $n$-tuple 
$\tilde{C}_1,\dots,\tilde{C}_n$ with $\tilde{C}_i \Subset C_i$, 
$i=1,\dots,n$, $\fFtoix(\tilde{C}_1,\dots,\tilde{C}_n) \cont \GzCC$.

\begin{Lemma}\label{lem:GzCCorthogonal}
Let $W$ be a wedge in Minkowski space and let $C_i \Subset W$, 
$i=1,\dots,n$ be spacelike cones. If $\Phi \in (\GzCC \Ooi)^\perp$, 
then
\begin{equation}
\angscalar{\Phi}{\aoix_{s_1}(G_1)\dots\aoix_{s_m}(G_m)\Ooi}=0
\end{equation}
for any $s_i \in \rPport$, $G_i \in \GzCC$, $i=1,\dots,m$.
\end{Lemma} 

\begin{proof}
We begin by showing that $\Phi \in (\GzCC \Ooi)^\perp$ implies 
\begin{equation*}
\Uscoix(s)\Phi \in (\GzCC \Ooi)^\perp, \qquad s \in \rPport.
\end{equation*} 
Let $\cN$ be a neighbourhood of the identity in $\rPport$ such that 
$\cN^{-1}\cdot C_i \subset W$, $i=1,\dots,n$, and let $G \in \GzCC$. 
Then there exists $\eps > 0$, depending on $s\in\cN$, such that 
$\angscalar{\Phi}{\aoix_{s\L_W(t)s^{-1}}(G)\Ooi}=0$ for $\abs{t}<\eps$. 
But by the above lemma
$t \mapsto \aoix_{s\L_W(t)s^{-1}}(G)\Ooi$ has an analytic 
continuation to a function on the strip $\{ -\pi<\Imag z < 0 \}$, so 
that
$\Uscoix(s \L_W(t)s^{-1})\Phi \in (\GzCC \Ooi)^\perp$ for any $t \in 
\bR$, $s \in \cN$. Then, iterating the argument, 
and using lemma~\ref{lem:poincareboost},
$\Uscoix(s)\Phi \in(\GzCC\Ooi)^\perp$ for each $s \in \rPport$.

If we now show that, for any $G_i \in \GzCC$, $s_i \in \rPport$, 
$i=1,\dots,m$, we have $\aoix_{s_1}(G_1)\dots\aoix_{s_m}(G_m)\Phi \in 
(\GzCC\Ooi)^\perp$, since $\GzCC$ is a $*$-algebra containing the 
identity operator, the conclusion of the lemma will follow, but this 
is proven easily by induction, using the first part of the proof.
\end{proof}

\begin{proof}[Proof of Theorem~\ref{thm:reehschliederwedge}]
By normal commutation relations, it is sufficient to show that $\Ooi$ 
is cyclic for $\cFtoix(W)$, i.e. $(\cFtoix(W)\Ooi)^\perp = \{0\}$. 
Let then $\Phi \in (\cFtoix(W)\Ooi)^\perp$ and $F_i \in 
\fFtoix(\tilde{C}_i)$, $i=1,\dots,n$, be arbitrary operators. For any 
$i=1,\dots,n$ there exists $s_i \in \rPport$ and a spacelike cone 
$C_i$ such that $s_i^{-1}\cdot\tilde{C}_i \Subset C_i \Subset W$. 
Then $\aoix_{s_i^{-1}}(F_i) \in \GzCC$ and, being $\Phi \in 
(\GzCC\Ooi)^\perp$,
\begin{equation*}
\angscalar{\Phi}{F_1\dots F_n \Ooi} = 
\angscalar{\Phi}{\aoix_{s_1}(\aoix_{s_1^{-1}}(F_1))\dots\aoix_{s_n}(\aoix_{s_n^{-1}}(F_n))\Ooi} 
= 0
\end{equation*}
by lemma~\ref{lem:GzCCorthogonal}, thus $\Phi$ is orthogonal to a 
total set of vectors in $\Hoix$, and then vanishes.
\end{proof}

\end{document}